\documentclass[prd,superscriptaddress,notitlepage,showpacs,nofootinbib,longbibliography]{revtex4-1}

\usepackage{graphicx}
\usepackage[utf8]{inputenc} 
\usepackage{placeins}
\usepackage{amsmath}
\usepackage{amssymb}
\usepackage{float}
\usepackage{xcolor,color,colortbl}
\usepackage{comment}
\usepackage{multirow}
\usepackage{ulem}
\usepackage{soul}

\def\beq{\begin{equation}}
\def\eeq{\end{equation}}
\def\bea{\begin{eqnarray}}
\def\eea{\end{eqnarray}}

\usepackage{amsbsy}
\usepackage{bm}
\usepackage{color}
\usepackage{fancyhdr}
\usepackage[pdftex]{epsfig}
\usepackage[colorlinks=true,linkcolor=blue,citecolor=blue,urlcolor=magenta,filecolor=blue]{hyperref}
\def \Y{\mathcal{Y}}
\setstcolor{red}

\begin{document}

\bigskip

\vspace{2cm}

\title{Analysing the charged scalar boson contribution to the charged-current $B$ meson anomalies}

\vskip 6ex

%------------------------
\author{Jonathan Cardozo}
\email{jcardozo@ut.edu.co}
\affiliation{Departamento de F\'{i}sica, Universidad del Tolima, C\'{o}digo Postal 730006299, Ibagu\'{e}, Colombia}
%------------------------
%------------------------
\author{J. H. Mu\~{n}oz}
\email{jhmunoz@ut.edu.co}
\affiliation{Departamento de F\'{i}sica, Universidad del Tolima, C\'{o}digo Postal 730006299, Ibagu\'{e}, Colombia}

\author{N\'{e}stor Quintero}
\email[]{nestor.quintero01@usc.edu.co}
\thanks{(Corresponding author)}
\affiliation{Facultad de Ciencias B\'{a}sicas, Universidad Santiago de Cali, Campus Pampalinda, Calle 5 No. 62-00, C\'{o}digo Postal 76001, Santiago de Cali, Colombia}
%------------------------
\author{Eduardo Rojas}
\email{eduro4000@gmail.com}
\affiliation{Departamento de Física, Universidad de Nari\~no, A.A. 1175, San Juan de Pasto, Colombia}
%------------------------

\bigskip

%*****************************************
\begin{abstract}
Experimental measurements collected by the BABAR, Belle, and LHCb experiments on different observables associated with the semileptonic transition $b \to c \tau \bar{\nu}_\tau$, indicate the existence of disagreement respect with the Standard Model predictions.  We analyse the charged scalar boson contributions to these charged-current $B$ meson anomalies within the framework of two Higgs doublet model  with the most general Yukawa couplings to quarks and leptons from the third generation,  involving left-handed and right-handed (sterile) neutrinos. We perform a phenomenological study of the Yukawa couplings parameter space that accomodates these anomalies. We consider the most recent data from HFLAV world-average and Belle combination, and the upper limits ${\rm BR}(B_c^- \to \tau^- \bar{\nu}_{\tau}) < 30\%$ and $10\%$.  In addition, we include in our study the prospect measurements on $R(D^{(\ast)})$ that the Belle II experiment could achieve and explore, for the first time, the future implications for the corresponding charged scalar Yukawa couplings. This analysis updates the existing literature and includes new important observables. Our  results show that current experimental $b\rightarrow c \tau \bar{\nu}_\tau$ data and  Belle II projection favor the interpretation of a charged scalar boson interacting with right-handed neutrinos. Furthermore, as a side analysis regarding the charged scalar boson interpretation, we revisit the relation between $R(D^\ast)$ and ${\rm BR}(B_c^- \to \tau^- \bar{\nu}_{\tau})$ by investigating whether the claim that pseudoscalar new physics interpretations of $R(D^{\ast})$ are implausible due to the $B_c$ lifetime is still valid, to the light of the recent data and Belle II prospects on $R(D^{\ast})$. Lastly, we reexamine addressing the $R(D^{(\ast)})$ anomalies in the context of the 2HDM of Type II. We show that with the current Belle combined data is possible to obtain an available parameter space on the plane $(M_{H^{\pm}}, \tan\beta)$ for a simultaneous explanation of the anomalies, in consistency with $B \to \tau \bar{\nu}_\tau$ and bounds from inclusive radiative $B$ decays. Moreover, projections at the Belle II experiment suggest that the 2HDM of Type II would be no longer disfavored.
\end{abstract}
%*****************************************
\maketitle

%&&&&&&&&&&&&&&&&&&&&&&&&&&&&&&&&&&&&&&&&&&&&&&&&&&&&&&&&&&&&&&&&&&&&&&&
\section{Introduction}
%&&&&&&&&&&&&&&&&&&&&&&&&&&&&&&&&&&&&&&&&&&&&&&&&&&&&&&&&&&&&&&&&&&&&&&&

The most recent experimental information accumulated by the BABAR, Belle, and LHCb experiments on the measurements of the observables $R(D^{(*)}) = {\rm BR}(B \to D^{(*)}\tau \bar{\nu}_\tau)/{\rm BR}(B \to D^{(*)} \ell \bar{\nu}_\ell)$, with $\ell = \mu$ or $e$, $R(J/\psi) = {\rm BR}(B_c \to J/\psi \tau \bar{\nu}_\tau)/{\rm BR}(B_c \to J/\psi \mu \bar{\nu}_\mu)$, the $\tau$ polarization asymmetry $P_\tau(D^\ast)$ and the longitudinal polarization of the $D^*$ meson $F_L(D^\ast)$ related with the channel $\bar{B} \to D^\ast \tau \bar{\nu}_\tau$ have shown deviations from their corresponding Standard Model (SM) estimations~\cite{Lees:2012xj,Lees:2013uzd,Huschle:2015rga,Sato:2016svk,Hirose:2017vbz,Aaij:2015yra,Aaij:2017deq,Aaij:2017uff,Abdesselam:2019dgh,Belle:2019rba,Amhis:2019ckw,HFLAVsummer,
Aaij:2017tyk,Watanabe:2017mip,Hirose:2017dxl,Hirose:2016wfn,Tanaka:2012nw,Abdesselam:2019wbt,Alok:2016qyh}. In Table~\ref{Table:1} we summarize the current experimental measurements and the SM predictions for these observables generated by the charged-current transition $b \to c \tau \bar{\nu}_\tau$. The SM average values reported by HFLAV take into account the recent theoretical progress on the calculations of $R(D^{(\ast)})$~\cite{Bigi:2016mdz,Bernlochner:2017jka,Jaiswal:2017rve,Bigi:2017jbd}.\footnote{For other recent works, see, for instance~\cite{Gambino:2019sif,Jaiswal:2020wer}}
For completeness, the inclusive ratio $R(X_c)={\rm BR}(B \to X_c \tau \bar{\nu}_\tau)/{\rm BR}(B \to X_c \ell \bar{\nu}_\ell)$, which is induced via the same transition $b \to c  \tau \bar{\nu}_\tau$~\cite{Kamali:2018bdp}, is also collected in Table~\ref{Table:1}. We can see that although the $R(D^{(*)})$ discrepancies have decreased with the latest measurements of Belle, it is still interesting and worthwhile  to analyse them in light of future data at Belle II, where it is expected statistical and experimental improvements on the observables $R(D^{(*)})$~\cite{Kou:2018nap}. These charged-current $B$ meson anomalies pose an interesting challenge, at theoretical level, in order to propose  possible scenarios of physics beyond SM according to current and future experimental results. \\

Several model-independent analyses of new physics (NP) explanations regarding the most general dimension-six effective Lagrangian with the current $b \rightarrow c \tau \bar{\nu}_{\tau}$ data have been explored~\cite{Asadi:2019xrc,Murgui:2019czp,Mandal:2020htr,Cheung:2020sbq,Sahoo:2019hbu,Shi:2019gxi,
Bardhan:2019ljo,Blanke:2018yud,Blanke:2019qrx,Alok:2019uqc,Huang:2018nnq}. One of the possible NP scenarios that could address the aforementioned anomalies is to consider sizeable scalar couplings that arise from a charged scalar boson. This is the case of the well known two-Higgs doublet model (2HDM) which has been widely studied~\cite{Crivellin:2012ye,Crivellin:2013wna,Celis:2012dk,Celis:2016azn,Ko:2012sv,HernandezSanchez:2012eg,
Crivellin:2015hha,Cline:2015lqp,Enomoto:2015wbn,Dhargyal:2016eri,Martinez:2018ynq,Wang:2016ggf,
Chen:2017eby,Iguro:2018fni,Iguro:2017ysu,Arbey:2017gmh,Chen:2018hqy,Hagiwara:2014tsa,Lee:2017kbi,
Iguro:2018qzf,Li:2018rax}. In the situation of the 2HDM of type II, this model is not favored by the experimental results reported by the BABAR experiment in 2012 and 2013~\cite{Lees:2012xj,Lees:2013uzd}. However, subsequent analysis performed by the Belle Collaboration in 2015 and 2016~\cite{Huschle:2015rga,Sato:2016svk} showed compatibility with the 2HDM of Type II. In particular, in Ref. \cite{Hirose:2017vbz} was discussed the compatibility of the Belle results reported for $R(D^{(*)})$ in 2016 with the 2HDM of Type II and found that these results seem to favor the parametric space with small values of $\tan \beta/M_{H^{\pm}}$, where $\tan\beta$ and $M_{H^{\pm}}$ are the ratio of vacuum expectation values and the charged Higgs boson mass, respectively. Thus, the 2HDM of Type II was ruled out as an interpretation of the anomalies and different versions of the 2HDM were put forward in the literature, such as, type III (generic), type X (lepton-specific), flipped, and aligned, that, in general, can provide an explanation to the $R(D^{(*)})$ anomalies under certain phenomenological assumptions~\cite{Crivellin:2012ye,Crivellin:2013wna,Celis:2012dk,Celis:2016azn,Ko:2012sv,HernandezSanchez:2012eg,
Crivellin:2015hha,Cline:2015lqp,Enomoto:2015wbn,Dhargyal:2016eri,Martinez:2018ynq,Wang:2016ggf,
Chen:2017eby,Iguro:2018fni,Iguro:2017ysu,Arbey:2017gmh,Chen:2018hqy,Hagiwara:2014tsa,Lee:2017kbi,Iguro:2018qzf,Li:2018rax}. Furthermore, the parametric space of charged scalar boson interpretations has to confront strong constraints from the upper limits on the branching ratio of the tauonic $B_c$ decay, ${\rm BR}(B_c^{-} \to \tau^{-} \bar{\nu}_\tau) \lesssim 30 \%$ and $10\%$, which are imposed from the lifetime of $B_c$ meson~\cite{Alonso:2016oyd} and the LEP data taken at the $Z$ peak~\cite{Akeroyd:2017mhr}, respectively.\\

%%%%%%%%%%%%%%%%%%%%%%%%%%%%%%%%%%%%%%%%%%%%%%%%%%%%%%%%
\begin{table}[!t]
\centering
\renewcommand{\arraystretch}{1.2}
\renewcommand{\arrayrulewidth}{0.8pt}
\begin{tabular}{ccc}
\hline
Observable & Expt. measurement & SM prediction \\
\hline
$R(D)$ & $0.307 \pm 0.037 \pm 0.016$ \ Belle-2019~\cite{Belle:2019rba} & 0.299 $\pm$ 0.003~\cite{Amhis:2019ckw,HFLAVsummer} \\
       & $0.326 \pm 0.034$ \ Belle combination~\cite{Belle:2019rba} & \\
       & $0.340 \pm 0.027 \pm 0.013$ \ HFLAV~\cite{Amhis:2019ckw} & \\
$R(D^\ast)$ & $0.283 \pm 0.018 \pm 0.014$ \ Belle-2019~\cite{Belle:2019rba} & 0.258 $\pm$ 0.005~\cite{Amhis:2019ckw,HFLAVsummer} \\
        & $0.283 \pm 0.018$ \ Belle combination~\cite{Belle:2019rba} & \\
        & $0.295 \pm 0.011 \pm 0.008$ HFLAV~\cite{Amhis:2019ckw} & \\
$R(J/\psi)$ & $0.71 \pm 0.17 \pm 0.18$~\cite{Aaij:2017tyk} & 0.283 $\pm$ 0.048~\cite{Watanabe:2017mip}  \\
$P_\tau(D^\ast)$ & $- 0.38 \pm 0.51 ^{+0.21}_{-0.16}$~\cite{Hirose:2017dxl,Hirose:2016wfn} &  $-0.497 \pm 0.013$~\cite{Tanaka:2012nw} \\
$F_L(D^\ast)$ & $0.60 \pm 0.08 \pm 0.035$~\cite{Abdesselam:2019wbt} & $0.46 \pm 0.04$~\cite{Alok:2016qyh}  \\
$R(X_c)$ & 0.223 $\pm$ 0.030~\cite{Kamali:2018bdp} & 0.216 $\pm$ 0.003~\cite{Kamali:2018bdp} \\
\hline
\end{tabular} \label{Table:1}
\caption{\small Experimental status and SM predictions on observables related to the charged transition $b \to c \tau \bar{\nu}_\tau$.}
\end{table}
%%%%%%%%%%%%%%%%%%%%%%%%%%%%%%%%%%%%%%%%%%%%%%%%%%%%%%%%

Another perspective to explain the charged-current $B$ meson anomalies is to assume that NP might be connected with right-handed neutrinos.  In this direction, several authors have incorporated a right-handed neutrino in the most general effective Hamiltonian for the $b \to c \tau \bar{\nu}_\tau$  transition with different mediators (a colorless charged scalar boson, a heavy charged vector boson or a leptoquark)  with the purpose of exploring scenarios of NP that could explain some observables related with this transition~\cite{Shi:2019gxi,Iguro:2018qzf,Iguro:2018vqb,Asadi:2018wea,Asadi:2018sym,Ligeti:2016npd,
Robinson:2018gza,Mandal:2020htr,Greljo:2018ogz,Azatov:2018kzb,Heeck:2018ntp,Babu:2018vrl,Bardhan:2019ljo,
He:2017bft,Gomez:2019xfw,Alguero:2020ukk,Dutta:2013qaa,Dutta:2017xmj,Dutta:2017wpq,Dutta:2018jxz}. With the assumption of a sterile right-handed neutrino with small mass, as a singlet of the gauge group of the SM, there is no interference between contributions of left-handed and right-handed neutrinos, so the branching ratio of $b \to c \tau \nu$ is given by an incoherent sum of these contributions:  ${\rm BR}(b \to c \tau \bar{\nu}_\tau) = {\rm BR}(b \to c \tau \bar{\nu}_L) + {\rm BR}(b \to c \tau \bar{\nu}_R) $. The majority of references that consider contributions from a colorless scalar boson and right-handed neutrinos   to the observables related to the $b \to c \tau \nu$ transition, studied the parametric space of Wilson coefficients.  However, an analysis of the parametric space conformed by the Yukawa couplings including recent measurements as polarizations of $D^*$ and the tau lepton, and the $R(J/\psi)$ observable  is missing in the literature.  This analysis is important because the Yukawa couplings  can be  related,  directly, with specific models and give information about   the maximum values for  these  couplings.\\

In this work, we perform a systematic and general  model-independent study about the impact of charged scalar contributions to the charged current transition $b \rightarrow c \tau \bar{\nu}_{\tau}$ including light right-handed neutrinos, focusing on the Yukawa couplings. In our study we  consider the observables $R(D^{(*)})$, $R(J/\psi)$, $P_{\tau}(D^*)$, $F_L(D^*)$, the inclusive $R(X_c)$, and constraints derived from the branching ratio  of  $B_c \rightarrow  \tau \bar{\nu}_\tau$.  In our work  we include  additional elements as the projected Belle II sensitivities and a complete analysis of the parametric space of the Yukawa couplings, which is not straightforward because there is no a trivial relation among the Wilson coefficients and the Yukawa couplings.  In particular, the future measurements at the ongoing Belle II experiment are a matter of importance to confirm or refute the tantalizing NP hints. \\

Recently, the impact of the mentioned scalar contributions with right-handed neutrinos on several observables related to  the $b \to c \tau \bar{\nu}_\tau$ transition was investigated in Refs. ~\cite{Mandal:2020htr,Iguro:2018qzf,Asadi:2018sym,Robinson:2018gza}. However, the analysis of Ref. ~\cite{Mandal:2020htr} was performed on the parametric space of the Wilson coefficients without including the future measurements that Belle II could achieve and the $R(J/\psi)$ observable.  The authors of Ref. ~\cite{Asadi:2018sym} considered the projected experimental results from Belle II but their analysis was also developed on the parametric space of the Wilson coefficients. The research in Ref. ~\cite{Robinson:2018gza} was  also  done on the Wilson coefficients without including the polarizations of $D^*$ and the tau lepton,  and $R(J/\psi)$. On the other hand, the study of Ref. ~\cite{Iguro:2018qzf} was carried out on the parametric space of the Yukawa couplings but it did not incorporate the forthcoming measurements of Belle II, the polarizations of $D^*$ and the tau lepton, and the $R(J/\psi)$ observable. Therefore, our work complements and extends the previous investigations performed in Refs.~\cite{Mandal:2020htr,Iguro:2018qzf,Asadi:2018sym,Robinson:2018gza} related with the effect of charged scalar contributions with right-handed neutrinos on the $b \to c \tau \nu$ transition.\\

 For completeness, we further scrutinize two important topics related with the charged scalar boson interpretation to the $R(D^{(\ast)})$ anomalies: (1) We reexamine the relation between the observable $R(D^*)$ and ${\rm BR}(B_c \to \tau \bar{\nu}_\tau)$, reported in Refs.~\cite{Alonso:2016oyd,Akeroyd:2017mhr},  in light of Belle combination~\cite{Belle:2019rba} and LHCb~\cite{Aaij:2017deq} experimental results, and the projection at Belle II experiment with an integrated luminosity of 50 $\rm ab^{-1}$ \cite{Kou:2018nap}. This analysis is very important because the constraint on ${\rm BR}(B_c \to \tau\bar{\nu}_\tau)$  affects substantially the contributions from scalar operators~\cite{Bardhan:2019ljo,Blanke:2018yud,Blanke:2019qrx}; (2) We also reanalyze the parametric space $(M_{H^{\pm}},\tan\beta)$ in order to determine if the  2HDM of Type II is still disfavored to explain the $R(D^{(\ast)})$ discrepancies considering the recent experimental results of Belle~\cite{Belle:2019rba} and the projected Belle II experiment~\cite{Kou:2018nap}.\\

This paper is organized as follows. In Sec.~\ref{model}, we present the expressions for the observables  $R(D^{(*)})$,    $R(J/\psi)$,  $P_\tau(D^\ast)$, $F_L(D^\ast)$, $R(X_c)$ and $BR(B_c \to \tau \nu)$, in terms of the Wilson coefficients associated to  scalar contributions considering left and right-handed neutrinos. In Sec.~\ref{analysis}, we perform a phenomenological analysis  on the parametric space of  Wilson coefficients and Yukawa couplings, considering the recent experimental results of Belle and the future projection at the Belle II experiment, to determine possible regions where scalar contributions with right-handed neutrinos could explain the charged $B$-meson anomalies. In Sec.~\ref{relation}, we dig into the relation between the $R(D^{*})$ anomaly and the ${\rm BR}(B_c \to \tau \bar{\nu}_\tau)$ considering constrains of $30\%$ and $10\%$ and the recent $b \to c \tau \bar{\nu}_{\tau}$ data and projected results at Belle II. In Sec.~\ref{2HDM-II}, we reanalyze the old discussion if the  2HDM of Type II is still rule out as an interpretation to the $R(D^{(*)})$ anomalies. Our main  conclusions are given in Sec.\ref{Conclusion}. Finally, in the appendix we perform a detailed phenomenological study of a general 2HDM in order to calculate  the effective Yukawa couplings and the corresponding Wilson coefficients.

%&&&&&&&&&&&&&&&&&&&&&&&&&&&&&&&&&&&&&&&&&&&&&&&&&&&&&&&&&&&&&&&&&&&&&&&
\section{Scalar contributions to the $b \to c \tau \bar{\nu}_{\tau}$ transition} 
\label{model}
%&&&&&&&&&&&&&&&&&&&&&&&&&&&&&&&&&&&&&&&&&&&&&&&&&&&&&&&&&&&&&&&&&&&&&&&

The effective Hamiltonian for the charged-current transition $b \to c \tau \bar{\nu}_{\tau}$ that includes all the four-fermion scalar operators, considering both left- and right-handed neutrinos, has the following form~\cite{Iguro:2018vqb,Asadi:2018wea,Asadi:2018sym,Ligeti:2016npd,Asadi:2018wea,Robinson:2018gza,Mandal:2020htr}
\bea\label{Leff_Total}
\mathcal{H}_{\rm eff}(b \to c \tau \bar{\nu}_{\tau}) &=& \frac{4 G_F}{\sqrt{2}} V_{cb}^{\rm CKM}	 \Big[(\bar{c} \gamma_\mu P_L b) (\bar{\tau} \gamma^\mu P_L \nu_{\tau}) + C_{S}^{LL}(\bar{c} P_L b) (\bar{\tau} P_L \nu_{\tau}) + C_{S}^{RL}(\bar{c} P_R b) (\bar{\tau} P_L \nu_{\tau}) \nonumber \\
&& + C_{S}^{LR}(\bar{c} P_L b) (\bar{\tau} P_R \nu_{\tau}) + C_{S}^{RR}(\bar{c} P_R b) (\bar{\tau} P_R \nu_{\tau})\Big]+ h.c., 
\eea
\noindent where $P_{R,L} = (1\pm \gamma_5)/2$, $G_F$ is the Fermi coupling constant and $V_{cb}^{\rm CKM}$ is the charm-bottom Cabbibo-Kobayashi-Maskawa (CKM) matrix element. The first term corresponds to the SM contribution from a virtual $W$ boson exchange, while the remaining four terms correspond to the all possible charged scalar contributions. The information of these NP operators is codify through the scalar Wilson coefficients (WCs) $C_{S}^{XY}$, where the first index $X= L, R$ represents the quark-current quirality projection, while the second one $Y= L,R$ is related with the leptonic-current quirality projection. Thus, Eq.~\eqref{Leff_Total} contains all of the dimension-six scalar operators involving both left-handed (LH) and right-handed (RH) neutrinos. We will assume that NP effects are only present in the third generation of leptons ($\tau, \nu_\tau$). This assumption is motivated by the absence of deviations from the SM for light lepton modes $\ell= e$ or $\mu$.

The ratios $R(M)$ ($M=D,D^\ast,J/\psi$), and the $D^\ast$ and $\tau$ longitudinal polarizations can be written in terms of the scalar WCs $C_{S}^{LL}, C_{S}^{RL}, C_{S}^{LR},$ and $C_{S}^{RR}$~\cite{Iguro:2018vqb,Asadi:2018wea,Asadi:2018sym}. The numerical expressions for these contributions are~\cite{Iguro:2018vqb,Asadi:2018wea,Asadi:2018sym}: 
\begin{eqnarray}
R(D) &=& R(D)_{\rm SM} \Big[1+ 1.49 \ \text{Re}(C_{S}^{RL}+ C_{S}^{LL})^\ast + 1.02 \big( |C_{S}^{RL}+ C_{S}^{LL}|^2 + |C_{S}^{LR}+ C_{S}^{RR}|^2 \big)\Big],  \label{RD_TypeII} \\ 
R(D^*) &=&   R(D^*)_{\rm SM} \Big[1+ 0.11 \ \text{Re}(C_{S}^{RL}- C_{S}^{LL})^\ast + 0.04 \big( |C_{S}^{RL}- C_{S}^{LL}|^2 + |C_{S}^{LR}- C_{S}^{RR}|^2 \big) \Big], \label{RDstar_TypeII}  \\
R(J/\psi) &=&  R(J/\psi)_{\text{SM}} \Big[1+ 0.12 \ \text{Re}(C_{S}^{RL}- C_{S}^{LL})^\ast + 0.04 \big( |C_{S}^{RL}-C_{S}^{LL}|^2 + |C_{S}^{LR}- C_{S}^{RR}|^2 \big)  \Big], \label{RJPSI_TypeII} \\
F_L(D^*) &=&  F_L(D^*)_{\rm SM} \ r_{D^\ast}^{-1} \Big[1+ 0.24 \ \text{Re}(C_{S}^{RL}- C_{S}^{LL})^\ast + 0.08 \big( |C_{S}^{RL}- C_{S}^{LL}|^2 + |C_{S}^{LR}- C_{S}^{RR}|^2 \big)\Big],\label{FLD_RPV} \\
P_\tau(D^*) &=&  P_\tau(D^*)_{\rm SM} \ r_{D^\ast}^{-1} \Big[1 - 0.22 \ \text{Re}(C_{S}^{RL}- C_{S}^{LL})^\ast - 0.07 \big( |C_{S}^{RL}- C_{S}^{LL}|^2 + |C_{S}^{LR}- C_{S}^{RR}|^2\big)\Big],
\label{PTAU_RPV}
\end{eqnarray}

\noindent with $r_{D^\ast} = R(D^*) / R(D^*)_{\rm SM}$. The numerical formula for $R(J/\psi)$ has been obtained by using the analytic expressions and form factors given in Ref.~\cite{Watanabe:2017mip}. Similarly, the tauonic decay $B_c^- \to \tau^- \bar{\nu}_{\tau}$ is also modified as follows~\cite{Iguro:2018vqb,Asadi:2018wea,Asadi:2018sym}
\begin{eqnarray} \label{BRBc_TypeII}
{\rm BR}(B_c^- \to \tau^- \bar{\nu}_{\tau}) &=& {\rm BR}(B_c^- \to \tau^- \bar{\nu}_{\tau})_{\text{SM}} \ \bigg[ \bigg| 1 + \frac{m_{B_c}^2}{m_\tau(m_b+m_c)} (C_{S}^{RL}- C_{S}^{LL})\bigg|^2 +  \bigg|\frac{m_{B_c}^2}{m_\tau(m_b+m_c)} (C_{S}^{LR}- C_{S}^{RR})\bigg|^2\bigg], \nonumber \\ 
\end{eqnarray}
\noindent where $m_{B_c}^2/m_\tau(m_b+m_c) = 4.33$. Finally, the ratio $R(X_c)$ of inclusive semileptonic $B$ decays in the $1S$ scheme can be written as~\cite{Kamali:2018bdp}
\begin{eqnarray}
R(X_c) &=& R(X_c)_{\rm SM} \Big[1+ 0.096 \ \text{Re}(C_{S}^{RL}- C_{S}^{LL})^\ast + 0.493 \ \text{Re}(C_{S}^{RL}+C_{S}^{LL})^\ast \nonumber \\
&& + 0.031 \big( |C_{S}^{RL} - C_{S}^{LL}|^2 + |C_{S}^{LR}- C_{S}^{RR}|^2\big) + 0.327 \big( |C_{S}^{RL}+C_{S}^{LL}|^2 + |C_{S}^{LR}+C_{S}^{RR}|^2\big) \Big] . \label{RXc}  
\end{eqnarray}

\noindent This formula was obtained by following the trick described in Ref.~\cite{Asadi:2018wea}, in which the contributions of scalar Wilson coefficients with RH neutrinos are calculated by using parity transformation ($L \leftrightarrows R$). Without loss of generality, in the following, we will restrict our analysis to the case of real scalar WCs\footnote{In Ref.~\cite{Murgui:2019czp} has been shown that the case of complex WCs do not provide an improvement in the description of the data.}.

%\noindent In the next section we will pay attention to these Wilson coefficients $C_V^{LL}$, $C_V^{RL}$, $C_V^{LR}$, and $C_V^{RR}$ given in terms effective couplings $\epsilon_{cb}^{L,R}$ and $\epsilon_{\tau\nu_\tau}^{L,R}$ and the ${W^\prime}$ boson mass, that can provide an explanation to the $b \to c \tau \bar{\nu}_{\tau}$ anomalies.

%*****************************
\section{Phenomenological analysis} \label{analysis}
%*****************************

\subsection{$b \to c \tau \bar{\nu}_{\tau}$ observables}
%--------------------------------------------------------------------------------------

Before to addressing the charged-current $B$ meson anomalies in terms of a charged scalar boson, it is necessary to discuss the present-day experimental measurements on the ratios $R(D)$ and $R(D^*)$ (see Table~\ref{Table:1}). The most recent world-average values reported by the Heavy Flavor Averaging Group (HFLAV) on the measurements of $R(D)$ and $R(D^*)$~\cite{Amhis:2019ckw,HFLAVsummer} exceed the SM predictions by 1.4$\sigma$ and 2.5$\sigma$, respectively. These averages include the preliminary Belle result presented at Moriond EW 2019~\cite{Abdesselam:2019dgh}. Later on, Belle reported their combined averages on $R(D^{(\ast)})$ which are in accordance with the SM within 0.8$\sigma$ and 1.4$\sigma$, respectively~\cite{Belle:2019rba}. Additionally, polarization observables associated with the channel $B \to D^\ast \tau \bar{\nu}_\tau$ have been observed in the Belle experiment, namely, the $\tau$ lepton polarization $P_\tau(D^\ast)$~\cite{Hirose:2017dxl,Hirose:2016wfn} and the $D^\ast$ longitudinal polarization $F_L(D^\ast)$~\cite{Abdesselam:2019wbt}. Thus, it is important to recognize the great Belle efforts during the last years to improve not only the measurements on $R(D^{(\ast)})$, but also to provide new $b \to c \tau \bar{\nu}_{\tau}$ observables. Given this current experimental situation, we will consider in our analysis two different sets of observables, namely

\begin{itemize}
    \item Set 1 (S1): $R(D^{(\ast)})$ HFLAV, $R(J/\psi), F_L(D^*), P_\tau(D^*), R(X_c)$,
    \item Set 2 (S2): $R(D^{(\ast)})$ Belle combination, $R(J/\psi), F_L(D^*), P_\tau(D^*), R(X_c)$,
\end{itemize}

\noindent where the corresponding theoretical and experimental values are given in Table~\ref{Table:1}. The purpose of these two sets is to observe the significance of the recent HFLAV world-average~\cite{Amhis:2019ckw,HFLAVsummer} and Belle combination data~\cite{Belle:2019rba} independently, as well as to provide a robust analysis by regarding the available experimental information on all of the charged transition $b \to c \tau \bar{\nu}_\tau$ observables, namely the ratios $R(J/\psi)$, $R(X_c)$, and the polarizations $P_\tau(D^\ast), F_L(D^\ast)$ reported by Belle~\cite{Hirose:2017dxl,Hirose:2016wfn,Abdesselam:2019wbt}. Moreover, we will consider into the analysis  the upper bounds ${\rm BR}(B_c^- \to \tau^- \bar{\nu}_{\tau}) < 30\%$~\cite{Alonso:2016oyd} and $10\%$~\cite{Akeroyd:2017mhr} to put constraints on the scalar NP scenarios. Keeping this in mind, we determine the regions in the parameter space favored by the experimental data for the set of observables S1 and S2. 

\subsection{Projected Belle II scenarios}
%--------------------------------------------------------------

Within the physics program of the Belle II experiment~\cite{Kou:2018nap} is expected that improvements at the level of $\sim 3\%$ and  $\sim 2\%$ will be achieved, for the statistical and systematic uncertainties of $R(D)$ and $R(D^{\ast})$, respectively. Taking into account in our analysis the projected uncertainties on $R(D^{(\ast)})$ when an integrated luminosity of $50 \ {\rm ab}^{-1}$ data will be accumulated~\cite{Kou:2018nap} and assuming the current Belle combination $R(D)$-$R(D^\ast)$ correlation ($\rho=-0.47$~\cite{Belle:2019rba}), we also examine the prospects of Belle II by considering two benchmark projected scenarios, i.e., plausible scenarios within the reach and capability of Belle II. These two scenarios are: 
\begin{enumerate}
\item \textbf{Belle II-P1}: Belle II measurements on $R(D^{(\ast)})$ keep the central values of Belle combination averages with the projected Belle II sensitivities for $50 \ {\rm ab}^{-1}$.  
\item \textbf{Belle II-P2}: Belle II measurements on $R(D^{(\ast)})$ are in agreement with the current SM predictions at the $0.1\sigma$ level with the projected Belle II sensitivities for $50 \ {\rm ab}^{-1}$.
\end{enumerate}

\noindent Similar scenarios were previously considered in Ref.~\cite{Asadi:2018sym}, but considering the HFLAV 2018 world-average values. Here, we complement this analysis by considering the most recent Belle data~\cite{Belle:2019rba}. Furthermore, we will study the phenomenological consequences of the Belle II experiment in the parametric space associated with the charged scalar Yukawa couplings. Such implications were not explored in the analysis of Ref.~\cite{Asadi:2018sym}, neither in other recent works.

\subsection{Fit procedure}
%-----------------------------------------

We carry out a standard $\chi^2$ analysis with the above-mentioned $\mathcal{O}_i$ observables associated with the transition $b \to c \tau\bar{\nu}_\tau$. The $\chi^2$ function is written as~\cite{PDG2020}
\begin{equation}
\chi^2 (C_S^{XY}) = \sum_{i,j}^{N_{\rm obs}} [\mathcal{O}^\text{exp}_i-\mathcal{O}^\text{th}_i(C_S^{XY})] \mathcal{C}_{ij}^{-1} [\mathcal{O}^\text{exp}_j-\mathcal{O}^\text{th}_j(C_S^{XY})],
\end{equation}

\noindent where $N_{\rm obs}$ is the number of observables, $\mathcal{O}^\text{exp}_i$ are the experimental measurements, and $\mathcal{O}^\text{th}_i$ are the theoretical observables, Eqs.~\eqref{RD_TypeII}-\eqref{PTAU_RPV} and~\eqref{RXc}, which are function of the scalar WCs $C_{S}^{XY}$ ($XY=LL,RL,LR,RR$). The covariance matrix $\mathcal{C}$ is the sum of the experimental and theoretical uncertainties, and includes the experimental correlation between $R(D)$ and $R(D^\ast)$. We will use in our analysis the correlation values $-0.38$ and $-0.47$, from HFLAV~\cite{Amhis:2019ckw,HFLAVsummer} and Belle combination~\cite{Belle:2019rba}, respectively.

We first get the minimum of the $\chi^2$ function ($\chi^2_{\rm min}$), and then we use it to assessing the $p$-value as a measured of goodness-of-fit. The $p$-value allow us to quantify the level of agreement between the data and the NP scenarios hypothesis~\cite{PDG2020}. The $p$-values are obtained as one minus cumulative function distribution for a certain number of degrees of freedom ($N_{\rm dof}$)~\cite{Shi:2019gxi,Blanke:2018yud,Blanke:2019qrx,PDG2020}. $N_{\rm dof}$ is equal to $N_{\rm dof} = N_{\rm obs} - N_{\rm par}$, where $N_{\rm par}$ is the number of parameters to be fitted. In our analysis we have $N_{\rm obs}=6$ for both set of observables S1 and S2. In addition, we also calculate the pull of the SM ($\rm{pull_{SM}}$) defined as the $p$-value corresponding to $\chi^2_{\rm SM} - \chi^2_{\rm min}$, with $\chi^2_{\rm SM} = \chi^2(0)$, and converted into an equivalent significance in units of standard deviation ($\sigma$)~\cite{Shi:2019gxi,Blanke:2018yud,Blanke:2019qrx,PDG2020}.

%%%%%%%%%%%%%%%%%%%%%%%%%%%%%%%%
\begin{table}[!t]
\centering
\renewcommand{\arraystretch}{1.2}
\renewcommand{\arrayrulewidth}{0.8pt}
\begin{tabular}{ccccc}
\hline
\multicolumn{5}{c}{Set S1 ($\chi_{\rm SM}^2 = 19.1$, $p$-value$_{\rm SM} =7.6 \times 10^{-4}$)} \\
\hline
Two scalar WCs & BFP & $p$-value (\%) & $\rm{pull_{SM}}$ & $1\sigma$ intervals \\                                                                                                                             
\hline
$(C^{LL}_S,C^{RL}_S)$ & (-1.28,-0.25) & 48.5 & 2.69 & $C^{LL}_S \in [-1.33,-1.22]$ \ $C^{RL}_S \in [-0.30,-0.18]$  \\
$(C^{LL}_S,C^{LR}_S)$  & (-0.91,-0.78) & 46.8 & 2.68 & $C^{LL}_S \in [-1.04,-0.68]$ \ $C^{LR}_S \in [-0.83,-0.71]$  \\
$(C^{LL}_S,C^{RR}_S)$  & (-0.91,-0.78) & 46.8 & 2.68 & $C^{LL}_S \in [-1.04,-0.68]$ \ $C^{RR}_S \in [-0.83,-0.71]$  \\
$(C^{LR}_S,C^{RR}_S)$  & (1.15,-0.82) & 44.1 & 2.64 & $C^{LR}_S \in [1.02,1.25]$ \ $C^{RR}_S \in [-1.04,-0.70]$  \\ 
\hline
\multicolumn{5}{c}{Set S2 ($\chi_{\rm SM}^2 = 11.2$, $p$-value$_{\rm SM} =2.4 \times 10^{-2}$)} \\
\hline
Two scalar WCs & BFP & $p$-value (\%) & $\rm{pull_{SM}}$ & $1\sigma$ intervals \\                                                                                                                             
\hline
$(C^{LL}_S,C^{RL}_S)$ & (-1.22,-0.21) & 50.3 & 1.31 & $C^{LL}_S \in [-1.29,-1.14]$ \ $C^{RL}_S \in [-0.30,-0.11]$  \\
$(C^{LL}_S,C^{LR}_S)$  & (-0.92,-0.68) & 49.3 & 1.30 & $C^{LL}_S \in [-1.07,-0.55]$ \ $C^{LR}_S \in [-0.75,-0.59]$  \\
$(C^{LL}_S,C^{RR}_S)$  & (-0.92,-0.68) & 49.3 & 1.30 & $C^{LL}_S \in [-1.07,-0.55]$ \ $C^{RR}_S \in [-0.75,-0.59]$  \\
$(C^{LR}_S,C^{RR}_S)$  & (0.95,-0.95) & 45.7 & 1.24 & $C^{LR}_S \in [0.63,1.18]$ \ $C^{RR}_S \in [-1.18,-0.63]$  \\
\hline
\end{tabular}
\caption{Best-fit point (BFP) values, $p$-value, $\rm{pull_{SM}}$, and $1\sigma$ allowed intervals by allowing two scalar WCs different from zero to fit the set of observables S1 and S2.} \label{fit}
\end{table} 
%%%%%%%%%%%%%%%%%%%%%%%%%%%%%%%%

%%%%%%%%%%%%%%%%%%%%%%%%%%%%%%%%
\begin{table}[!t]
\centering
\renewcommand{\arraystretch}{1.2}
\renewcommand{\arrayrulewidth}{0.8pt}
\begin{tabular}{cc}
\hline
\multicolumn{2}{c}{Belle II-P1} \\
\hline
Two scalar WCs & $1\sigma$ intervals \\                                                                                                                             
\hline
$(C^{LL}_S,C^{RL}_S)$ &  $C^{LL}_S \in [-1.11,-1.06]$ \ $C^{RL}_S \in [-0.41,-0.36]$  \\
$(C^{LL}_S,C^{LR}_S)$ &  $C^{LL}_S \in [-0.69,-0.49]$ \ $C^{LR}_S \in [-0.75,-0.71]$  \\
$(C^{LL}_S,C^{RR}_S)$ &  $C^{LL}_S \in [-0.69,-0.49]$ \ $C^{RR}_S \in  [-0.75,-0.71]$  \\
$(C^{LR}_S,C^{RR}_S)$ &  $C^{LR}_S \in [0.69,0.93]$ \ $C^{RR}_S \in [-0.84,-0.61]$  \\
\hline
\multicolumn{2}{c}{Belle II-P2} \\
\hline
Two scalar WCs & $1\sigma$ intervals \\                                                                                                                             
\hline
$(C^{LL}_S,C^{RL}_S)$ &  $C^{LL}_S \in [-0.77,-0.72]$ \ $C^{RL}_S \in [-0.73,-0.69]$  \\
$(C^{LL}_S,C^{LR}_S)$ &  $C^{LL}_S \in [-0.04,8\times 10^{-4}]$ \ $C^{LR}_S \in [-0.26,-0.05]$  \\
$(C^{LL}_S,C^{RR}_S)$ &  $C^{LL}_S \in [-0.02,8\times 10^{-4}]$ \ $C^{RR}_S \in [-0.26,-0.05]$  \\
$(C^{LR}_S,C^{RR}_S)$ &  $C^{LR}_S \in [-5\times 10^{-3},0.38]$ \ $C^{RR}_S \in [-0.32,0.06]$  \\
\hline
\end{tabular} 
\caption{Projections Belle II-P1 and Belle-P2 of the $1\sigma$ allowed intervals for two scalar WCs different from zero.}\label{BelleII}
\end{table}

\subsection{Analysis on the parametric space of the scalar Wilson coefficients} 
\label{WC_analysis}
%----------------------------------------

For completeness of our analysis, we first begin studying the allowed scalar WCs parametric space. In this direction, similar recent analyses have been performed by considering the most recent data~\cite{Asadi:2019xrc,Murgui:2019czp,Mandal:2020htr,Cheung:2020sbq,Sahoo:2019hbu,Shi:2019gxi,
Bardhan:2019ljo,Blanke:2018yud,Blanke:2019qrx,Alok:2019uqc,Huang:2018nnq}. However, in contrast to these previous works, we consider the future implications that could be obtained from the Belle II experiment. Thus, our study provides complementary information to the ones discussed in Refs.~\cite{Asadi:2019xrc,Murgui:2019czp,Mandal:2020htr,Cheung:2020sbq,Sahoo:2019hbu,Shi:2019gxi,
Bardhan:2019ljo,Blanke:2018yud,Blanke:2019qrx,Alok:2019uqc,Huang:2018nnq}. \\

 We fit the set of observables S1 and S2 by allowing two scalar WCs different from zero (and setting the others two equal to zero).  Depending on the choices for the chiral charges,  there are three different scenarios, namely, operators with only LH neutrinos ($C_S^{LL},C_S^{RL}$), mixed operators with LH $+$ RH neutrinos ($C_S^{LL},C_S^{LR}$) and ($C_S^{LL},C_S^{RR}$), and operators with only RH neutrinos ($C_S^{LR},C_S^{RR}$).
\noindent In Table~\ref{fit} we report our results of the best-fit point (BFP) values, $p$-value, $\rm{pull_{SM}}$, and $1\sigma$ allowed intervals. In these two scalar WCs scenarios $N_{\rm par}=2$, thus $N_{\rm dof}=4$. For the SM we obtained $\chi^2_{\rm SM} = 19.1$ ($11.2$) for the set S1 (S2), corresponding to a $p$-value$_{\rm SM} =7.6 \times 10^{-4}$ ($2.4 \times 10^{-2}$). The largest $p$-value is obtained for the benchmark scenario $(C_S^{LL},C_S^{RL})$, however, scenarios with RH neutrinos have also a favorable $p$-value. In general, these scenarios provide good quality to adjust the experimental $b \to c \tau\bar{\nu}_\tau$ anomalies. We have checked that smaller $p$-values of the order $\sim 24 \%$, $\sim 33 \%$ and $\sim 18 \%$ are obtained for the cases of one, three or four non-zero scalar WCs, respectively, and they do not provide good fits of the data. Furthermore, we show in Table~\ref{BelleII} the Belle II-P1 and Belle-P2 projections of the $1\sigma$ allowed intervals for two scalar WCs scenarios. The Belle II-P1 projection would still allow sizeable couplings, thus, leaving room for significant NP contributions. While for Belle II-P2, these scenarios would be, in general, strongly constrained.

\begin{figure*}[!t]
\centering
\includegraphics[scale=0.27]{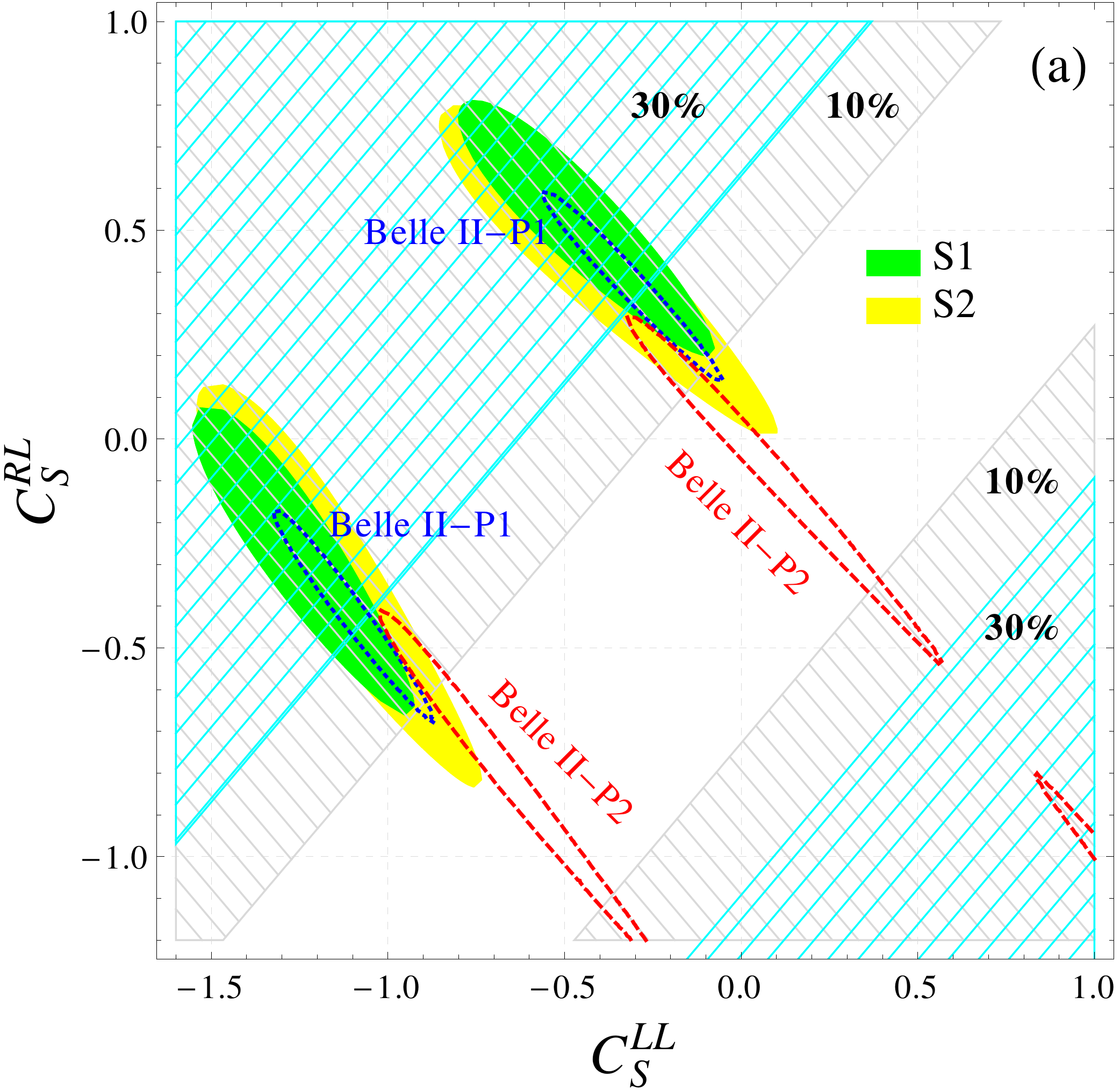} \
\includegraphics[scale=0.27]{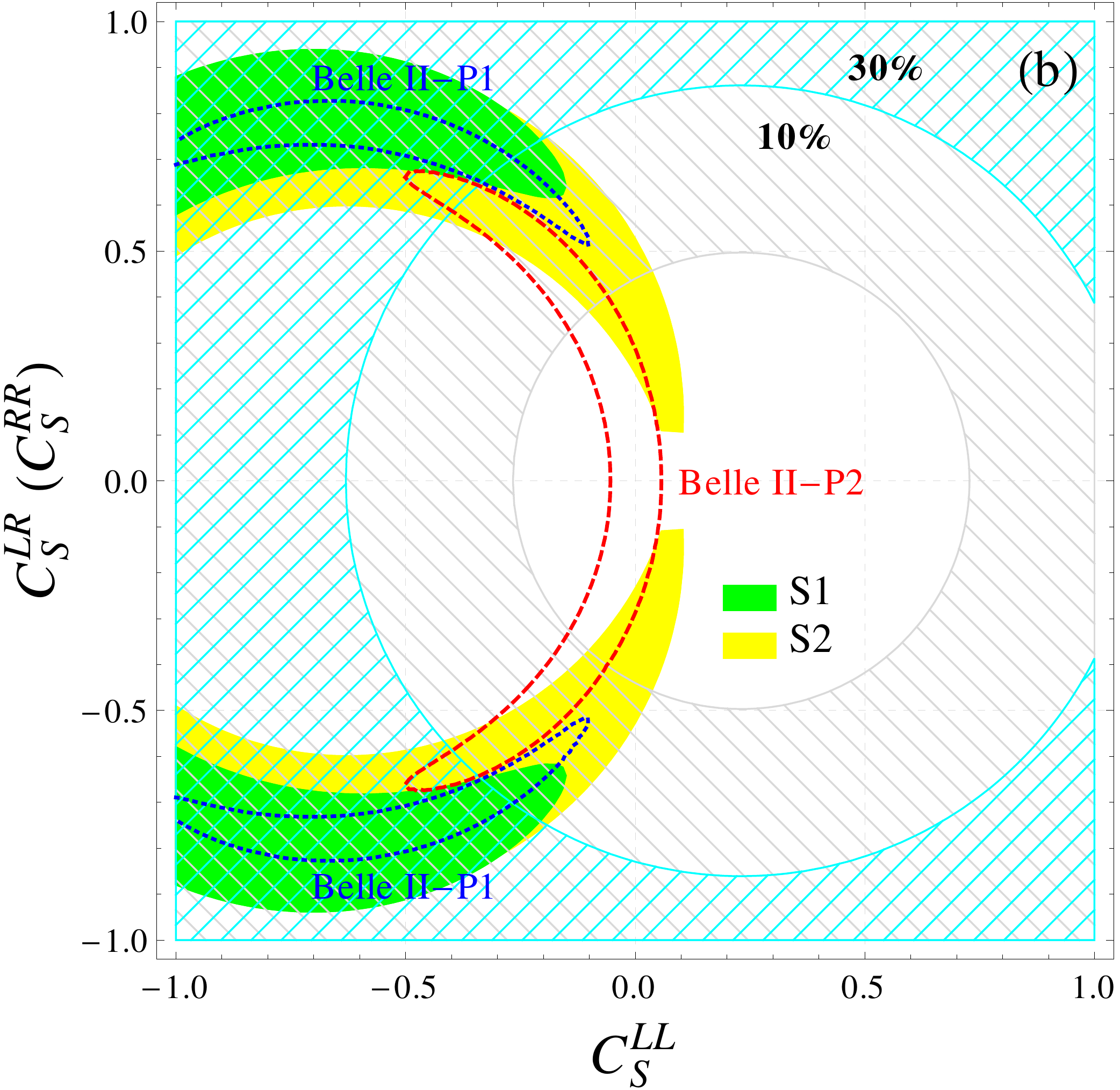} \
\includegraphics[scale=0.27]{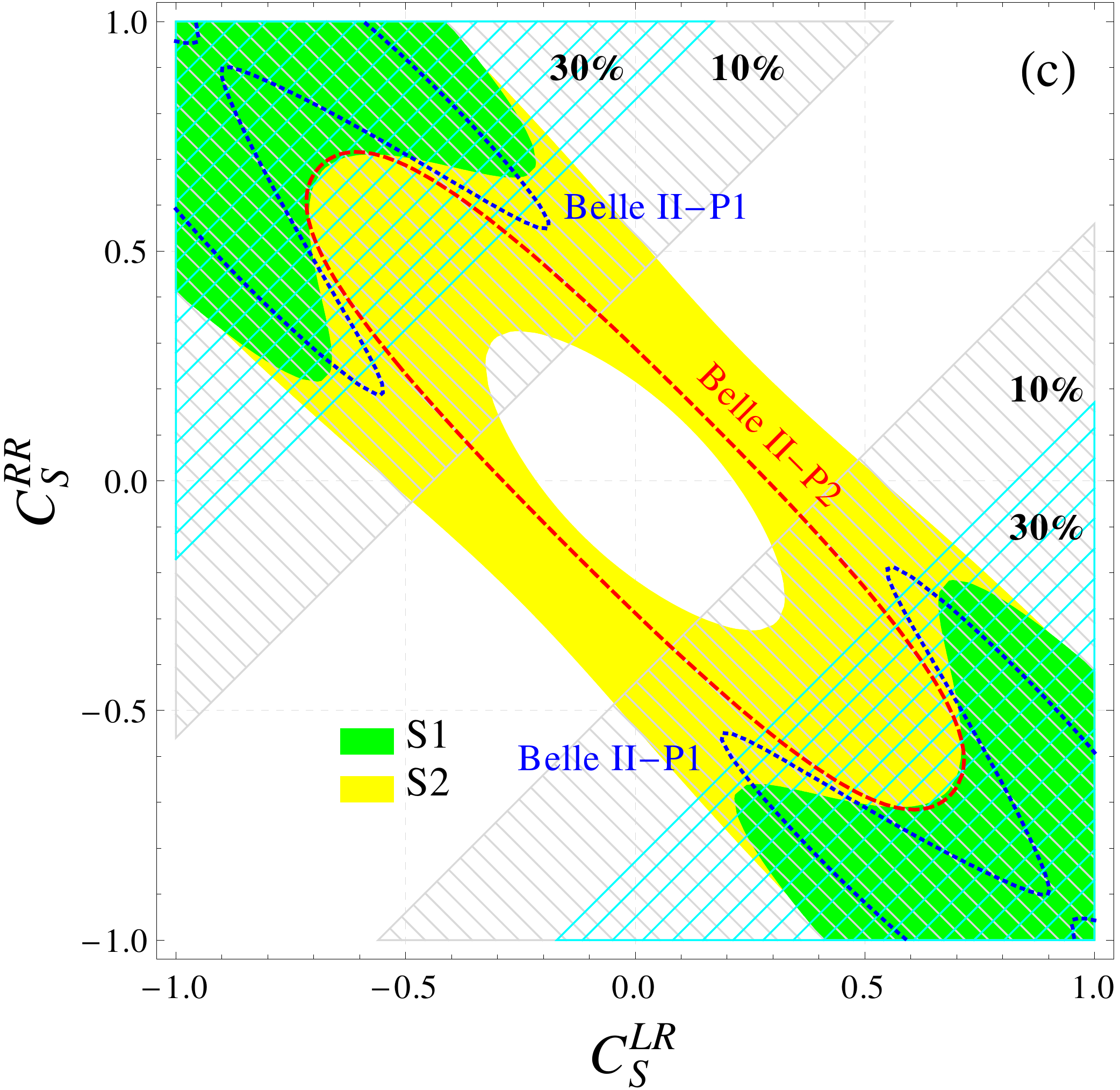}
\caption{\small The 95\% C.L. allowed parameter space for the set of observables S1 [green region] and S2 [yellow region] in the planes: (a) ($C_S^{LL},C_S^{RL}$), (b) ($C_S^{LL},C_S^{LR}$) or ($C_S^{LL},C_S^{RR}$), and (c) ($C_S^{LR},C_S^{RR}$). The cyan and gray hatched regions represent the excluded regions by the 30\% and 10\% upper limits on ${\rm BR}(B_c \to \tau \bar{\nu}_{\tau})$, respectively. The projections Belle II-P1 and Belle II-P2 for an integrated luminosity of $50 \ {\rm ab}^{-1}$ are illustrated by the blue dotted and red dashed contour lines, respectively (See the text for details).}
\label{ParameterSpace1}
\end{figure*}

%None of the previous studies have included the Belle II prospects in their fits.
%Nuevos resultados de Belle requieren un revisón del análisis y su proyección en Belle II necesitan
To further discussion, we plot in Fig.~\ref{ParameterSpace1} the 95\% confidence level (C.L.) allowed parameter space in the planes: (a) ($C_S^{LL},C_S^{RL}$), (b) ($C_S^{LL},C_S^{LR}$) or ($C_S^{LL},C_S^{RR}$), and (c) ($C_S^{LR},C_S^{RR}$). The green and yellow regions are obtained by considering the set of observables S1 and S2, respectively. The cyan (gray) hatched region shows the disallowed parameter space by ${\rm BR}(B_c^- \to \tau^- \bar{\nu}_{\tau}) < 30\%$ ($10\%$). The projections Belle II-P1 and Belle II-P2 for an integrated luminosity of $50 \ {\rm ab}^{-1}$ are illustrated by the blue dotted and red dashed contour lines, respectively. 
In all of the two WCs scenarios considered, the allowed regions by the set of observables S1 (dominated by $R(D^{(\ast)})$ HFLAV) are ruled out by ${\rm BR}(B_c^- \to \tau^- \bar{\nu}_{\tau}) < 30\%$ and $10\%$. This is in agreement with recent analyses~\cite{Murgui:2019czp,Mandal:2020htr,Shi:2019gxi,Blanke:2018yud,Blanke:2019qrx}. On the contrary, it is observed that there are small allowed regions by the set of observables S2 without violating ${\rm BR}(B_c^- \to \tau^- \bar{\nu}_{\tau}) < 30\%$ and $10\%$. On the other hand, regarding the Belle II experiment, the available regions from projection Belle II-P1 indicate that these WCs scenarios would be excluded. In contrast, the projection Belle II-P2 would provide stronger constraints than ${\rm BR}(B_c^- \to \tau^- \bar{\nu}_{\tau}) < 30\%$ and $10\%$, but still allowing a small window for NP contributions. Let us notice that projection Belle II-P2 would cover a similar region to the one set from the analysis of the mono-tau signature $pp \to \tau_h X + \rm{MET}$ at the LHC, as well as the projected sensitivity at the high-luminosity LHC~\cite{Shi:2019gxi}.

%This result is in disagreement with the very recent analysis of Ref.~\cite{Mandal:2020htr}, in which this RH neutrino scalar interpretation cannot be satisfied by $R(D^\ast)$ and $F_L(D^\ast)$. It is worth to notice that $N_{\rm obs}$ and $N_{\rm dof}$ considered in our numerical analysis are different to the one performed in Ref.~\cite{Mandal:2020htr}.

%still leaves room for significant new physics (BSM) contributions to this ratio

%%%%%%%%%%%%%%%%%%%%%%%%%%%%%%%%
%%%%%%%%%%%%%%%%%%%%%%%%%%%%%%%%

\subsection{Analysis on the parametric space of Yukawa couplings}
%----------------------------------------

In this section we are going to discuss  the main purpose of this work. We explore the implications on the parametric space associated with the charged Higgs Yukawa couplings to the quarks and leptons in the generic 2HDM that can accomodate the charged-current $B$ meson anomalies. In the literature, 2HDMs with a more generic flavor structure have been extensively explored as an explanation to the $R(D^{(\ast)})$ discrepancies~\cite{Crivellin:2012ye,Crivellin:2013wna,Celis:2012dk,Celis:2016azn,Ko:2012sv,HernandezSanchez:2012eg,
Crivellin:2015hha,Cline:2015lqp,Enomoto:2015wbn,Dhargyal:2016eri,Martinez:2018ynq,Wang:2016ggf,
Chen:2017eby,Iguro:2018fni,Iguro:2017ysu,Arbey:2017gmh,Chen:2018hqy,Hagiwara:2014tsa,Lee:2017kbi}. In all these works, the neutrinos have been considered to be LH. Only in Refs.~\cite{Cline:2015lqp,Iguro:2018qzf,Li:2018rax}, a generic 2HDM involving RH neutrinos has been studied to address the anomalies. Most of these models were implemented by considering the 2018 HFLAV averages and, in addition, none of them include the experimental measurements of polarizations of the $D^*$ meson and the $\tau$ lepton, the $R(J/\psi)$ observable, the latest Belle results and the  projected  measurements at the ongoing Belle II experiment. In the following analysis all these ingredients are taken into account. \\

We study the scenarios for a charged scalar boson with general Yukawa couplings involving LH and RH neutrinos. By ``general" we will refer to Yukawa couplings without additional assumptions, such as Cheng-Sher ansatz (see, for instance~\cite{Wang:2016ggf,Chen:2017eby}). In order to provide an explanation to the $b \to c \tau \bar{\nu}_{\tau}$ anomalies, we will adopt the phenomenological assumption in which the charged scalar boson ($H^\pm$) couples only to the bottom-charm quarks and the third generation of leptons $\tau$-$\nu_{\tau}$, i.e., the corresponding Yukawa couplings are differente from zero, while the other ones are taken to be zero. Therefore, NP effects are negligible for light lepton modes ($e$ or $\mu$). In addition, we consider these Yukawa couplings as real (charge-parity conserving) arbitrary free parameters.

The most general  Lagrangian for the $b\rightarrow c \tau \bar{\nu}_\tau$  transition induced by the Yukawa couplings of a charged scalar boson $H^\pm$ is given by (see Eq.~\eqref{EqApp})
\begin{align}  \label{L_Yukawa}
\mathcal{L}_{H^\pm}(b\rightarrow c \tau \bar{\nu}_\tau)  
                = &- H^+\left( \bar{c}     X^{D}_{c b}         P_R b
                    -            \bar{c}     X^{U*}_{bc}         P_L b 
                    +            \bar{\nu}_\tau   X^{E}_{\nu_\tau \tau}    P_R \tau                    
                    -            \bar{\nu}_\tau   X^{N*}_{\tau\nu_\tau} P_L \tau\right)\notag\\
                  &-H^-\left(   \bar{b}    X^{D*}_{cb}         P_L c
                                  -\bar{b}   X^{U}_{bc}          P_R c
                                +\bar{\tau}  X^{E*}_{\nu_\tau \tau}   P_L \nu_\tau 
                                -\bar{\tau}  X^{N}_{\tau\nu_\tau }  P_R \nu_\tau 
                                 \right)  ,
\end{align}

\noindent where $X^{U}_{bc}$, $X^{D}_{cb}$, $X^{E}_{\nu_\tau \tau}$, and $X^{N}_{\tau\nu_\tau}$ are the Yukawa couplings to the up-quarks, down-quarks, charged leptons and neutrinos, respectively, with $X_{gh}^{f}= (X_{hg}^{f})^*$. We will use the shorthand notation $X^{E}_{\tau} \equiv X^{E}_{\nu_\tau \tau}$ and  $X^{N}_{\tau} \equiv X^{N}_{\tau \nu_\tau }$. In particular, we want to emphasize that the fourth and eighth terms in the previous expression correspond to the neutrino Yukawa coupling $X^{N}_{\tau}$, which describes the interaction between the RH neutrino and a charged scalar boson. While the third and sixth terms correspond to the charged lepton (electron-like) Yukawa coupling  $X^{E}_{\tau}$ that usually appears for LH neutrinos. In Appendix~\ref{AppA}, we provide details on the derivation of the Yukawa Lagrangian, Eq.~\eqref{L_Yukawa}, within a general 2HDM with the inclusion of RH neutrinos. To avoid dangerous tree-level flavor-changing neutral currents, we will impose the aligned condition on the down-type quarks~\cite{Li:2018rax}. 

After integrating out $H^\pm$, the scalar WCs from the effective four-fermion Lagrangian, Eq.~\eqref{Leff_Total}, are written as~(see appendix~\ref{sec:appendixB})
\begin{eqnarray}
C^{LL}_S &=&+\frac{\sqrt{2}}{4G_FV^{\rm CKM}_{cb}}  
\frac{\left(X^{U*}_{bc}\right)\left(X^{E*}_{\tau}\right)}{M_{H^{\pm}}^2} , \\
C^{RL}_S &=& -\frac{\sqrt{2}}{4G_FV^{\rm CKM}_{cb}} 
\frac{\left(X^{D}_{cb}\right)\left(X^{E*}_{\tau}\right)}{M_{H^{\pm}}^2}, \\
C^{LR}_S &=&-\frac{\sqrt{2}}{4G_FV^{\rm CKM}_{cb}}   
\frac{\left(X^{U*}_{bc}\right)\left(X^{N}_{\tau}\right)}{M_{H^{\pm}}^2}, \\
 C^{RR}_S &=& +\frac{\sqrt{2}}{4G_FV^{\rm CKM}_{cb}}  
 \frac{\left(X^{D}_{cb}\right)\left(X^{N}_{\tau}\right)}{M_{H^{\pm}}^2},
\end{eqnarray}

\noindent with $M_{H^\pm}$ being the $H^\pm$ charged scalar boson mass. Thus, the $b\rightarrow c \tau \bar{\nu}_\tau$ observables, Eqs.~\eqref{RD_TypeII} to~\eqref{RXc}, can be expressed in terms of the effective scalar and pseudoscalar contributions, namely 
\begin{eqnarray}
C^{RL}_S \pm C^{LL}_S &=&\frac{\sqrt{2}}{4G_FV^{\rm CKM}_{cb}}  
\frac{\left(\pm X^{U*}_{bc} - X^{D}_{cb} \right)\left(X^{E*}_{\tau}\right)}{M_{H^{\pm}}^2} , \\
C^{LR}_S \pm C^{RR}_S &=&\frac{\sqrt{2}}{4G_FV^{\rm CKM}_{cb}}   
\frac{\left(-X^{U*}_{bc} \pm X^{D}_{cb} \right)\left(X^{N}_{\tau}\right)}{M_{H^{\pm}}^2}, 
\end{eqnarray}

\noindent for LH and RH neutrinos, respectively. In the most general case, three Yukawa couplings are always involved in a non trivial way. Keeping this in mind, we perform a $\chi^2$ analysis by allowing three Yukawa couplings different from zero, i.e., $(X^{U}_{bc},X^{D}_{cb},X^{E}_{\tau})$ for LH neutrinos scenarios and $(X^{U}_{bc},X^{D}_{cb},X^{N}_{\tau})$ for RH neutrinos scenarios, respectively. We found that simplified scenarios regarding two Yukawa couplings (up- or down-quark and lepton or neutrino) cannot simultaneously accomodate the $R(D)$ and $R(D^\ast)$ data (even relaxing the uncertainties at the $2\sigma$ level) yielding to $p$-values $\lesssim 8 \%$.

%%%%%%%%%%%%%%%%%%%%%%%%%%%%%%%%
\begin{table}[!t]
\centering
\renewcommand{\arraystretch}{1.2}
\renewcommand{\arrayrulewidth}{0.8pt}
\begin{tabular}{cccc} 
\hline\hline
\multicolumn{4}{c}{Set S1 ($\chi_{\rm SM}^2 = 19.1$, $p$-value$_{\rm SM} =7.6 \times 10^{-4}$)} \\
\hline
Yukawa couplings & BFP & $p$-value (\%) & $\rm{pull_{SM}}$ \\                                                                                                                             
\hline
$(X^{U}_{bc},X^{D}_{cb},X^{E}_{\tau})$ &$ (0.33,0.38,-0.47)$ & 28.8 & 2.95   \\
$(X^{U}_{bc},X^{D}_{cb},X^{N}_{\tau})$  & $(-0.35,-0.25,-1.09)$ & 28.9 & 2.95  \\
\hline
\multicolumn{4}{c}{Set S2 ($\chi_{\rm SM}^2 = 11.2$, $p$-value$_{\rm SM} =2.4 \times 10^{-2}$)} \\
\hline
Yukawa couplings & BFP & $p$-value (\%) & $\rm{pull_{SM}}$ \\                                                                                                                             
\hline
$(X^{U}_{bc},X^{D}_{cb},X^{E}_{\tau})$ & $(0.23,0.25,-0.64)$ & 28.4 & 1.56   \\
$(X^{U}_{bc},X^{D}_{cb},X^{N}_{\tau})$  & $(-0.38,0.31,-0.90)$ & 27.7 & 1.54   \\
\hline\hline
\end{tabular}
\caption{BFP values, $p$-value, and $\rm{pull_{SM}}$ by allowing three Yukawa couplings different from zero to fit the set of observables S1 and S2.}
\label{Yukawafit}
\end{table} 
%%%%%%%%%%%%%%%%%%%%%%%%%%%%%%%%

\begin{figure*}[!t]
\centering 
\includegraphics[scale=0.3]{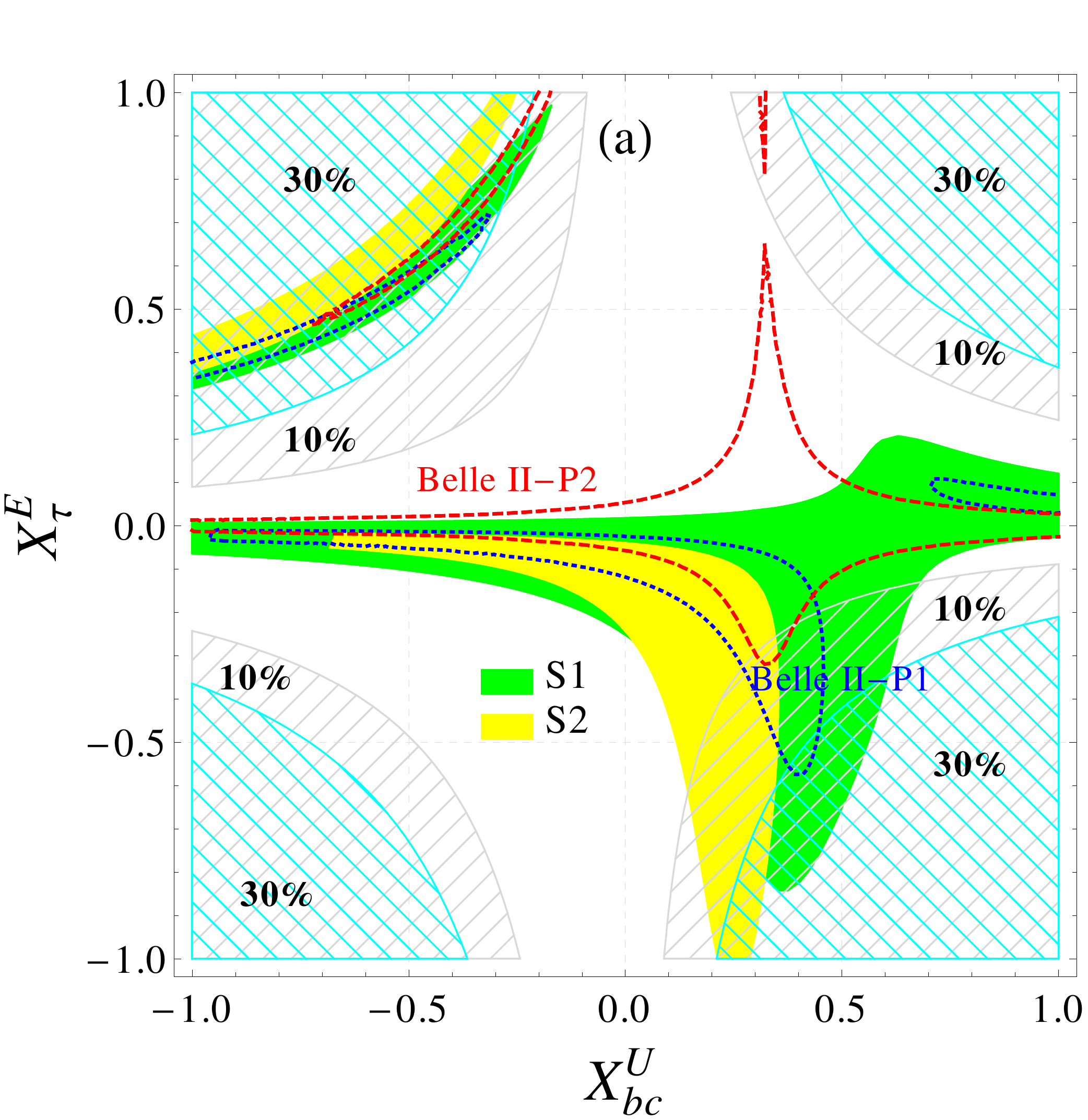} \
\includegraphics[scale=0.3]{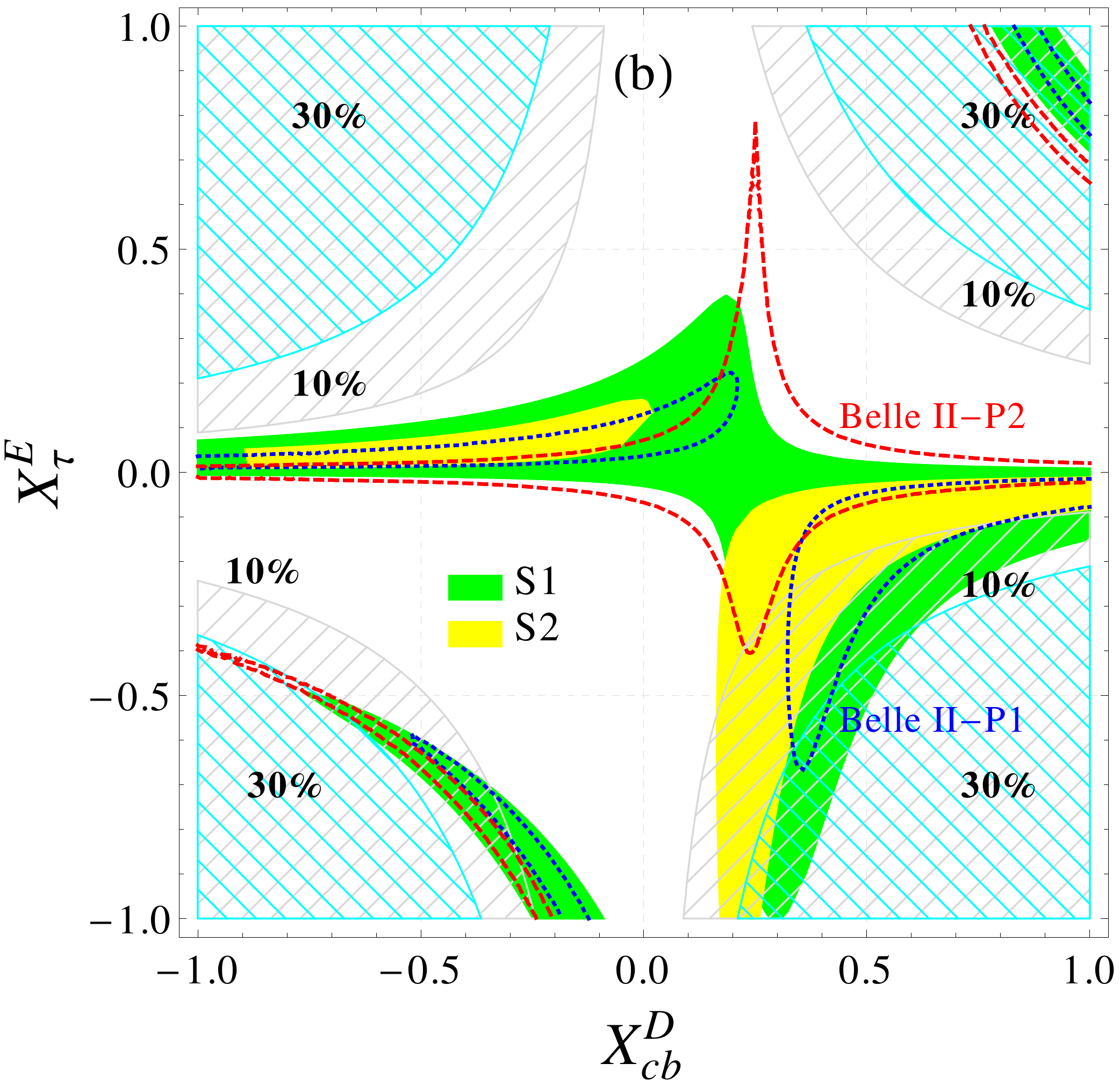} \medskip

\includegraphics[scale=0.3]{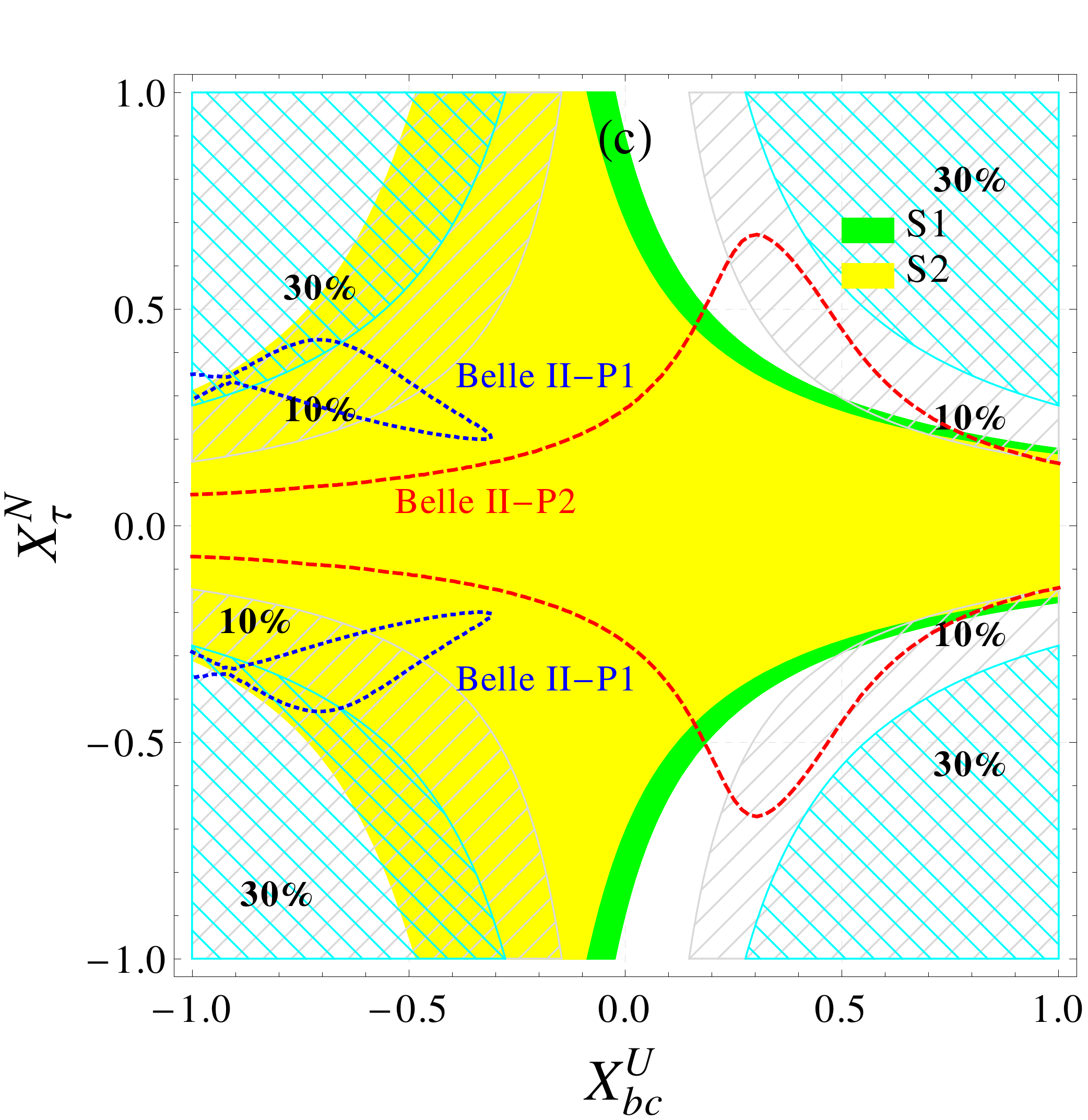} \
\includegraphics[scale=0.3]{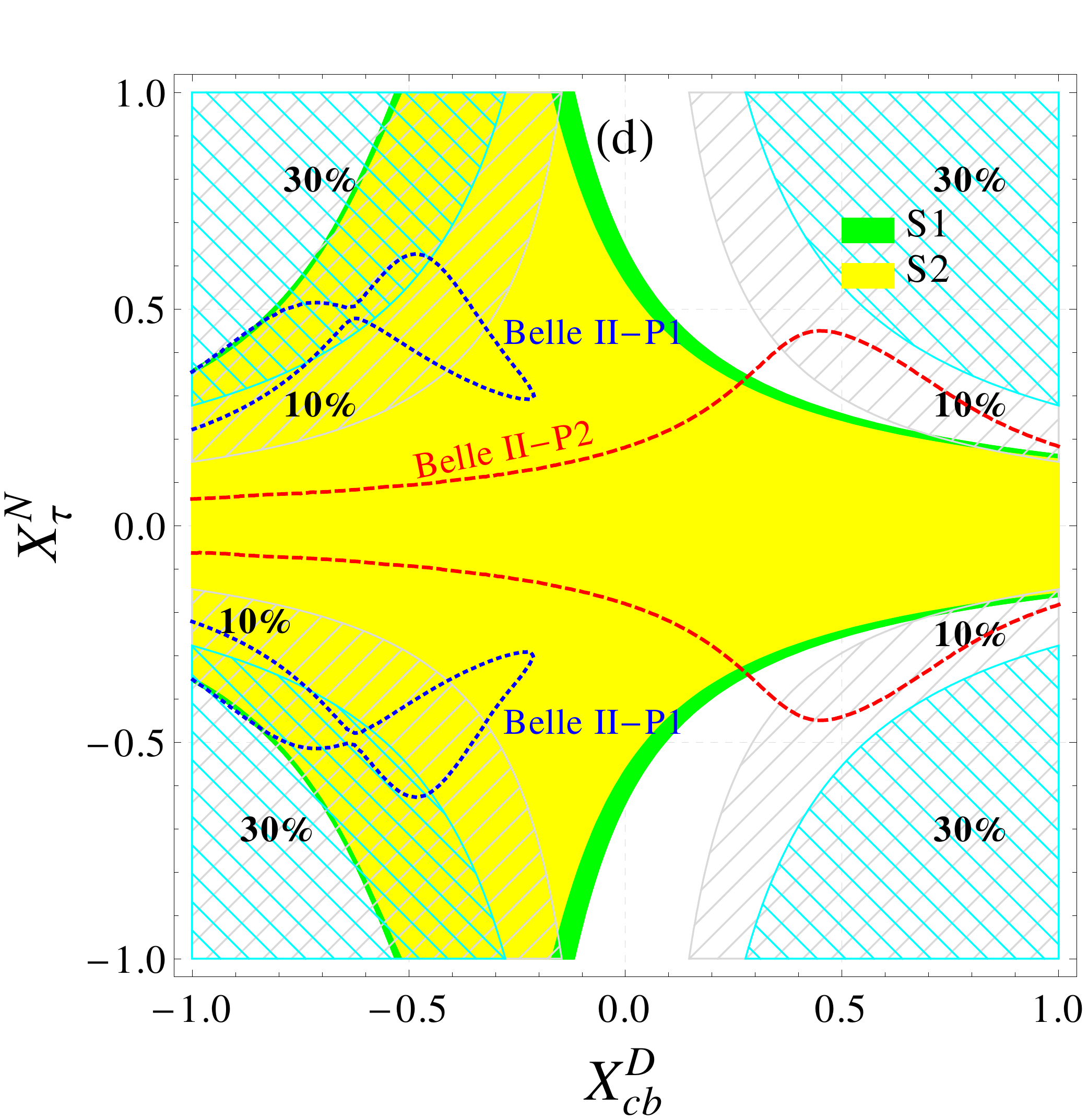}
\caption{\small The 95\% C.L. allowed two dimensional parameter space for the sets S1 [green region] and S2 [yellow region] in the Yukawa coupling planes: (a) ($X^U_{bc},X^E_{\tau}$), (b) ($X^U_{bc},X^N_{\tau}$), (c) ($X^D_{cb},X^E_{\tau}$) and (d) ($X^D_{cb},X^N_{\tau}$), for a charged Higgs mass of $M_{H^\pm} = 500$~GeV. The hatched regions in gray and cyan represent the excluded regions by the 10\% and 30\% upper limits on ${\rm BR}(B_c \to \tau \bar{\nu}_{\tau})$, respectively. The projections Belle II-P1 and Belle II-P2 for an integrated luminosity of $50 \ {\rm ab}^{-1}$~\cite{Kou:2018nap} are illustrated by the blue dotted and red dashed contour lines, respectively.  In each case, we vary the Yukawa couplings on the plane while keeping the remaining one at the BFP.}
\label{ParameterSpace2}
\end{figure*}

We display in Table~\ref{Yukawafit} the BFP values, $p$-value, and $\rm{pull_{SM}}$ by allowing three Yukawa couplings different from zero to fit the set of observables S1 and S2. Particularly, we observe that the neutrino Yukawa couplings must be as large as one ($|X^N_{\tau}| \sim 1$) to reproduce the $b\rightarrow c \tau \bar{\nu}_\tau$ data. Furthermore, the 95\% C.L. allowed two dimensional parameter space in the Yukawa couplings planes: (a) ($X^U_{bc},X^E_{\tau}$), (b) ($X^D_{cb},X^E_{\tau}$), (c) ($X^U_{bc},X^N_{\tau}$) and (d) ($X^D_{cb},X^N_{\tau}$), are shown in Fig.~\ref{ParameterSpace2}. The green and yellow regions represent the allowed parameter space that simultaneously can accommodate the set of observables S1 and S2, respectively. These plots have been obtained for a representative charged Higgs mass of $M_{H^\pm} = 500$~GeV, which corresponds to a benchmark mass value usually considered in the literature~\cite{Chen:2017eby,Iguro:2017ysu,Li:2018rax}. In each case, we vary the Yukawa couplings on the plane while keeping the remaining one at the BFP. The hatched regions in gray and cyan refer to the excluded regions by the 10\% and 30\% upper limits on ${\rm BR}(B_c \to \tau \bar{\nu}_{\tau})$, respectively. The prospects Belle II-P1 and Belle II-P2 for an integrated luminosity of $50 \ {\rm ab}^{-1}$~\cite{Kou:2018nap} are illustrated by the blue dotted and red dashed contour lines, respectively. According to our analysis, we get the following remarks:
\begin{enumerate}
\item For the Yukawa couplings planes ($X^U_{bc},X^E_{\tau}$) and ($X^D_{cb},X^E_{\tau}$) related with LH neutrinos, panels~\ref{ParameterSpace2}(a) and~\ref{ParameterSpace2}(b), Yukawa couplings regions with absolute values of the order $\mathcal{O}(10^{-1})$ are allowed by the set of observables S1 and S2, as well as by the prospects Belle II-P1 and Belle II-P2.
\item As concerns with the neutrino Yukawa couplings planes ($X^U_{bc},X^N_{\tau}$) and ($X^D_{cb},X^N_{\tau}$) (RH neutrino solutions), panels~\ref{ParameterSpace2}(c) and~\ref{ParameterSpace2}(d), both for set of observables S1 and S2 there is a wide allowed region Yukawa couplings with absolute values of the order $\mathcal{O}(10^{-1})$, that would be reduced by the projection Belle II-P2. The case of projection Belle II-P1 would be almost excluded because of bounds ${\rm BR}(B_c^- \to \tau^- \bar{\nu}_{\tau}) < 30\%$ and $10\%$. Our results imply that current experimental $b\rightarrow c \tau \bar{\nu}_\tau$ data favors the interpretation of a charged scalar boson with right-handed neutrinos.
\end{enumerate}

 %%%%%%%%%%%%%%%%%%%%%%%%%%%%%%%%
%%%%%%%%%%%%%%%%%%%%%%%%%%%%%%%%

%*****************************************
\section{Digging into the relation between $R(D^\ast)$ and ${\rm BR}(B_c^- \to \tau^- \bar{\nu}_{\tau})$} \label{relation}
%*****************************************

The importance of the relation between $R(D^\ast)$ and ${\rm BR}(B_c^- \to \tau^- \bar{\nu}_{\tau})$ was first pointed out in Ref.~\cite{Alonso:2016oyd}, where an upper limit of ${\rm BR}(B_c^{-} \to \tau^{-} \bar{\nu}_\tau) \leq 30 \%$ is imposed by considering the lifetime of $B_c$ meson. This bound puts strong constraints on the pseudoscalar interpretation to $R(D^{\ast})$ generated by the effective coupling $\epsilon_P = C_S^{RL} - C_S^{LL}$~\cite{Alonso:2016oyd}. Later on, a stronger bound of ${\rm BR}(B_c^{-} \to \tau^{-} \bar{\nu}_\tau) \leq 10 \%$ was obtained in Ref.~\cite{Akeroyd:2017mhr} from the LEP data taken at the $Z$ peak. Recently, these limits have been critically examined and relaxed bounds of $\leq 39 \%$~\cite{Bardhan:2019ljo}  and $\leq 60 \%$~\cite{Blanke:2018yud,Blanke:2019qrx} have been obtained.
In the following, for completeness, we revisit whether the claim that pseudoscalar NP interpretations of $R(D^\ast)$ are implausible~\cite{Alonso:2016oyd} is still valid (or not) to the light of the recent measurements,
\begin{eqnarray}
R(D^\ast) &=& 0.295 \pm 0.014 \ \ \text{HFLAV~\cite{Amhis:2019ckw,HFLAVsummer}}, \nonumber \\
R(D^\ast) &=& 0.284 \pm 0.018 \ \ \text{Belle combination~\cite{Belle:2019rba}}, \\
R(D^\ast) &=& 0.291 \pm 0.035 \ \ \text{LHCb~\cite{Aaij:2017deq}}, \nonumber
\end{eqnarray}

\noindent by considering ${\rm BR}(B_c^- \to \tau^- \bar{\nu}_{\tau}) < 30\%$~\cite{Alonso:2016oyd} and $10\%$~\cite{Akeroyd:2017mhr}. We also include in our analysis the Belle II prospects Belle II-P1 and Belle II-P2 described in Sec.~\ref{analysis}, to see the future implications that could be achieved at Belle II for an integrated luminosity of 50 $\rm ab^{-1}$~\cite{Kou:2018nap}. 

%%%%%%%%%%%%%%%%%%%%%%%%%%%%%%%%
\begin{figure*}[!t]
\centering
\includegraphics[scale=0.28]{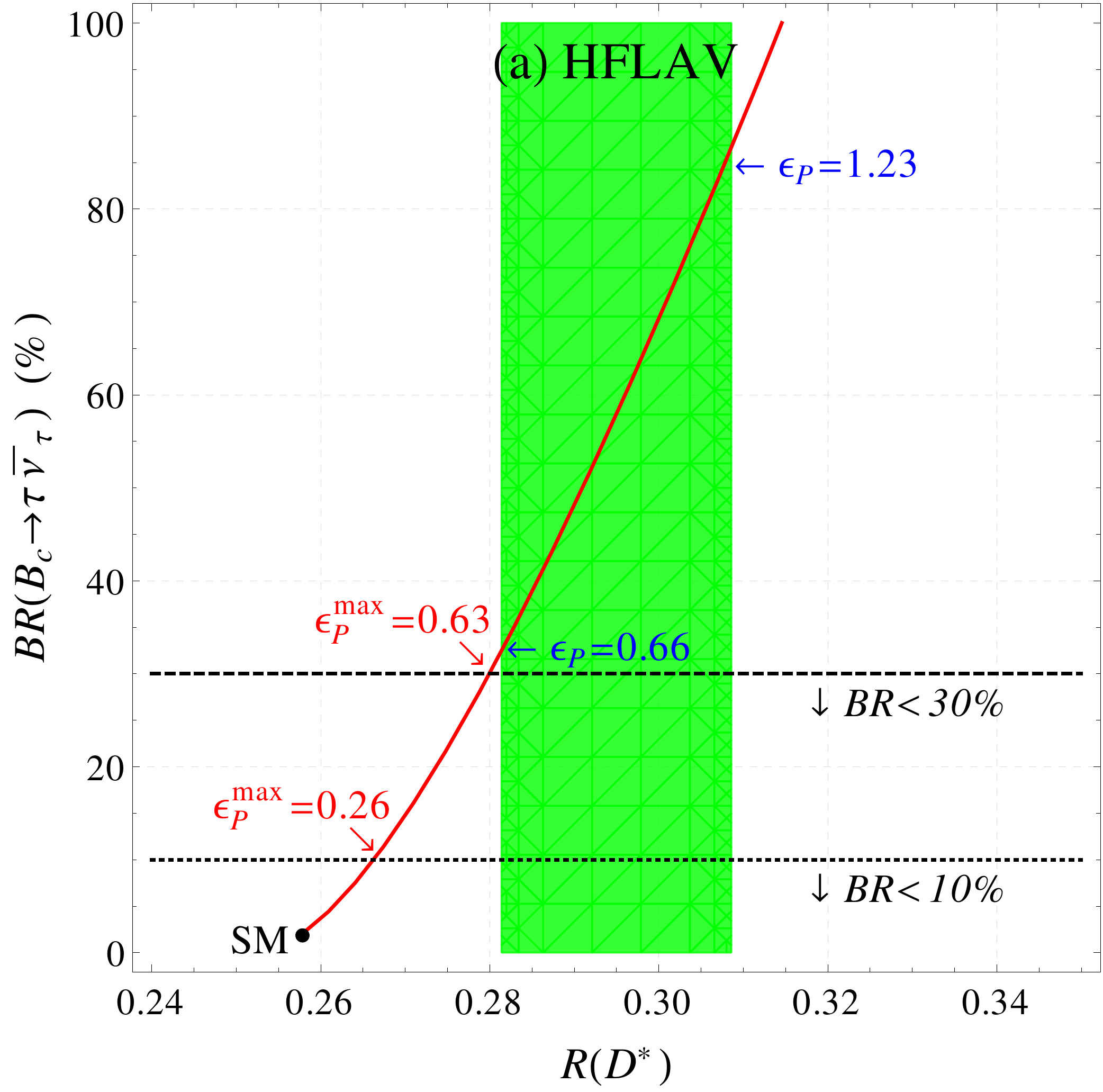}  \
\includegraphics[scale=0.28]{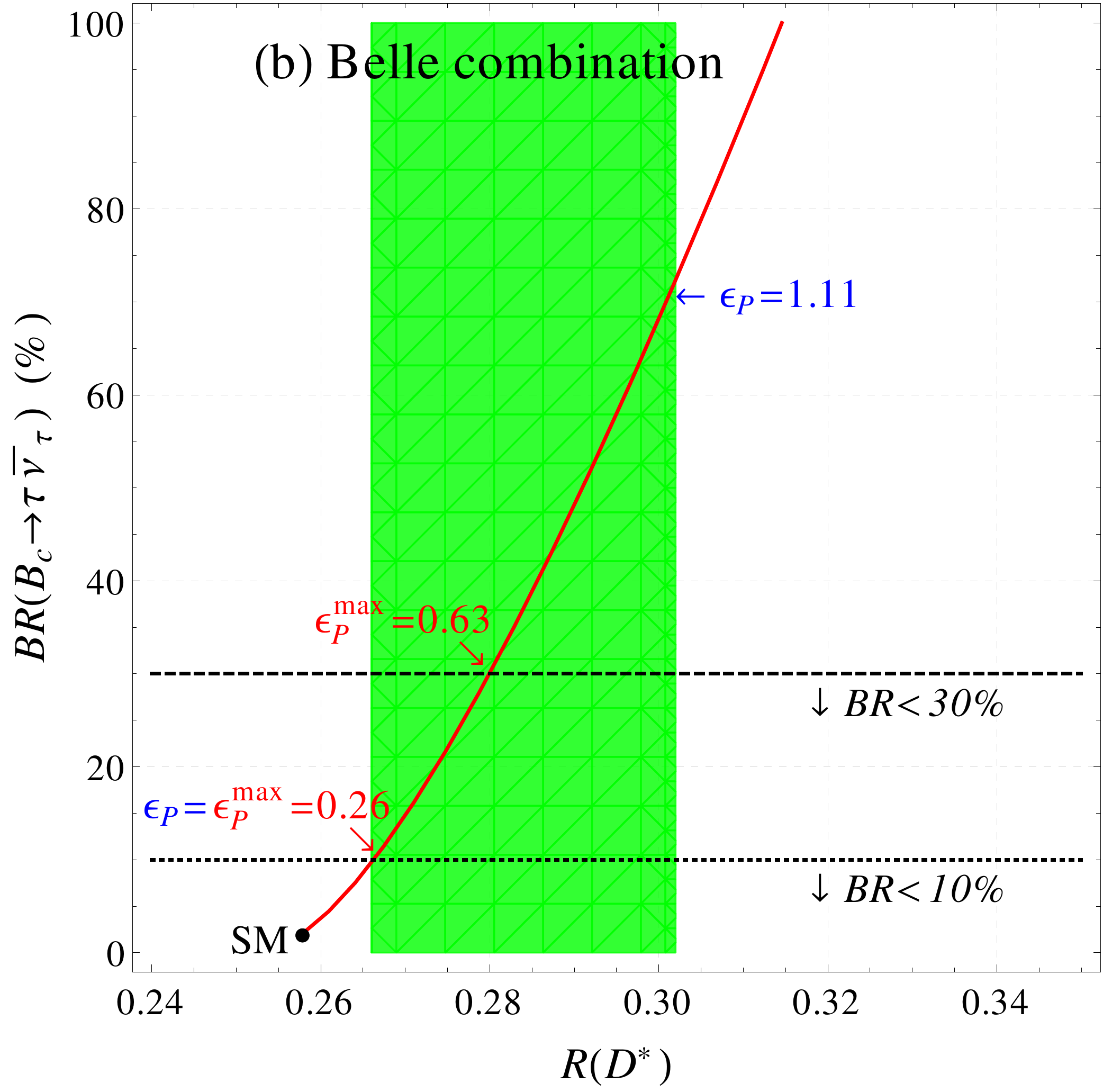} \medskip

\includegraphics[scale=0.28]{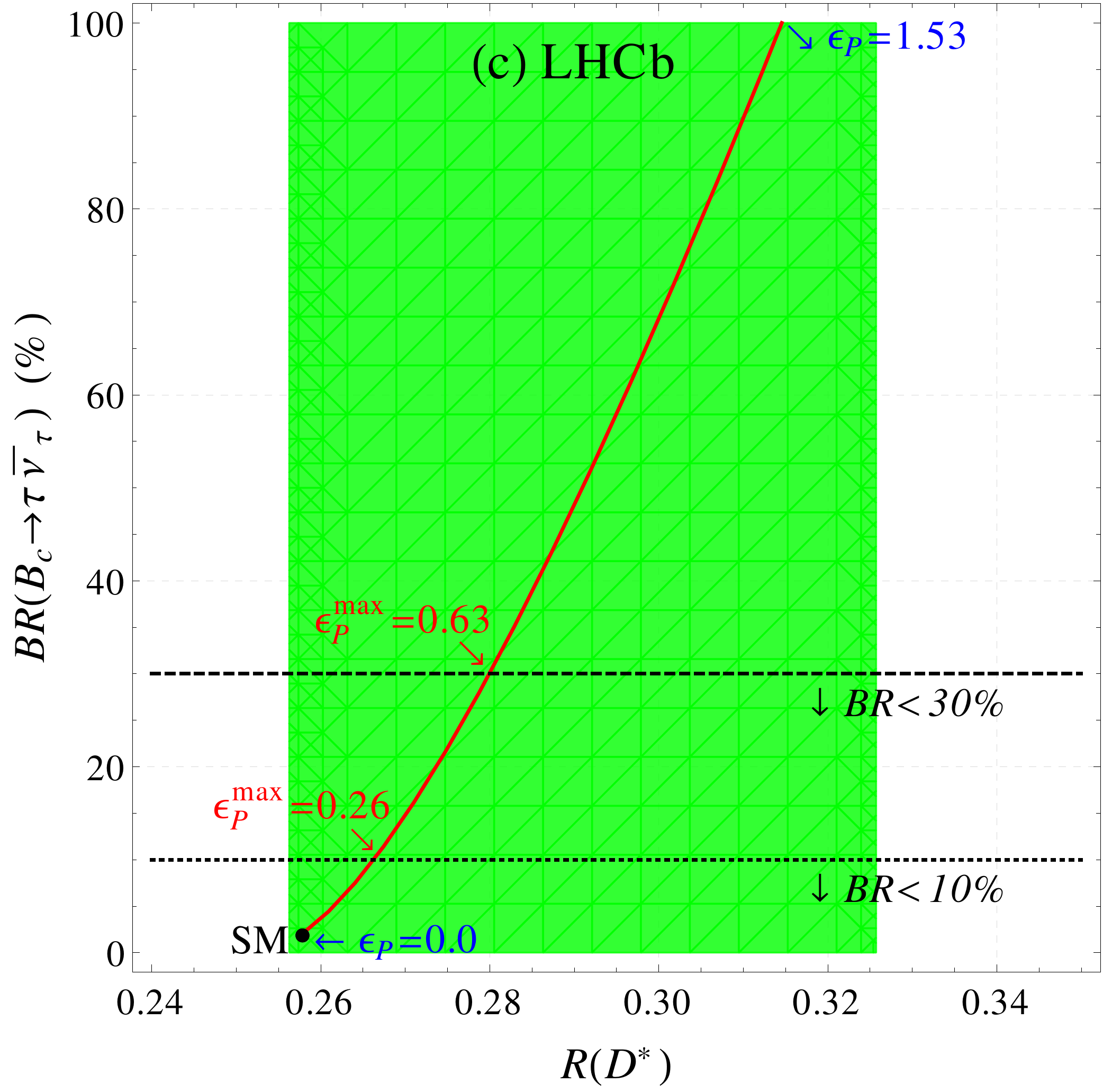} \
\includegraphics[scale=0.28]{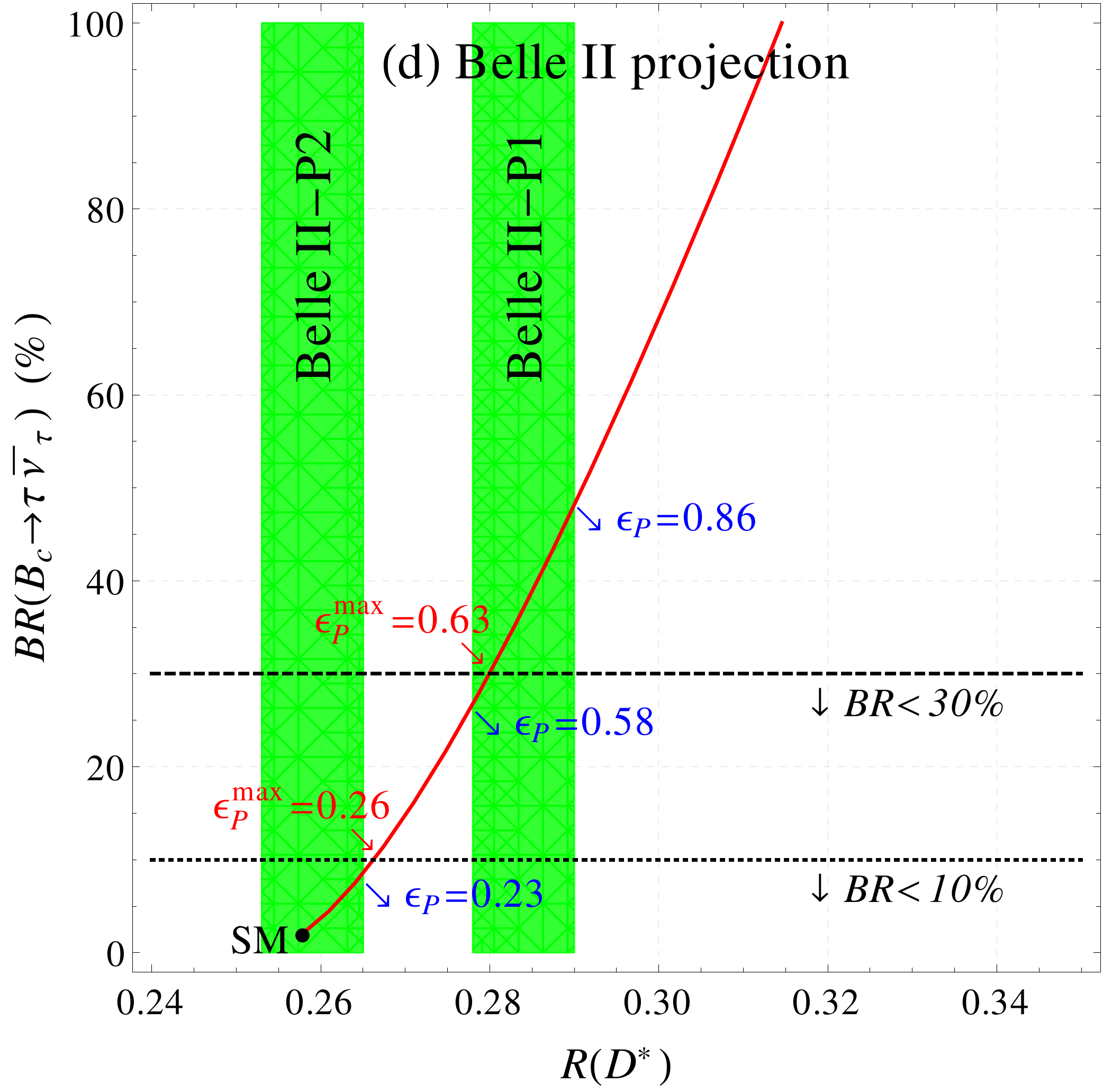} 
\caption{\small Relation between $R(D^\ast)$ and ${\rm BR}(B_c^- \to \tau^- \bar{\nu}_{\tau})$ for $R(D^\ast)$ from (a)  HFLAV~\cite{Amhis:2019ckw,HFLAVsummer}, (b) Belle combination~\cite{Belle:2019rba}, (c) LHCb~\cite{Aaij:2017deq}, and (d) Belle II prospects Belle II-P1 and Belle II-P2; represented by the green $1\sigma$ band, respectively. The red solid line shows the parametric dependence of $R(D^\ast)$ and ${\rm BR}(B_c^- \to \tau^- \bar{\nu}_{\tau})$ on the effective coupling $\epsilon_P$. The dashed (dotted) horizontal line represents ${\rm BR}(B_c^- \to \tau^- \bar{\nu}_{\tau}) < 30\%$ ($10\%$).}
\label{RDstarvsBctaunu}
\end{figure*}
%%%%%%%%%%%%%%%%%%%%%%%%%%%%%%%%

In Fig.~\ref{RDstarvsBctaunu} we plot the relation between $R(D^\ast)$ and ${\rm BR}(B_c^- \to \tau^- \bar{\nu}_{\tau})$ for $R(D^\ast)$ from (a) HFLAV~\cite{Amhis:2019ckw,HFLAVsummer}, (b) Belle combination~\cite{Belle:2019rba}, (c) LHCb~\cite{Aaij:2017deq}, and (d) Belle II prospects Belle II-P1 and Belle II-P1 for 50 $\rm ab^{-1}$. In all the cases, the band represents the $1\sigma$ experimental value. The red solid line shows the parametric dependence of $R(D^\ast)$ and ${\rm BR}(B_c^- \to \tau^- \bar{\nu}_{\tau})$ on the effective coupling $\epsilon_P$, while the dashed and dotted horizontal lines represent the upper limit ${\rm BR}(B_c^- \to \tau^- \bar{\nu}_{\tau}) < 30\%$ and $10\%$, respectively. The SM value is represented by the black circle. It is found that the $1\sigma$ allowed solutions are
\begin{eqnarray}
{\rm HFLAV} &:& \ \epsilon_P = [0.66,1.23], \nonumber \\
{\rm Belle \ combination} &:& \ \epsilon_P = [0.26,1.11], \nonumber \\
{\rm LHCb} &:& \ \epsilon_P = [0.0,1.53], \\
\text{Belle II-P1} &:& \ \epsilon_P = [0.58,0.86], \nonumber \\
\text{Belle II-P2} &:& \ \epsilon_P = [0.0,0.23], \nonumber
\end{eqnarray}

\noindent respectively, as depicted in Fig.~\ref{RDstarvsBctaunu}. In order to fulfill the bound ${\rm BR}(B_c^- \to \tau^- \bar{\nu}_{\tau}) < 30\%$ ($10\%)$, the maximum value required is $\epsilon^{\rm max}_P = 0.63$ ($0.26$), corresponding to a value of $R(D^\ast)=0.281$ ($0.266$). For the case of HFLAV, the allowed $\epsilon_P$ values violate the bound ${\rm BR}(B_c^- \to \tau^- \bar{\nu}_{\tau}) < 30\%$. In contrast, the allowed $\epsilon_P$ values for Belle combination satisfy both ${\rm BR}(B_c^- \to \tau^- \bar{\nu}_{\tau}) < 30\%$ and $10\%$. 
On the other hand, the experimental uncertainties of the LHCb data are large enough  to be  consistent with the bounds ${\rm BR}(B_c^- \to \tau^- \bar{\nu}_{\tau}) < 30\%$ and $10\%$, as well as with the SM. As for the Belle II-P1 would only respect the limit of $30\%$, while for Belle II-P2, small values of $\epsilon_P$ would be favored in fulfillment with the limits of $30\%$ and $10\%$. In addition, Belle II-P2 would also be able to prove an aggressive bound of  ${\rm BR}(B_c^- \to \tau^- \bar{\nu}_{\tau}) < 5\%$.

Therefore, ${\rm BR}(B_c^- \to \tau^- \bar{\nu}_{\tau}) < 30\%$~\cite{Alonso:2016oyd} still disfavors the $\epsilon_P$ pseudoscalar explanation of the $R(D^\ast)$ HFLAV average, while this is no longer the case for the $R(D^\ast)$ data from Belle combination~\cite{Belle:2019rba} and LHCb~\cite{Aaij:2017deq}. The projection Belle II-P1 would not be in conflict with ${\rm BR}(B_c^- \to \tau^- \bar{\nu}_{\tau}) < 30\%$, whereas the projection Belle II-P2 would lead to stronger constraints than those obtained from ${\rm BR}(B_c^- \to \tau^- \bar{\nu}_{\tau})$. Certainly, future measurements by the Belle II~\cite{Kou:2018nap} (as well as LHCb) experiment are required in order to clarify this situation.

 %%%%%%%%%%%%%%%%%%%%%%%%%%%%%%%%
\begin{figure*}[!h]
\centering
\includegraphics[scale=0.4]{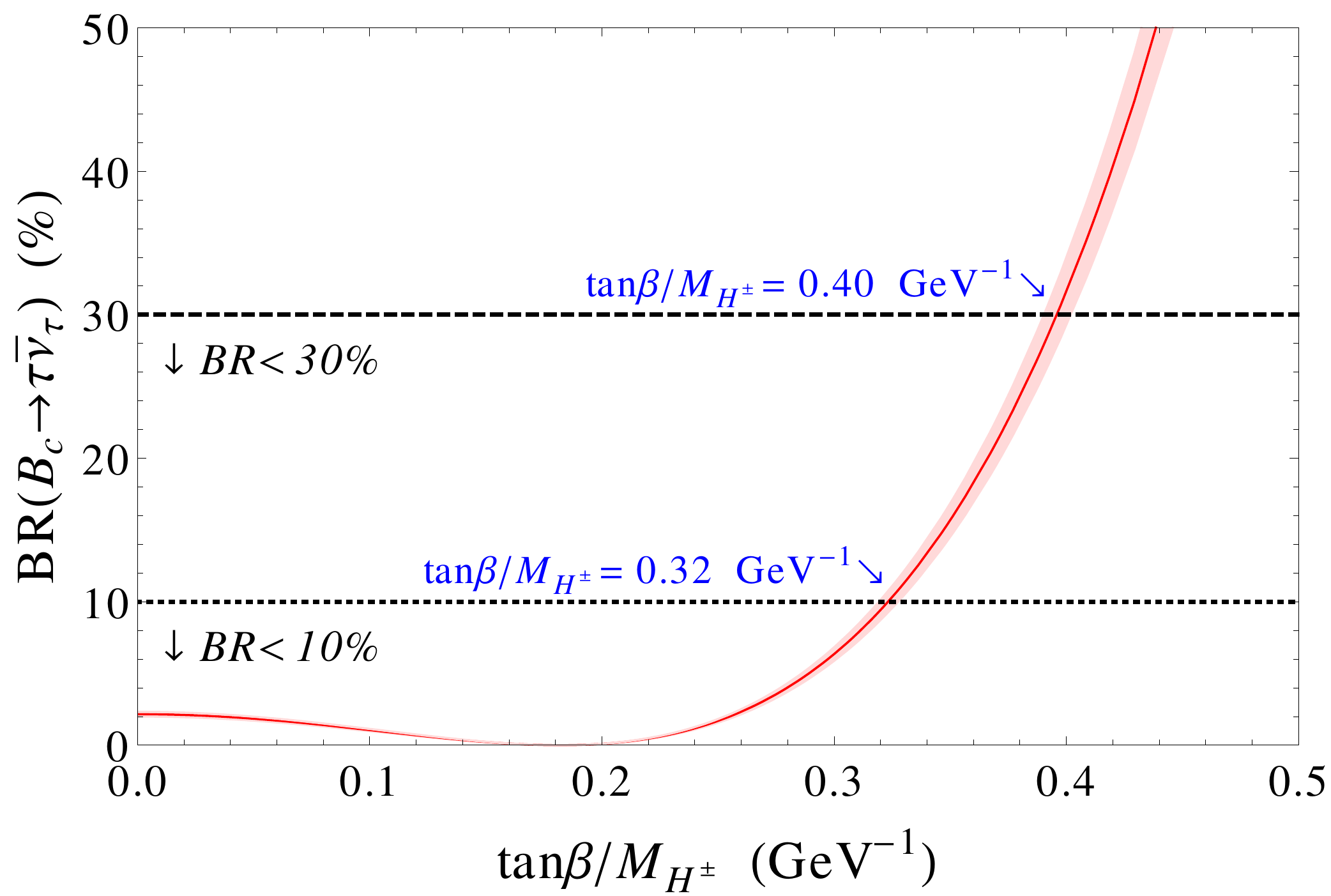}
\caption{\small ${\rm BR}(B_c^- \to \tau^- \bar{\nu}_{\tau})$ in the 2HDM of type II as a function of $\tan \beta/M_{H^\pm}$ (red solid line). The light-red band represents the $1\sigma$ error, while the dashed and dotted horizontal lines represent ${\rm BR}(B_c^- \to \tau^- \bar{\nu}_{\tau}) < 30\%$ and $10\%$, respectively.}
\label{BRvsxbeta}
\end{figure*}
%%%%%%%%%%%%%%%%%%%%%%%%%%%%%%%%\\

%--------------------------------
\section{Reexamining the explanation from 2HDM of Type II} \label{2HDM-II}
%--------------------------------

Although our principal interest is to study the effects of a charged scalar boson Higgs with right-handed neutrinos to the $b \to c \tau \bar{\nu}_\tau$ transition, in this section we discuss  whether  the 2HDM of Type II can still explain the $R(D^{(*)})$ anomalies in light of current measurements. This analysis is relevant because this model is an important NP scenario in the study of the charged-current $B$ anomalies.

The  2HDM of Type II provides one of the simplest scenarios with charged scalar bosons ($H^\pm$). Within this framework, the NP effects of a charged Higgs boson depend on the mass of charged scalar boson $M_{H^\pm}$ and $\tan\beta$ (defined as the ratio of vacuum expectation values $v_2/v_1$), which are described in terms of a single parameter, $\tan \beta/M_{H^\pm}$. In the 2HDM of Type II, tree-level charged Higgs boson contributions to the $B$ meson processes induced by the semileptonic transition $b \to c\tau\bar{\nu}_\tau$  have been investigated (See, for instance~\cite{Hou:1992sy,Tanaka:1994ay,Tanaka:2010se,Kamenik:2008tj,Nierste:2008qe,Fajfer:2012vx}). In 2012 and 2013, the BABAR Collaboration with their full data sample reported a disagreement on the measurements of the ratio of semileptonic $B$ decays, $R(D)$ and $R(D^\ast)$, with respect to the SM predictions~\cite{Lees:2012xj,Lees:2013uzd}. According to the BABAR results, the 2HDM of Type II cannot explain simultaneously the $R(D^{(\ast)})$ discrepancies~\cite{Lees:2012xj,Lees:2013uzd}. Since then (and to date), the 2HDM of Type II interpretation was ruled out and 2HDM models with a more generic flavor structure were considered in the literature~\cite{Crivellin:2012ye,Crivellin:2013wna,Celis:2012dk,Celis:2016azn,Ko:2012sv,HernandezSanchez:2012eg,
Crivellin:2015hha,Cline:2015lqp,Enomoto:2015wbn,Dhargyal:2016eri,Martinez:2018ynq,Wang:2016ggf,
Chen:2017eby,Iguro:2018fni,Iguro:2017ysu,Arbey:2017gmh,Chen:2018hqy,Hagiwara:2014tsa,Lee:2017kbi,
Iguro:2018qzf,Li:2018rax}. It is worth noting that subsequent analysis performed by Belle Collaboration in 2015~\cite{Huschle:2015rga} and 2016~\cite{Sato:2016svk} showed compatibility with the 2HDM of Type II in the $\tan \beta/M_{H^\pm}$ regions around $0.45 \ {\rm GeV}^{-1}$~\cite{Huschle:2015rga} and $[0.65,0.76] \ {\rm GeV}^{-1}$~\cite{Sato:2016svk}, respectively, in contradiction with the BABAR measurements~\cite{Lees:2012xj,Lees:2013uzd}. Thereby, given the current experimental situation on the $R(D^{(\ast)})$ anomalies, HFLAV~\cite{Amhis:2019ckw,HFLAVsummer} and Belle combination~\cite{Belle:2019rba}, and the Belle II future sensitivity~\cite{Kou:2018nap}, in the following we reexamine whether the 2HDM of Type II is still ruled out (or not) as an explanation to the $R(D^{(\ast)})$ anomalies. In addition, it is also important to confront this model not only to the $R(D^{(\ast)})$ measurements but also to all $b \to c\tau\bar{\nu}_\tau$ observables.\\

By construction, in the 2HDM of Type II the neutrinos are considered to be LH, therefore, the scalar WCs that contribute to the charged-current $b \to c\tau \bar{\nu}_{\tau}$ are written as~\cite{Sato:2016svk,Tanaka:2010se,Tanaka:2012nw}
\begin{eqnarray}
C_S^{RL} &=& -\frac{m_b m_\tau \tan\beta}{M_{H^\pm}^2} , \\
C_S^{LL} &=&-\frac{m_c m_\tau}{M_{H^\pm}^2},
\end{eqnarray}

\noindent where $m_b$, $m_c$, and $m_\tau$ are the masses of the bottom quark, charm quark, and $\tau$ lepton, respectively. In the literature~\cite{Sato:2016svk,Tanaka:2010se,Tanaka:2012nw}, the coefficient $C_S^{LL}$ is usually neglected ($C_S^{LL} \simeq 0$), thus, the charged Higgs boson effect is only dominated by $C_S^{RL}$ which is driven by $\tan \beta/M_{H^\pm}$. We begin our analysis by considering the constraints on the parameter $\tan \beta/M_{H^\pm}$ from the upper limits ${\rm BR}(B_c^- \to \tau^- \bar{\nu}_{\tau}) < 30\%$~\cite{Alonso:2016oyd} and $10\%$~\cite{Akeroyd:2017mhr}. In Fig.~\ref{BRvsxbeta} we show the ${\rm BR}(B_c^- \to \tau^- \bar{\nu}_{\tau})$ in the 2HDM of type II as a function of $\tan \beta/M_{H^\pm}$ (red solid line), where the dashed and dotted horizontal lines represent ${\rm BR}(B_c^- \to \tau^- \bar{\nu}_{\tau}) < 30\%$ and $10\%$, respectively. We get the following strong bounds
\begin{eqnarray}
\tan \beta/M_{H^\pm} < 0.40 \ (0.32),
\end{eqnarray}

\noindent for ${\rm BR}(B_c^- \to \tau^- \bar{\nu}_{\tau}) < 30\%$ ($10\%$), implying that large values of $\tan \beta/M_{H^\pm}$ are excluded. This plot has been obtained by using the SM estimation ${\rm BR}(B_c^- \to \tau^- \bar{\nu}_{\tau})_{\rm SM} = (2.16 \pm 0.16 )\%$~\cite{Gomez:2019xfw}. In addition, in Table~\ref{interval2HDM} we present the $2\sigma$ allowed regions by each of the $b \to c \tau \bar{\nu}_{\tau}$ observables on the parameter $\tan \beta / M_{H^{\pm}}$, namely, $R(D^{(\ast)})$ HFLAV and Belle combination, $R(J/\psi), F_L(D^*), P_\tau(D^*)$, and $R(X_c)$. Regarding the Belle II future scenario described in Sec.~\ref{analysis}, we also show the prospects Belle II-P1 and Belle II-P2 that could be achieved at Belle II for an integrated luminosity of 50 $\rm ab^{-1}$~\cite{Kou:2018nap}. 
As a result, it is possible to find a common small values region, $\tan \beta/M_{H^\pm} = [0.0,0.32]$,  without conflicting with the strongest bound imposed by ${\rm BR}(B_c^- \to \tau^- \bar{\nu}_{\tau}) < 10\%$; with the exception of $R(D^\ast)$ HFLAV and $R(J/\psi)$ that allows large $\tan \beta/M_{H^\pm}$ values, which are in tension with ${\rm BR}(B_c^- \to \tau^- \bar{\nu}_{\tau}) < 30\%$ and $10\%$. Besides, the projected Belle II-P1 values would be rule out by the bounds of $30\%$ and $10\%$, while the projection Belle II-P2 would point out to small values of $\tan \beta/M_{H^\pm}$.

%----------------------------------
\begin{table}[!t]
\centering
\renewcommand{\arraystretch}{1.2}
\renewcommand{\arrayrulewidth}{0.8pt}
\begin{tabular}{l|c}
\hline\hline
\multicolumn{1}{c|}{Observable}           & Allowed regions ($2\sigma$) on $\tan \beta / M_{H^{\pm}}$ ($\rm GeV^{-1}$) \\ 
\hline \hline

$R(D)$ HFLAV                               & $[0.0,0.08]\cup[0.44,0.47]$ \\ 
$R(D)$ Belle combination             & $[0.0,0.11]\cup[0.43,0.47]$ \\ 
$R(D)$ Belle II-P1                           & $[0.45,0.46]$ \\ 
$R(D)$ Belle II-P2                           & $[0.0,0.06]\cup[0.44,0.45]$ \\ 
\hline
$R(D^*)$ HFLAV                            & $[0.64,0.75]$        \\ 
$R(D^\ast)$ Belle combination      & $[0.0,0.24]\cup[0.56,0.75]$        \\ 
$R(D^\ast)$ Belle II-P1                      & $[0.67,0.70]$        \\ 
$R(D^\ast)$ Belle II-P2                        & $[0.0,0.17]\cup[0.58,0.63]$        \\ 
\hline
$R(J/\psi)$                                & $[1.0,1.05]$           \\ 
$F_L(D^\ast)$                                & $[0.0,0.09]\cup[0.44,0.67]$ \\ 
$P_{\tau}(D^\ast)$                           & $[0.0,0.32] \cup[0.56,1.03]$            \\
$R(X_c)$ &  $[0.0,0.17]\cup[0.44,0.72]$  \\
\hline\hline
\end{tabular}
\caption{The $2\sigma$ allowed regions on the parameter $\tan \beta / M_{H^{\pm}}$ of the 2HDM of Type II, obtained for the different $b \to c \tau \bar{\nu}_{\tau}$ observables. The Belle II future projections for an integrated luminosity of 50 $\rm ab^{-1}$ are also included.}
\label{interval2HDM}
\end{table}
%----------------------------------

 %%%%%%%%%%%%%%%%%%%%%%%%%%%%%%%%
\begin{figure*}[!t]
\centering
\includegraphics[scale=0.35]{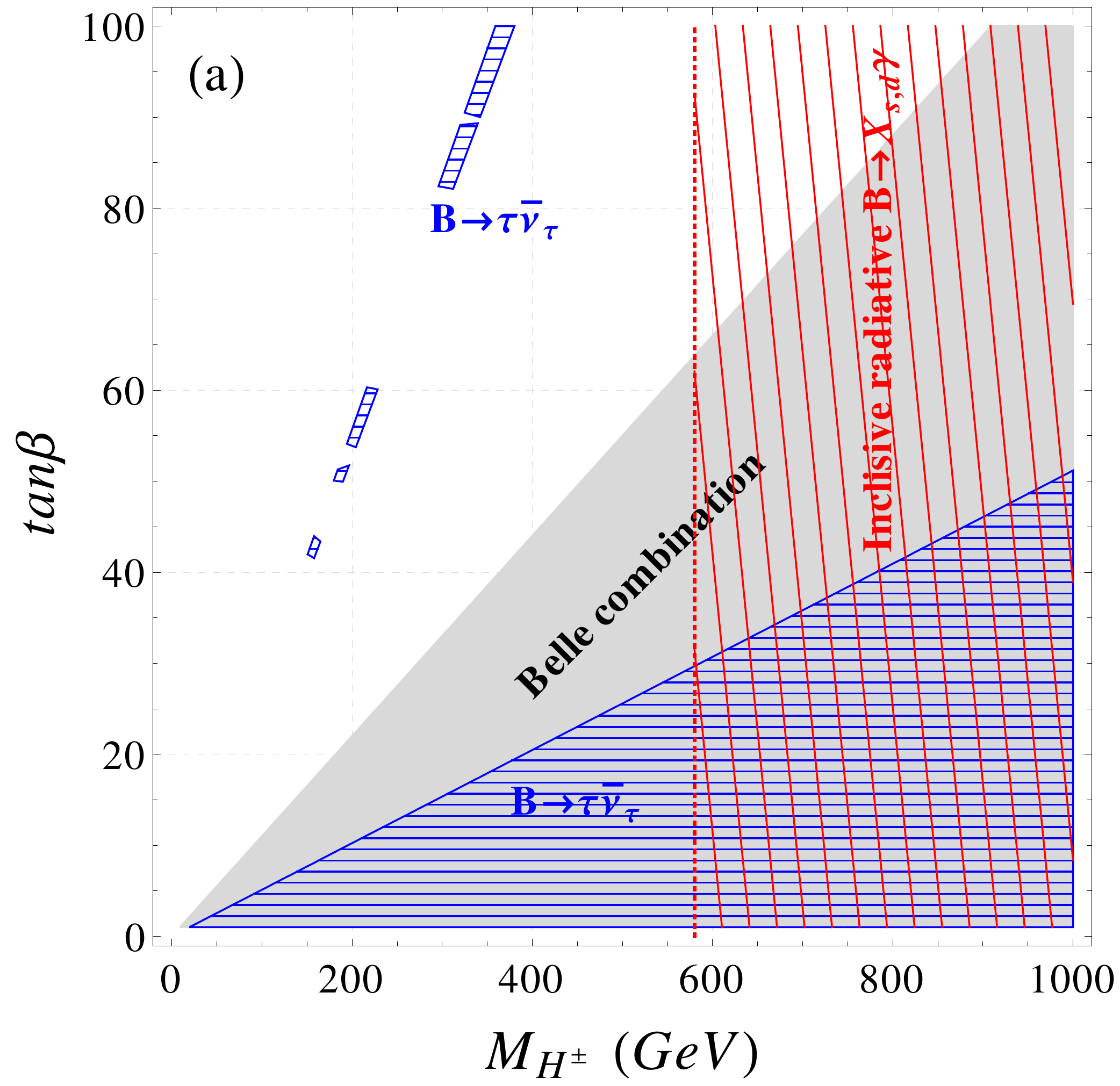} \ 
\includegraphics[scale=0.35]{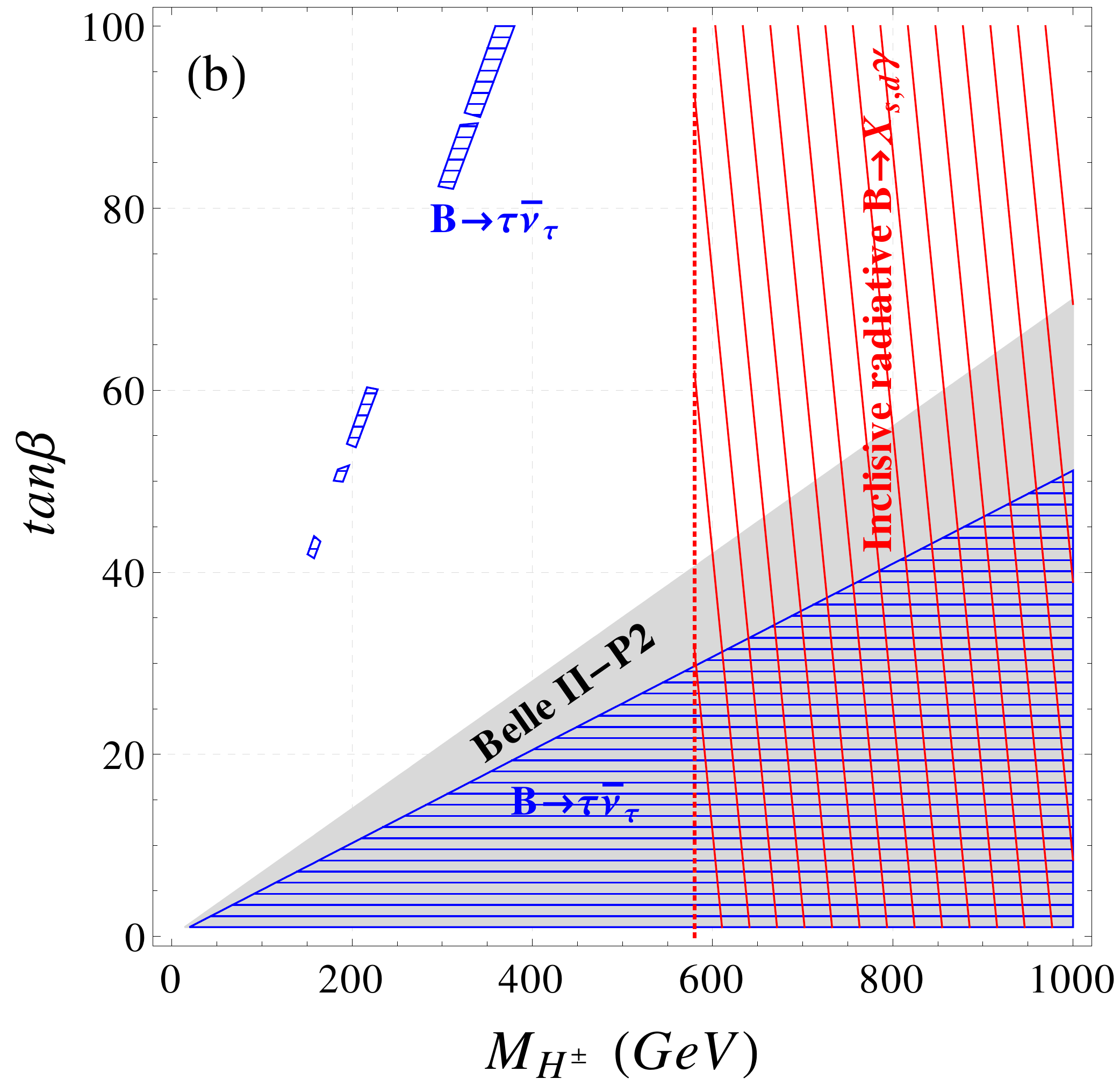}
\caption{\small Allowed parameter space in the plane $(M_{H^{\pm}}, \tan\beta)$ of the 2HDM Type II for (a) Belle  combination and (b) Belle II-P2 for 50 $\rm ab^{-1}$. The allowed regions by $B \to \tau \bar{\nu}_\tau$ and the inclusive radiative decays $B \to X_{s,d} \gamma$~\cite{Misiak:2017bgg} are represented by the blue and red hatched regions, respectively.}
\label{PS2}
\end{figure*}
%%%%%%%%%%%%%%%%%%%%%%%%%%%%%%%%

To further discussion, we now translate these results into the plane $(M_{H^{\pm}}, \tan\beta)$ of the 2HDM of Type II, as is shown in Figs.~\ref{PS2}(a) and~\ref{PS2}(b) by the gray region for the $R(D^{(\ast)})$ solutions from Belle combined and Belle II 50 $\rm ab^{-1}$, respectively. The red hatched region corresponds to the 95\% C.L. strong bound on the charged Higgs mass $M_{H^{\pm}}>$ 580 GeV (independent of $ \tan\beta$), obtained from inclusive radiative decays of the $B$ meson, $B \to X_{s,d}\gamma$~\cite{Misiak:2017bgg}. In addition, we also considered the tree-level contributions from the charged Higgs boson exchange in the tauonic decay $B \to \tau \bar{\nu}_\tau$~\cite{Hou:1992sy} 
\bea
{\rm BR}(B^-  \to  \tau^- \bar{\nu}_\tau) &=& {\rm BR}(B^-  \to  \tau^- \bar{\nu}_\tau)_{\rm SM} \ \Big(1- \tan^2\beta \frac{m_B^2}{M_{H^{\pm}}^2}\Big)^2 .
\eea

\noindent with
\bea \label{SM_leptonic}
{\rm BR}(B^-  \to  \tau^- \bar{\nu}_\tau)_{\rm SM} &=&\tau_{B} \dfrac{G_F^2}{8\pi} |V_{ub}|^2 f_{B}^2 m_{B} m_{\tau}^2  \Big(1- \dfrac{m_{\tau}^2}{m_{B}^2}\Big)^2 ,
\eea

\noindent where $V_{ub}$ denotes the CKM matrix element involved, and $f_{B}$ and $\tau_{B}$ are the $B^-$ meson decay constant and lifetime, respectively. By using $f_B=(190.0\pm1.3)$ MeV and $V_{ub} = (3.94 \pm 0.36)\times 10^{-3}$ from Particle Data Group (PDG)~\cite{PDG2020}, we get a SM prediction of ${\rm BR}(B^-  \to  \tau^- \bar{\nu}_\tau)_{\rm SM} = (9.89 \pm 0.13)\times 10^{-5}$ that is in agreement ($0.4 \sigma$) with the experimental value ${\rm BR}(B^-  \to  \tau^- \bar{\nu}_\tau)_{\rm Exp} = (10.9 \pm 2.4)\times 10^{-5}$ reported by PDG~\cite{PDG2020}. The allowed regions from $B \to \tau \bar{\nu}_\tau$ are represented by the blue hatched regions in Figs.~\ref{PS2}(a) and~\ref{PS2}(b). For the Belle combination we found that for $30 \lesssim \tan\beta \lesssim 70$, it is possible to get a large region on the parameter space $(M_{H^{\pm}}, \tan\beta)$ to account for a joint explanation to the $R(D)$ and $R(D^{\ast})$ anomalies, in consistency with $B \to \tau \bar{\nu}_\tau$ and bounds from inclusive radiative $B$ decays ($M_{H^{\pm}}>$ 580 GeV). On the other hand, the projection Belle II-P2 suggests that the 2HDM of Type II would be no longer disfavored.

%&&&&&&&&&&&&&&&&&&&&

%&&&&&&&&&&&&&&&&&&&&

\FloatBarrier

%\FloatBarrier
%&&&&&&&&&&&&&&&&&&&&
\section{Summary and Conclusions} \label{Conclusion}
%&&&&&&&&&&&&&&&&&&&&

Given the present-day 2020 experimental $b \to c \tau \bar{\nu}_\tau$ data, we analysed the so-called charged-current $B$ meson anomalies in terms of a charged scalar boson within the framework of a generic 2HDM with right-handed neutrinos. Our study is composed of two set of observables that include the HFLAV world-average and Belle combination measurements on $R(D^{(\ast)})$, along with $R(J/\psi)$, the polarizations $P_\tau(D^\ast)$ and $F_L(D^\ast)$, the inclusive ratio $R(X_c)$, as well as taking into account the upper limits ${\rm BR}(B_c^- \to \tau^- \bar{\nu}_{\tau}) < 30\%$ and $10\%$~\cite{Alonso:2016oyd,Akeroyd:2017mhr}. We have also investigated the impact of future measurements at Belle II for $R(D^{(\ast)})$, by regarding two well-motivated projected scenarios that could be achieved for an integrated luminosity of $50 \ {\rm ab}^{-1}$ (referred by us as Belle II-P1 and Belle II-P2). For completeness of our work, we first explored the associated scalar WCs by paying special attention to those including RH neutrinos, namely $C_S^{LR}$ and $C_S^{RR}$.  As the main outcome, the projection Belle II-P2 would provide stronger constraints than ${\rm BR}(B_c^- \to \tau^- \bar{\nu}_{\tau}) < 30\%$ and $10\%$, but still allowing a small window for NP contributions. These prospects would cover a similar region to the one set from the analysis of the mono-tau signature $pp \to \tau_h X + \rm{MET}$ at the LHC and the projected sensitivity at the high-luminosity LHC.

In the second part of our analysis, we present the main purpose of our work. We performed a phenomenological study of the parameter space  associated with the charged Higgs Yukawa couplings to the charm-bottom and leptons from third generation in the generic 2HDM that can accommodate the charged-current $B$ meson anomalies. By focusing on RH neutrino scenarios, it is found that for parameter spaces with neutrino Yukawa couplings, there are allowed regions with absolute values of the  Yukawa couplings of the order $\mathcal{O}(10^{-1})$, for a benchmark charged Higgs mass of $M_{H^\pm} = 500$~GeV. Moreover, the prospects at the Belle II scenario would allow enough parametric space to explain anomalies, without conflicting with ${\rm BR}(B_c^- \to \tau^- \bar{\nu}_{\tau}) < 30\%$ and $10\%$. In general, our results imply that current experimental $b\rightarrow c \tau \bar{\nu}_\tau$ data favors the interpretation of a charged scalar boson with RH neutrinos.

As an important by product of our analysis regarding the charged scalar boson explanation, we revisited whether the claim that pseudoscalar NP ($\epsilon_P$) interpretations of $R(D^{\ast})$ are implausible due to the $B_c$ lifetime~\cite{Alonso:2016oyd}, is still valid to the light of the recent $R(D^{\ast})$ measurements. We found that ${\rm BR}(B_c^- \to \tau^- \bar{\nu}_{\tau}) < 30\%$ still disfavors the $\epsilon_P$ pseudoscalar explanation of the $R(D^\ast)$ HFLAV average value, while this is no longer the case for the $R(D^\ast)$ data from Belle combination~\cite{Belle:2019rba} and LHCb~\cite{Aaij:2017deq}. The projections Belle II-P1 and Belle II-P2 would not be in conflict with ${\rm BR}(B_c^- \to \tau^- \bar{\nu}_{\tau}) < 30\%$. Future measurements at Belle II experiment are required in order to clarify this situation.

Finally, we also reexamined whether the 2HDM of Type II is still disfavored as an explanation to the $R(D^{(\ast)})$ anomalies. As the main outcome from this analysis, it is found that for Belle combination is possible to get a large region on the parameter space $(M_{H^{\pm}}, \tan\beta)$ to account for a joint explanation to the $R(D)$ and $R(D^{\ast})$ anomalies, in consistency with $B \to \tau \bar{\nu}_\tau$ and bounds from inclusive radiative $B$ decays ($M_{H^{\pm}}>$ 580 GeV). This allowed parameter space is not in conflict with the strong bounds $\tan \beta/M_{H^\pm} < 0.40$ and 0.32 that can be set from ${\rm BR}(B_c^- \to \tau^- \bar{\nu}_{\tau}) < 30\%$ and $10\%$, respectively, implying that large values of $\tan \beta/M_{H^\pm}$ are excluded. Moreover, the projection Belle II-P2 point out that this model would be no longer disfavored. Thus, after almost eight years, there are good prospects that the 2HDM of Type II explanation will rise from the ashes.

\acknowledgments
%We are grateful with William A. Ponce and Carlos E. Vera for their contribution at the early stage of this work.  
J. Cardozo and J. H. Mu\~{n}oz are grateful with the Oficina de Investigaciones - Universidad del Tolima by financial support. N. Quintero acknowledges support from Direcci\'{o}n General de Investigaciones - Universidad Santiago de Cali under Project No. 935-621120-G01. 
E. Rojas  acknowledges support from ``Vicerrectoría de Investigaciones e Interacción Social VIIS de la Universidad de Nariño'',  project numbers 1928 and 2172.

\appendix

\section{General two  Higgs doublet  model} \label{AppA}
%------------------------------------------
The most general Lagrangian for the interaction of two Higgs  doublets $\Phi_1$, $\Phi_2$  with the fermions of the SM is given by
\begin{align}\label{eq:yukawa}
\mathcal{L}= & -\bar{q}_L^{\prime i} \Phi_1 y_{ij}^{1D}d^{\prime j}_{R}     -\bar{q}_L^{\prime i} \Phi_2 y_{ij}^{2D}d^{\prime j}_{R}
              -\bar{q}_L^{\prime i} \tilde\Phi_1 y_{ij}^{1U}u^{\prime j}_{R}-\bar{q}_L^{\prime i} \tilde\Phi_2 y_{ij}^{2U}u^{\prime j}_{R}\notag \\
             & -\bar{l}_L^{\prime i} \Phi_1 y_{ij}^{1E}e^{\prime j}_{R}-\bar{l}_L^{\prime i} \Phi_2 y_{ij}^{2E}e^{\prime j}_{R}
              -\bar{l}_L^{\prime i} \tilde\Phi_1 y_{ij}^{1N}\nu^{\prime j}_{R}-\bar{l}_L^{\prime i}\tilde\Phi_2 y_{ij}^{2N}\nu^{\prime j}_{R}\notag\\
              =& -\bar{q}_L^{\prime i} \Phi_\alpha y_{ij}^{\alpha D}d^{\prime j}_{R}-\bar{q}_L^{\prime i} \tilde\Phi_\alpha y_{ij}^{\alpha U}u^{\prime j}_{R}
                 -\bar{l}_L^{\prime i} \Phi_\alpha  y_{ij}^{\alpha E}e^{\prime j}_{R}-\bar{l}_L^{\prime i} \tilde\Phi_\alpha  y_{ij}^{\alpha N}\nu^{\prime j}_{R},  
\end{align}
where a sum is assumed on repeated indices.
Here  $i,j$ run over $1,2,3$ and $\alpha$ over $1,2$. 
The super index  $U$ refers to  up-like  quarks~(the same is true for the super indices $D$, $E$, $N$ which refer to down-like, electron-like, neutrino-like fermions, respectively).
The Higgs  boson doublet fields are parametrized as follows:
\begin{align}
\Phi_\alpha = 
\begin{pmatrix}
\phi_\alpha^{+}\\
\frac{v_\alpha+\phi_{\alpha}^{0}+iG_\alpha^{0}}{\sqrt{2}}
\end{pmatrix},
\hspace{1cm}
\tilde\Phi_\alpha=i\sigma_2 \Phi_\alpha^{*}.
\end{align}
It is necessary to rotate to the Georgi basis, i.e.,
\begin{align}
\begin{pmatrix}
H_1 \\
H_2 
\end{pmatrix}
= 
\begin{pmatrix}
 \cos \beta & \sin \beta\\
-\sin \beta & \cos \beta
\end{pmatrix}
\begin{pmatrix}
\Phi_1\\
\Phi_2
\end{pmatrix}
\equiv R_{\beta\alpha}\Phi_\alpha , 
\end{align}
where $\tan \beta =\frac{v_2}{v_1}$. 
This basis is chosen in such a way that only the neutral component of $ H_1 $ acquires a vacuum expectation value  $v/\sqrt{2}$ with  $v=\sqrt{v_1^2+v_2^2}$.
In this way $\Phi_\alpha y_{ij}^{\alpha F}= y_{ij}^{\alpha F}(R_{\alpha\beta}^T) R_{\beta \gamma} \Phi_{\gamma}= \Y_{ij}^{\beta F} H_\beta $.
Where we have defined
\begin{align}
H_\beta=  R_{\beta\alpha}\Phi_{\alpha},\hspace{0.2cm}\text{and}\hspace{0.2cm}\Y^{\beta F}_{ij}=R_{\beta \alpha}y_{ij}^{\alpha F}.\\
\end{align}
By writing $\Y^{\alpha F}_{ij}$ explicitly we can classify the different Two Higgs Doublet Model~(2HDM) types
\begin{align}\label{eq:mathcaly}
 \Y^{1F}_{ij}=& +\cos \beta y^{1F}_{i,j}+\sin \beta y^{2F}_{i,j}\notag\\
 \Y^{2F}_{ij}=& -\sin \beta y^{1F}_{i,j}+\cos \beta y^{2F}_{i,j}.
\end{align}
With these definitions equation~(\ref{eq:yukawa}) becomes
\begin{align}
\mathcal{L} = -\bar{q}_L^{\prime i} H_\beta  \Y_{ij}^{\beta D}d^{\prime j}_{R}-\bar{q}_L^{\prime i} \tilde H_\beta \Y_{ij}^{\beta U}u^{\prime j}_{R}
              -\bar{l}_L^{\prime i} H_\beta \Y_{ij}^{\beta E}e^{\prime j}_{R}-\bar{l}_L^{\prime i} \tilde H_\beta  \Y_{ij}^{\beta N}\nu^{\prime j}_{R}.   
\end{align}
It is necessary to rotate to the mass eigenstates of the fermion mass, i.e.,
\begin{align}
f_{L,R}= U^F_{L,R}f'_{L,R}. 
\end{align}
From the Lagrangian for the charged currents 
\begin{align}
\mathcal{L}_{CC}&=-\frac{g}{\sqrt{2}} \bar{u}_{Li}'\gamma^{\mu} d_{Li}'W^+ -\frac{g}{\sqrt{2}} \bar{e}_{Li}'\gamma^{\mu} \nu_{Li}'W^-+\text{h.c}\notag\\
                &=-\frac{g}{\sqrt{2}} \bar{u}_{Li}\gamma^{\mu}V_{_{CKM}} d_{Li}W^+ -\frac{g}{\sqrt{2}} \bar{e}_{Li}\gamma^{\mu}V_{_{PMNS}} \nu_{Li}W^-+\text{h.c},
\end{align}
it is possible to obtain the Cabibbo-Kobayashi-Maskawa (CKM) and the Pontecorvo-Maki-Nakagawa-Sakata (PMNS) mixing matrices  $V_{_{CKM}}= U^{U}_L U^{D \dagger}_L$ and  $V_{_{PMNS}}= U^{E}_L U^{\nu \dagger}_L$ by rotating to the fermion mass eigenstates to obtain an expression for  the  Yukawa couplings closely related to the observables and the Wilson coefficients

\begin{align}
\mathcal{L}_{H^\pm} =&
              -\bar{u}_L^{\prime i}   H_\beta^+ \Y_{ij}^{\beta D}              d^{\prime j}_{R}
              +\bar{d}_L^{\prime i}   H_\beta^{+\dagger} \Y_{ij}^{\beta U}     u^{\prime j}_{R}
              -\bar{\nu}_L^{\prime i} H_\beta^+ \Y_{ij}^{\beta E}              e^{\prime j}_{R}
              +\bar{e}_L^{\prime i}   H_\beta^{+\dagger} \Y_{ij}^{\beta N}     \nu^{\prime j}_{R}
              +\text{h.c}.
\end{align}

%                =&- H_l^+         \bar{u}_L^{ h}  (U^{u}_{L})^{hi}   Y^{ij}_{ld}  (U^{d\dagger}_{R})^{jk}  d^{k}_{R}
%                   +H_l^{+\dagger}\bar{d}_L^{h}   (U^{d}_{L})^{hi}   Y^{ij}_{lu}  (U^{u\dagger}_{R})^{jk}  u^{k}_{R}\notag\\
%                &  -H_l^+         \bar{\nu}_L^{h} (U^{\nu}_{L})^{hi} Y^{ij}_{le}  (U^{e\dagger}_{R})^{jk}  e^{k}_{R}
%                   +H_l^{+\dagger}\bar{e}_L^{h}   (U^{e}_{L})^{hi}   Y^{ij}_{l\nu}(U^{\nu\dagger}_{R})^{jk}\nu^{k}_{R}
%                +\text{h.c}\notag\\   
%               =&- H_l^+         \bar{u}_L^{ h}  \left(U^{u}_{L}   \Y_{ld}   U^{d\dagger}_{R}  \right)^{hk}  d^{k}_{R}
%                   +H_l^{+\dagger}\bar{d}_L^{h}  \left(U^{d}_{L}   \Y_{lu}   U^{u\dagger}_{R}  \right)^{hk}  u^{k}_{R}\notag\\
%                &  -H_l^+         \bar{\nu}_L^{h}\left(U^{\nu}_{L} \Y_{le}   U^{e\dagger}_{R}  \right)^{hk}  e^{k}_{R}
%                   +H_l^{+\dagger}\bar{e}_L^{h}  \left(U^{e}_{L}   \Y_{l\nu} U^{\nu\dagger}_{R}\right)^{hk} \nu^{k}_{R}
%               +\text{h.c}\notag\\
The charged Higgs fields $H^{\pm}_{1}$ are absorbed by the $W^{\pm}$ bosons in such a way the unique charged scalars are $H_2^{\pm}$, which from now on we will simply denote as $H^{\pm}$
\begin{align}
    \mathcal{L}(H^{\pm})    
                  = &- H^+\left( \bar{u}_L^{h}X^{D}_{u^h d^k}      d^{k}_{R}
                   -            \bar{u}_R^{h}   X^{U*}_{d^hu^k}      d^{k}_{L} 
                   +            \bar{\nu}_L^{h} X^{E}_{\nu^he^k}     e^{k}_{R}
                   -            \bar{\nu}_R^{h} X^{N*}_{e^h\nu^k}  e^{k}_{L}\right)\notag\\
                 &-H^-\left(  \bar{d}_R^{ h}  X^{D*}_{u^hd^k}      u^{k}_{L}
                                +\bar{e}_R^{h}  X^{E*}_{\nu^he^k}    \nu^{k}_{L} 
                                -\bar{d}_L^{h}  X^{U}_{d^hu^k}       u^{k}_{R}
                                -\bar{e}_L^{h}  X^{N}_{e^h\nu^k}   \nu^{k}_{R} 
                                 \right).  
                            \end{align}
Where  $X_{n^hp^k}^{P} = \left(U^{N}_{L}   \Y^{2P}   U^{P\dagger}_{R}  \right)^{hk} $  and    $X_{p^hn^k}^{N} = \left(U^{P}_{L}   \Y^{2N}   U^{N\dagger}_{R}  \right)^{hk} $,
here we label the up and down isospin components with $P$  and $N$, respectively. 
The corresponding couplings  for the boson $ H_1 $ are denoted by $Y_{n^hp^k}^{P}$ and $Y_{p^hn^k}^{N}$.
% \begin{align}
% \mathcal{L}_{H^\pm}(H_2) 
%                = &- H_l^+\left( \bar{u}_L^{h}   Y_{u^h d^k}^{l}      d^{k}_{R}
%                    +            \bar{\nu}_L^{h} Y_{\nu^he^k}^{l}     e^{k}_{R}
%                    -            \bar{u}_R^{h}   Y^{l*}_{d^hu^k}   d^{k}_{L} 
%                    -            \bar{\nu}_R^{h} Y^{l*}_{e^h\nu^k} e^{k}_{L}\right)\notag\\
%                  &-H_l^-\left(  \bar{d}_R^{ h}  Y^{l*}_{u^hd^k}   u^{k}_{L}
%                                 +\bar{e}_R^{h}  Y^{l*}_{\nu^he^k} \nu^{k}_{L} 
%                                 -\bar{d}_L^{h}  Y_{d^hu^k}^{l}       u^{k}_{R}
%                                 -\bar{e}_L^{h}  Y_{e^h\nu^k}^{l}     \nu^{k}_{R} 
%                                  \right)  
% \end{align}
We are interested in  beyond the SM contributions to the effective Lagrangian, which can explain the  charged current anomalies, i.e., 
\begin{align} \label{EqApp}
\mathcal{L}_{H^\pm}(b\rightarrow c \tau \bar{\nu}_\tau)  
                = &- H^+\left( \bar{c}     X^{D}_{c b}         P_R b
                    -            \bar{c}     X^{U*}_{bc}         P_L b 
                    +            \bar{\nu}   X^{E}_{\nu \tau}    P_R \tau                    
                    -            \bar{\nu}   X^{N*}_{\tau\nu} P_L \tau\right)\notag\\
                  &-H^-\left(   \bar{b}    X^{D*}_{cb}         P_L c
                                  -\bar{b}   X^{U}_{bc}          P_R c
                                +\bar{\tau}  X^{E*}_{\nu \tau}   P_L \nu 
                                -\bar{\tau}  X^{N}_{\tau\nu }  P_R \nu 
                                 \right).  
\end{align}
%The rotations matrices depend on   the form of the Yukawa's, however,  for phenomenological applications is a custom  to choose $U_L^u=1_{3\times 3}$,
%$U_L^D=V_{_{CKM}}^{\dagger}$, $U_L^E=1_{3\times 3}$, $U_L^N=V_{_{PMNS}}^{\dagger}$.
%In the standard model, the right-handed fermions are isospin singlets hence these fields could be redefined to absorb the right-handed diagonalizing matrices $U_{R}^{F}$. Under these assumptions 
%The Yukawa couplings can be written as:  
%\begin{align}
%X_{du}^{U} =&  V_{_{CKM}}^{\dagger} \Y_{2U}, %\hspace{0.5cm}  
%X_{ud}^{D} =          \Y_{2D} \notag\\
%X_{ e \nu}^{N} =&   \Y_{2U}   , %\hspace{1.5cm}    
%X_{\nu e}^{E} = V_{_{PMNS}}^{\dagger}\Y_{2E}      
%\end{align}
%If well it is true that these choices are not unique, they deliver a good approximation compared with concrete models, where it is possible to build the diagonalizing matrices. 

\subsection{Effective Lagrangian}
\label{sec:appendixB}
%--------------------------------------------
Since the quark level the process $\bar{B} \rightarrow D^{(*)}\tau^{-}\bar{\nu}_\tau$ is given by $b \rightarrow c\tau^{-}\bar{\nu}_\tau$, it can be mediated, at tree level, by the additional Higss $H^{-}$ with the following Feynman rules
 \begin{align}
 b_R \longrightarrow c_L, H^{-},\hspace{1cm} -iX^{D}_{cb}\notag\\
 b_L \longrightarrow c_R, H^{-},\hspace{1cm} +iX^{U*}_{bc}\notag\\
      H^{-}\longrightarrow \tau^{-}_R, \bar{\nu}_{\tau L}, \hspace{1cm} -iX^{E*}_{\nu \tau}\notag\\
      H^{-}\longrightarrow \tau^{-}_L, \bar{\nu}_{\tau R}, \hspace{1cm} +iX^{N}_{\tau\nu }.
 \end{align}
The process $b \rightarrow c\tau^{-}\bar{\nu}_\tau\ $ can be realized in four different ways 
\begin{align}
b_R \rightarrow c_L, H^{-}\left(\rightarrow\tau^{-}_R, \bar{\nu}_{\tau L}\right),\hspace{0.5cm}i\mathcal{L}_1= \bar{c}_L\left(-iX^{D}_{cb}\right)b_R\frac{i}{M_{H^{\pm}}^2}\bar{\tau}_R\left(-iX^{E*}_{\nu \tau}\right)\nu_{L}^{(+)}\notag\\
b_R \rightarrow c_L, H^{-}\left(\rightarrow\tau^{-}_L, \bar{\nu}_{\tau R}\right),\hspace{0.5cm}i\mathcal{L}_2= \bar{c}_L\left(-iX^{D}_{cb}\right)b_R\frac{i}{M_{H^{\pm}}^2}\bar{\tau}_L\left(+iX^{N}_{\tau\nu}\right)\nu_{R}^{(+)}\notag\\ 
b_L \rightarrow c_R, H^{-}\left(\rightarrow\tau^{-}_R, \bar{\nu}_{\tau L}\right),\hspace{0.5cm}i\mathcal{L}_3= \bar{c}_R\left(+iX^{U*}_{bc}\right)b_L\frac{i}{M_{H^{\pm}}^2}\bar{\tau}_R\left(-iX^{E*}_{\nu \tau}\right)\nu_{L}^{(+)}\notag\\
b_L \rightarrow c_R, H^{-}\left(\rightarrow\tau^{-}_L, \bar{\nu}_{\tau R}\right),\hspace{0.5cm}i\mathcal{L}_4= \bar{c}_R\left(+iX^{U*}_{bc}\right)b_L\frac{i}{M_{H^{\pm}}^2}\bar{\tau}_L\left(+iX^{N}_{\tau\nu}\right)\nu_{R}^{(+)}.
\end{align}

Thus, the effective Lagrangian for $b \rightarrow c\tau^{-}\bar{\nu}_\tau$ transition is written as 
\bea
\mathcal{L}_{\rm eff}^{H^\pm}(b \to c \tau \bar{\nu}_{\tau}) &=& \frac{4 G_F}{\sqrt{2}} V_{cb}^{\rm CKM}	 \Big[C_{S}^{LL}(\bar{c} P_L b) (\bar{\tau} P_L \nu_{\tau}) + C_{S}^{RL}(\bar{c} P_R b) (\bar{\tau} P_L \nu_{\tau}) \nonumber \\
&& + C_{S}^{LR}(\bar{c} P_L b) (\bar{\tau} P_R \nu_{\tau}) + C_{S}^{RR}(\bar{c} P_R b) (\bar{\tau} P_R \nu_{\tau})\Big], 
\eea

\noindent where

\begin{eqnarray}
C^{LL}_S &=&+\frac{\sqrt{2}}{4G_FV^{\rm CKM}_{cb}}  
\frac{\left(X^{U*}_{bc}\right)\left(X^{E*}_{\nu \tau}\right)}{M_{H^{\pm}}^2} , \\
C^{RL}_S &=& -\frac{\sqrt{2}}{4G_FV^{\rm CKM}_{cb}} 
\frac{\left(X^{D}_{cb}\right)\left(X^{E*}_{\nu \tau}\right)}{M_{H^{\pm}}^2}, \\
C^{LR}_S &=&-\frac{\sqrt{2}}{4G_FV^{\rm CKM}_{cb}}   
\frac{\left(X^{U*}_{bc}\right)\left(X^{N}_{\tau\nu}\right)}{M_{H^{\pm}}^2}, \\
 C^{RR}_S &=& +\frac{\sqrt{2}}{4G_FV^{\rm CKM}_{cb}}  
 \frac{\left(X^{D}_{cb}\right)\left(X^{N}_{\tau\nu}\right)}{M_{H^{\pm}}^2}.
\end{eqnarray}

We can obtain explicit expressions for these coefficients with the results of the appendix~\ref{sec:appendix-b}
\begin{table}
 \begin{center}
 \def\arraystretch{2.0}
 \begin{tabular}{|c|l|l|}
 \hline
     & $\hspace{1.3cm}C^{LL}_S$ & $\hspace{1.3cm}C^{RL}_S$  \\ 
 \hline
I &
$+V_{_{PMNS}}^{\nu_{\tau}\tau}\frac{m_c m_\tau}{M_{H^{\pm}}^2} \cot^2\beta $
&
$-V_{_{PMNS}}^{\tau\nu_{\tau}}\frac{m_b m_\tau}{M_{H^{\pm}}^2} \cot^2\beta $ \\
 \hline
II &
            $-V_{_{PMNS}}^{\tau\nu_{\tau}}\frac{m_c m_\tau}{M_{H^{\pm}}^2}$
&  
            $-V_{_{PMNS}}^{\tau\nu_{\tau}}\frac{m_b m_\tau}{M_{H^{\pm}}^2} \tan^2\beta $ \\ 
 \hline
X & 
            $-V_{_{PMNS}}^{\tau\nu_{\tau}}\frac{m_c m_\tau}{M_{H^{\pm}}^2}$ 
&  
            $+V_{_{PMNS}}^{\tau\nu_{\tau}}\frac{m_b m_\tau}{M_{H^{\pm}}^2}$  \\ 
 \hline
Y & 
$+V_{_{PMNS}}^{\tau\nu_{\tau}}\frac{m_c m_\tau}{M_{H^{\pm}}^2} \cot^2\beta $  
&
            $+V_{_{PMNS}}^{\tau\nu_{\tau}}\frac{m_b m_\tau}{M_{H^{\pm}}^2}$ \\
\hline
 \end{tabular}
  \caption{Wilson coefficients for several 2HDMs in the literature. For a precise definition of these models see  appendix~\ref{sec:appendixc}.}
 \end{center}
 \end{table}
 
\subsection{Couplings for the 2HDMs} \label{sec:appendixc}
%------------------------------------------------------------------------------

From the equation ~(\ref{eq:mathcaly}) we have the following expressions for the 
couplings
\begin{align}\label{eq:yx}
Y_{g^hf^k}^{F} = \left(U^{G}_{L}   \Y^{1F}   U^{F\dagger}_{R}  \right)^{hk}  =
\left(U^{G}_{L} \left[+\cos \beta y^{1F}+\sin \beta y^{2F}\right]U^{F\dagger}_{R}\right)^{hk} \notag\\
 X_{g^hf^k}^{F} = \left(U^{G}_{L}   \Y^{2F}   U^{F\dagger}_{R}  \right)^{hk} =
\left(U^{G}_{L} \left[-\sin \beta y^{1F}+\cos \beta y^{2F}\right] U^{F\dagger}_{R}\right)^{hk}.  
\end{align}
In these expressions  $g$ and $f$ run over the  Yukawa matrices superscripts 
$u$, $d$, $e$  and $\nu$ in such a way that  for a charged Higgs $(g^h,h^k,F)\in \{(u^h,d^k,D), (d^h,u^k,U), (\nu^h,e^k,N),(e^h,\nu^k,E)\}$.  From these definitions the following 2HDM types are well known in the literature
\begin{itemize}
 \item Type I: all masses of quarks and leptons are given by $\Phi_2$, i.e.,  $y^{1D}_{ij}=y^{1U}_{ij}=y^{1E}_{ij}=0$.  
 \item Type II: the up-type quarks obtain their masses from $\Phi_2$, while the down-type quarks and charged leptons from $\Phi_1$,
 i.e., $y^{2D}_{ij}=y^{1U}_{ij}=y^{2E}_{ij}=0$.
 \item Type X: the quarks obtain their masses from $\Phi_2$, and the leptons from $\Phi_1$, i.e., $y^{1D}_{ij}=y^{1U}_{ij}=y^{2E}:{ij}=0$
 \item Type  Y: the down-like quarks obtain their masses from $\Phi_1$ and the remaining SM fermions acquire their masses from $\Phi_2$,
 i.e., $y^{2D}_{ij}=y^{1U}_{ij}=y^{1E}_{ij}=0$.
\end{itemize}
As we will see later these models avoid flavor changing neutral currents~(FCNC), which it is quite convenient for the phenomenological analysis.
\subsection{Neutral Current couplings}
We can obtain  the coupling of the neutral Higgs boson to the SM fermion from Eq.~(\ref{eq:yx})  by doing   $f=g$,  in this case, it is more convenient to define 
 $Y_{hk}^{F}\equiv Y_{f^hf^k}^{F}$
and  $X_{hk}^{F}\equiv X_{f^hf^k}^{F}$
\begin{align}\label{eq:yxnc}
Y_{hk}^{F} = \left(U^{F}_{L}   \Y^{1F}   U^{F\dagger}_{R}  \right)^{hk}  =
\left(U^{F}_{L} \left[+\cos \beta y^{1F}+\sin \beta y^{2F}\right]U^{F\dagger}_{R}\right)^{hk} \notag\\
 X_{hk}^{F} = \left(U^{F}_{L}   \Y^{2F}   U^{F\dagger}_{R}  \right)^{hk} =
\left(U^{F}_{L} \left[-\sin \beta y^{1F}+\cos \beta y^{2F}\right] U^{F\dagger}_{R}\right)^{hk}.  
\end{align}
For some applications is good to put the Yukawa couplings in terms of the fermion mass.
Since $Y^F_{hk}$ is diagonal in the fermion mass eigenstate we can write these Yukawa matrices as follows 
\begin{align}
Y_{hk}^{F}= \frac{\sqrt{2}}{v}m^F_k\delta_{hk},
\end{align}
clearing $U^{F}_{L} y^{1F}U^{f\dagger}_{R}$ from the first expression in Eq.~(\ref{eq:yx}) 
and replacing in the second one, we get   
\begin{align}\label{eq:y2}
 X_{hk}^{F} =
-\frac{\sqrt{2}}{v}\tan\beta m^F_k\delta_{hk}+\frac{1}{\cos\beta}\left(U^{F}_{L} y^{2F}U^{F\dagger}_{R}\right)^{hk},
\end{align}
Contrary case, if  we clear $U^{F}_{L} y^{2F}U^{f\dagger}_{R}$  we get
 \begin{align}\label{eq:y1}
 X_{hk}^{F} =
\frac{\sqrt{2}}{v}\cot\beta m^F_k\delta_{hk}-\frac{1}{\sin\beta}\left(U^{F}_{L} y^{1F}U^{f\dagger}_{R}\right)^{hk}. 
\end{align}
Equations (\ref{eq:y2}) and (\ref{eq:y1}) are equivalent, however, their usefulness depends on the model.
For $h\neq k$ these identities imply that if $y^{2F}$ is zero~(since $\delta_{hk}=0$), the component $(U_{L}^F y^{1F}U_{R}^{F \dagger})_{hk}$ is also zero.
From this result, we can conclude that  $U_{L}^F y^{1F}U_{R}^{F\dagger}$ must be diagonal. This result is also clear fom Eq.~\eqref{eq:yx}.
These results imply that all the 2HDM types mentioned above avoid flavor changing neutral currents~(FCNC).

\subsection{Charged currents}
\label{sec:appendix-b}
In order to explain the $R_D$ and $R_{D*}$ anomalies, flavor violating couplings are needed. We are particularly interested  in the effective interaction Lagrangian between a charged Higgs boson and the standard model  charged currents. In order to determine the explicit form of the charged current matrices $Y_{g^hf^k}^{F}$  and $X_{g^hf^k}^{F}$, it is convenient to put all in terms of the neutral charged currents 
\begin{align}
Y_{g^hf^k}^{F} =& \left(U^{G}_{L}                            \Y^{1F}   U^{F\dagger}_{R}  \right)^{hk}  
               = \left(U^{G}_{L} U^{F\dagger}_{L}U^{F}_{L}  \Y^{1F}   U^{F\dagger}_{R}  \right)^{hk}
               = \left(U^{G}_{L} U^{F\dagger}_{L}  Y_{h k}^{F}  \right)^{hk}.
\end{align}
 In particular cases this expression means
\begin{align}
Y_{g^hf^k}^{F} = &\left(U^{G}_{L} U^{F\dagger}_{L} \frac{\sqrt{2}}{v}{\bf m^F}    \right)^{hk}
= 
\begin{cases}
 \left(V_{_{CKM}} \frac{\sqrt{2}}{v}{\bf m^D}    \right)^{hk}=\frac{\sqrt{2}}{v}(V_{_{CKM}})^{hk}m^{D}_k & \text{for  $G=U$
 and $F=D$} \notag\\
\left(V_{_{CKM}}^{\dagger} \frac{\sqrt{2}}{v}{\bf m^U}    \right)^{hk}=\frac{\sqrt{2}}{v}(V_{_{CKM}}^\dagger)^{hk}m^{U}_k& \text{for  $G=D$
 and $F=U$} \notag\\
 \left(V_{_{PMNS}} \frac{\sqrt{2}}{v}{\bf m^E}    \right)^{hk}=\frac{\sqrt{2}}{v}(V_{_{PMNS}})^{hk}m^{E}_k& \text{for  $G=N$
 and $F=E$} \notag\\ 
  \left(V_{_{PMNS}}^{\dagger} \frac{\sqrt{2}}{v}{\bf m^N}    \right)^{hk}=\frac{\sqrt{2}}{v}(V_{_{PMNS}}^\dagger)^{hk}m^{N}_k& \text{for  $G=E$
 and $F=N$} \notag\\ 
 \end{cases}.
\end{align}
Procceding in identical way for the couplings to the additional Higgs doublet

\begin{align}
X_{g^hf^k}^{F} =&\left(U^{G}_{L} U^{F\dagger}_{L}  X_{h k}^{F}  \right)^{hk}\notag\\
               =&-\tan\beta Y_{g^hf^k}^{F} +\frac{1}{\cos\beta}\left(U^{G}_{L} y_{2F}U^{F\dagger}_{R}\right)^{hk}\notag\\
               =&+\cot\beta Y_{g^hf^k}^{F} -\frac{1}{\sin\beta}\left(U^{F}_{L} y_{1F}U^{F\dagger}_{R}\right)^{hk}.  
\end{align}
This equation allows determining the couplings for each  of the models.
% \begin{center}
% \begin{tabular}{|c|c|c|c|c|}
% \hline
%     & $X_{cb}^{D}$ & $X_{bc}^{U*}$ &$X_{\nu\tau}^{E*}$ &$X_{\tau\nu_\tau}^{N}$\\ 
% \hline
% I   & $+\cot\beta \left(V_{_{CKM}} \frac{\sqrt{2}}{v}{\bf m^D}\right)^{cb}$  
%     & $+\cot\beta \left(V_{_{CKM}}^{\dagger} \frac{\sqrt{2}}{v}{\bf m^U}    \right)^{bc*}$ 
%     & $+\cot\beta \left(V_{_{PMNS}} \frac{\sqrt{2}}{v}{\bf m^E}    \right)^{\nu_\tau\tau *}$ & 0\\ 
% \hline 
% II  & $-\tan\beta \left(V_{_{CKM}} \frac{\sqrt{2}}{v}{\bf m^D}\right)^{cb}$  
%     & $+\cot\beta \left(V_{_{CKM}}^{\dagger} \frac{\sqrt{2}}{v}{\bf m^U}    \right)^{bc*}$ 
%     & $-\tan\beta \left(V_{_{PMNS}} \frac{\sqrt{2}}{v}{\bf m^E}    \right)^{\nu_\tau\tau *}$ & 0\\  
% \hline 
% X   & $+\cot\beta \left(V_{_{CKM}} \frac{\sqrt{2}}{v}{\bf m^D}\right)^{cb}$  
%     & $+\cot\beta \left(V_{_{CKM}}^{\dagger} \frac{\sqrt{2}}{v}{\bf m^U}    \right)^{bc*}$ 
%     & $-\tan\beta \left(V_{_{PMNS}} \frac{\sqrt{2}}{v}{\bf m^E}    \right)^{\nu_\tau\tau *}$ & 0\\    
% \hline  
% Y   & $-\tan\beta \left(V_{_{CKM}} \frac{\sqrt{2}}{v}{\bf m^D}\right)^{cb}$  
%     & $+\cot\beta \left(V_{_{CKM}}^{\dagger} \frac{\sqrt{2}}{v}{\bf m^U}    \right)^{bc*}$ 
%     & $+\cot\beta \left(V_{_{PMNS}} \frac{\sqrt{2}}{v}{\bf m^D}    \right)^{\nu_\tau\tau *}$ & 0\\ 
% \hline  
% \end{tabular}
% \end{center}

%\frac{\sqrt{2}}{v}V_{_{CKM}}^{cb}m_b

\begin{table}
 \begin{center}
 \def\arraystretch{1.5}
 \begin{tabular}{|c|c|c|c|c|}
 \hline
     & $X_{cb}^{D}$ & $X_{bc}^{U*}$ &$X_{\nu\tau}^{E*}$ &$X_{\tau\nu_\tau}^{N}$\\ 
 \hline
 I   & $+\cot\beta \frac{\sqrt{2}}{v}V_{_{CKM}}^{cb}m_b$  
     & $+\cot\beta \frac{\sqrt{2}}{v}V_{_{CKM}}^{cb}m_c$ 
     & $+\cot\beta \frac{\sqrt{2}}{v}V_{_{PMNS}}^{\tau\nu_{\tau}}m_\tau$ & 0\\ 
 \hline 
 II  & $-\tan\beta \frac{\sqrt{2}}{v}V_{_{CKM}}^{cb}m_b$  
     & $+\cot\beta \frac{\sqrt{2}}{v}V_{_{CKM}}^{cb}m_c$ 
     & $-\tan\beta \frac{\sqrt{2}}{v}V_{_{PMNS}}^{\tau\nu_{\tau}}m_\tau$ & 0\\  
 \hline 
 X   & $+\cot\beta \frac{\sqrt{2}}{v}V_{_{CKM}}^{cb}m_b$  
     & $+\cot\beta \frac{\sqrt{2}}{v}V_{_{CKM}}^{cb}m_c$ 
     & $-\tan\beta \frac{\sqrt{2}}{v}V_{_{PMNS}}^{\tau\nu_{\tau}}m_\tau$ & 0\\    
 \hline  
 Y   & $-\tan\beta \frac{\sqrt{2}}{v}V_{_{CKM}}^{cb}m_b$  
     & $+\cot\beta \frac{\sqrt{2}}{v}V_{_{CKM}}^{cb}m_c$ 
     & $+\cot\beta \frac{\sqrt{2}}{v}V_{_{PMNS}}^{\tau\nu_{\tau}}m_\tau$ & 0\\ 
 \hline  
 \end{tabular}
 \end{center}
 \caption{Yukawa couplings for several 2HDMs in the literature.} 
\end{table}

%\bibliographystyle{h-physrev4}
%\bibliographystyle{JHEP-2}
%\bibliographystyle{JHEP}
%\bibliographystyle{utphys}
%\bibliographystyle{apsrev4-1}
%\bibliography{newrefs}

\begin{thebibliography}{99}

%----- Experimental measurements ---------------------

%\cite{Lees:2012xj}
\bibitem{Lees:2012xj} 
  J.~P.~Lees {\it et al.} [BaBar Collaboration],
  Evidence for an excess of $\bar{B} \to D^{(*)} \tau^-\bar{\nu}_\tau$ decays,
  Phys.\ Rev.\ Lett.\  {\bf 109}, 101802 (2012)
  %doi:10.1103/PhysRevLett.109.101802
  [arXiv:1205.5442 [hep-ex]].
  %%CITATION = doi:10.1103/PhysRevLett.109.101802;%%
  %788 citations counted in INSPIRE as of 08 Apr 2020

%\cite{Lees:2013uzd}
\bibitem{Lees:2013uzd} 
  J.~P.~Lees {\it et al.} [BaBar Collaboration],
  Measurement of an Excess of $\bar{B} \to D^{(*)}\tau^- \bar{\nu}_\tau$ Decays and Implications for Charged Higgs Bosons,
  Phys.\ Rev.\ D {\bf 88}, no. 7, 072012 (2013)
  %doi:10.1103/PhysRevD.88.072012
  [arXiv:1303.0571 [hep-ex]].
  %%CITATION = doi:10.1103/PhysRevD.88.072012;%%
  %651 citations counted in INSPIRE as of 08 Apr 2020


%\cite{Huschle:2015rga}
\bibitem{Huschle:2015rga} 
  M.~Huschle {\it et al.} [Belle Collaboration],
  Measurement of the branching ratio of $\bar{B} \to D^{(\ast)} \tau^- \bar{\nu}_\tau$ relative to $\bar{B} \to D^{(\ast)} \ell^- \bar{\nu}_\ell$ decays with hadronic tagging at Belle,
  Phys.\ Rev.\ D {\bf 92}, no. 7, 072014 (2015)
  %doi:10.1103/PhysRevD.92.072014
  [arXiv:1507.03233 [hep-ex]].
  %%CITATION = doi:10.1103/PhysRevD.92.072014;%%
  %601 citations counted in INSPIRE as of 08 Apr 2020


%\cite{Sato:2016svk}
\bibitem{Sato:2016svk} 
  Y.~Sato {\it et al.} [Belle Collaboration],
  Phys.\ Rev.\ D {\bf 94}, no. 7, 072007 (2016)
  %doi:10.1103/PhysRevD.94.072007
  [arXiv:1607.07923 [hep-ex]].
  %%CITATION = doi:10.1103/PhysRevD.94.072007;%%
  %275 citations counted in INSPIRE as of 08 Apr 2020
  
%\cite{Hirose:2017vbz}
\bibitem{Hirose:2017vbz} 
  S.~Hirose [Belle Collaboration],
  $\bar{B} \rightarrow D^{(*)} \tau^- \bar{\nu}_\tau$ and Related Tauonic Topics at Belle,
  arXiv:1705.05100 [hep-ex].
  %%CITATION = ARXIV:1705.05100;%%
  %1 citations counted in INSPIRE as of 08 Apr 2020
  
%\cite{Aaij:2015yra}
\bibitem{Aaij:2015yra} 
  R.~Aaij {\it et al.} [LHCb Collaboration],
  Measurement of the ratio of branching fractions $\mathcal{B}(\bar{B}^0 \to D^{*+}\tau^{-}\bar{\nu}_{\tau})/\mathcal{B}(\bar{B}^0 \to D^{*+}\mu^{-}\bar{\nu}_{\mu})$,
  Phys.\ Rev.\ Lett.\  {\bf 115}, no. 11, 111803 (2015)
  Erratum: [Phys.\ Rev.\ Lett.\  {\bf 115}, no. 15, 159901 (2015)]
  %doi:10.1103/PhysRevLett.115.159901, 10.1103/PhysRevLett.115.111803
  [arXiv:1506.08614 [hep-ex]].
  %%CITATION = doi:10.1103/PhysRevLett.115.159901, 10.1103/PhysRevLett.115.111803;%%
  %743 citations counted in INSPIRE as of 08 Apr 2020


%\cite{Aaij:2017deq}
\bibitem{Aaij:2017deq} 
  R.~Aaij {\it et al.} [LHCb Collaboration],
  Test of Lepton Flavor Universality by the measurement of the $B^0 \to D^{*-} \tau^+ \nu_{\tau}$ branching fraction using three-prong $\tau$ decays,
  Phys.\ Rev.\ D {\bf 97}, no. 7, 072013 (2018)
  %doi:10.1103/PhysRevD.97.072013
  [arXiv:1711.02505 [hep-ex]].
  %%CITATION = doi:10.1103/PhysRevD.97.072013;%%
  %158 citations counted in INSPIRE as of 08 Apr 2020


%\cite{Aaij:2017uff}
\bibitem{Aaij:2017uff} 
  R.~Aaij {\it et al.} [LHCb Collaboration],
  Measurement of the ratio of the $B^0 \to D^{*-} \tau^+ \nu_{\tau}$ and $B^0 \to D^{*-} \mu^+ \nu_{\mu}$ branching fractions using three-prong $\tau$-lepton decays,
  Phys.\ Rev.\ Lett.\  {\bf 120}, no. 17, 171802 (2018)
  %doi:10.1103/PhysRevLett.120.171802
  [arXiv:1708.08856 [hep-ex]].
  %%CITATION = doi:10.1103/PhysRevLett.120.171802;%%
  %189 citations counted in INSPIRE as of 08 Apr 2020

%\cite{Abdesselam:2019dgh}
\bibitem{Abdesselam:2019dgh}
A.~Abdesselam \textit{et al.} [Belle], Measurement of $\mathcal{R}(D)$ and $\mathcal{R}(D^{\ast})$ with a semileptonic tagging method, arXiv:1904.08794 [hep-ex].
%127 citations counted in INSPIRE as of 12 Aug 2020

%\cite{Belle:2019rba}
\bibitem{Belle:2019rba}
G.~Caria \textit{et al.} [Belle Collaboration], Measurement of $\mathcal{R}(D)$ and $\mathcal{R}(D^*)$ with a Semileptonic Tagging Method, 
Phys. Rev. Lett. \textbf{124} (2020) no.16, 161803
%doi:10.1103/PhysRevLett.124.161803
[arXiv:1910.05864 [hep-ex]].

%\cite{Hirose:2017dxl}
\bibitem{Hirose:2017dxl} 
  S.~Hirose {\it et al.} [Belle Collaboration],
  Measurement of the $\tau$ lepton polarization and $R(D^*)$ in the decay $\bar{B} \rightarrow D^* \tau^- \bar{\nu}_\tau$ with one-prong hadronic $\tau$ decays at Belle,
  Phys.\ Rev.\ D {\bf 97}, no. 1, 012004 (2018)
  %doi:10.1103/PhysRevD.97.012004
  [arXiv:1709.00129 [hep-ex]].
  %%CITATION = doi:10.1103/PhysRevD.97.012004;%%
  %113 citations counted in INSPIRE as of 08 Apr 2020


%\cite{Hirose:2016wfn}
\bibitem{Hirose:2016wfn} 
  S.~Hirose {\it et al.} [Belle Collaboration],
  Measurement of the $\tau$ lepton polarization and $R(D^*)$ in the decay $\bar{B} \to D^* \tau^- \bar{\nu}_\tau$,
  Phys.\ Rev.\ Lett.\  {\bf 118}, no. 21, 211801 (2017)
  %doi:10.1103/PhysRevLett.118.211801
  [arXiv:1612.00529 [hep-ex]].
  %%CITATION = doi:10.1103/PhysRevLett.118.211801;%%
  %359 citations counted in INSPIRE as of 08 Apr 2020
  
%\cite{Amhis:2019ckw}
\bibitem{Amhis:2019ckw} 
  Y.~S.~Amhis {\it et al.} [HFLAV Collaboration],
  Averages of $b$-hadron, $c$-hadron, and $\tau$-lepton properties as of 2018,
  arXiv:1909.12524 [hep-ex].
  %%CITATION = ARXIV:1909.12524;%%
  %63 citations counted in INSPIRE as of 08 Apr 2020
  
\bibitem{HFLAVsummer}
For updated results see HFLAV average of $R(D^{(\ast)})$ for Spring 2019 in \url{https://hflav-eos.web.cern.ch/hflav-eos/semi/spring19/html/RDsDsstar/RDRDs.html}.

%\cite{Aaij:2017tyk}
\bibitem{Aaij:2017tyk} 
  R.~Aaij {\it et al.} [LHCb Collaboration],
  Measurement of the ratio of branching fractions $\mathcal{B}(B_c^+\,\to\,J/\psi\tau^+\nu_\tau)$/$\mathcal{B}(B_c^+\,\to\,J/\psi\mu^+\nu_\mu)$,
  Phys.\ Rev.\ Lett.\  {\bf 120}, no. 12, 121801 (2018)
  %doi:10.1103/PhysRevLett.120.121801
  [arXiv:1711.05623 [hep-ex]].
  %%CITATION = doi:10.1103/PhysRevLett.120.121801;%%
  %186 citations counted in INSPIRE as of 08 Apr 2020

%\cite{Abdesselam:2019wbt}
\bibitem{Abdesselam:2019wbt} 
  A.~Abdesselam {\it et al.} [Belle Collaboration],
  Measurement of the $D^{\ast-}$ polarization in the decay $B^0 \to D^{\ast -}\tau^+\nu_{\tau}$,
  arXiv:1903.03102 [hep-ex].
  %%CITATION = ARXIV:1903.03102;%%
  %33 citations counted in INSPIRE as of 08 Apr 2020
  
%--------------------------------------------------------------

%\cite{Watanabe:2017mip}
\bibitem{Watanabe:2017mip} 
  R.~Watanabe,
  New Physics effect on $B_c \to J/\psi \tau\bar\nu$ in relation to the $R_{D^{(*)}}$ anomaly,
  Phys.\ Lett.\ B {\bf 776}, 5 (2018)
  %doi:10.1016/j.physletb.2017.11.016
  [arXiv:1709.08644 [hep-ph]].
  %%CITATION = doi:10.1016/j.physletb.2017.11.016;%%
  %60 citations counted in INSPIRE as of 08 Apr 2020


%\cite{Tanaka:2012nw}
\bibitem{Tanaka:2012nw} 
  M.~Tanaka and R.~Watanabe,
  New physics in the weak interaction of $\bar B\to D^{(*)}\tau\bar\nu$,
  Phys.\ Rev.\ D {\bf 87}, no. 3, 034028 (2013)
  %doi:10.1103/PhysRevD.87.034028
  [arXiv:1212.1878 [hep-ph]].
  %%CITATION = doi:10.1103/PhysRevD.87.034028;%%
  %226 citations counted in INSPIRE as of 08 Apr 2020


%\cite{Alok:2016qyh}
\bibitem{Alok:2016qyh} 
  A.~K.~Alok, D.~Kumar, S.~Kumbhakar and S.~U.~Sankar,
  $D^{*}$ polarization as a probe to discriminate new physics in $\bar{B}\to D^{*} \tau \bar{\nu}$,
  Phys.\ Rev.\ D {\bf 95}, no. 11, 115038 (2017)
  %doi:10.1103/PhysRevD.95.115038
  [arXiv:1606.03164 [hep-ph]].
  %%CITATION = doi:10.1103/PhysRevD.95.115038;%%
  %75 citations counted in INSPIRE as of 08 Apr 2020

%--- Recent R(D) and R(D*) prediction-------------
\bibitem{Bigi:2016mdz} 
D.~Bigi and P.~Gambino, Revisiting $B\to D \ell \nu$, Phys.\ Rev.\ D {\bf 94}, 094008 (2016).
\href{http://arxiv.org/abs/1606.08030}{[arXiv:1606.08030 [hep-ph]]}
%doi:10.1103/PhysRevD.94.094008

\bibitem{Bernlochner:2017jka} 
F.~U.~Bernlochner, Z.~Ligeti, M.~Papucci, and D.~J.~Robinson, Combined analysis of semileptonic $B$ decays to $D$ and $D^*$: $R(D^{(*)})$, $|V_{cb}|$, and new physics, Phys.\ Rev.\ D {\bf 95}, 115008 (2017). \href{http://arxiv.org/abs/1703.05330}{arXiv:1703.05330 [hep-ph]}
%doi:10.1103/PhysRevD.95.115008

\bibitem{Jaiswal:2017rve} 
S.~Jaiswal, S.~Nandi, and S.~K.~Patra, Extraction of $|V_{cb}|$ from $B\to D^{(*)}\ell\nu_\ell$ and the Standard Model predictions of $R(D^{(*)})$, JHEP {\bf 1712}, 060 (2017) \href{http://arxiv.org/abs/1707.09977}{[arXiv:1707.09977 [hep-ph]]}.
 %doi:10.1007/JHEP12(2017)060
 
\bibitem{Bigi:2017jbd}
D.~Bigi, P.~Gambino ,and S.~Schacht, $R(D^*)$, $|V_{cb}|$, and the Heavy Quark Symmetry relations between form factors,   JHEP {\bf 11}, 061 (2017). \href{http://arxiv.org/abs/1707.09509}{[arXiv:1707.09509 [hep-ph]]}
%doi:10.1007/JHEP11(2017)061 

%\cite{Gambino:2019sif}
\bibitem{Gambino:2019sif}
P.~Gambino, M.~Jung and S.~Schacht, The $V_{cb}$ puzzle: An update, Phys. Lett. B \textbf{795}, 386-390 (2019)
%doi:10.1016/j.physletb.2019.06.039
[arXiv:1905.08209 [hep-ph]].
%29 citations counted in INSPIRE as of 28 Jun 2020

\bibitem{Jaiswal:2020wer}
S.~Jaiswal, S.~Nandi and S.~K.~Patra,
%``Updates on SM predictions of $|V_{cb}|$ and $R(D^{*})$ in $B\to D^{*}\ell\nu_\ell$ decays,''
[arXiv:2002.05726 [hep-ph]].
%4 citations counted in INSPIRE as of 28 Jun 2020

%---------------------------------------------------------

%\cite{Kamali:2018bdp}
\bibitem{Kamali:2018bdp} 
  S.~Kamali,
  New physics in inclusive semileptonic $B$ decays including nonperturbative corrections,
  Int.\ J.\ Mod.\ Phys.\ A {\bf 34}, no. 06n07, 1950036 (2019)
  %doi:10.1142/S0217751X19500362
  [arXiv:1811.07393 [hep-ph]].
  %%CITATION = doi:10.1142/S0217751X19500362;%%
  %2 citations counted in INSPIRE as of 08 Apr 2020
  
%\cite{Kou:2018nap}
\bibitem{Kou:2018nap} 
  E.~Kou {\it et al.} [Belle-II Collaboration],
  The Belle II Physics Book,
  PTEP {\bf 2019}, no. 12, 123C01 (2019)
  Erratum: [PTEP {\bf 2020}, no. 2, 029201 (2020)]
  %doi:10.1093/ptep/ptz106, 10.1093/ptep/ptaa008
  [arXiv:1808.10567 [hep-ex]].
  %%CITATION = doi:10.1093/ptep/ptz106, 10.1093/ptep/ptaa008;%%
  %352 citations counted in INSPIRE as of 08 Apr 2020


%\cite{Asadi:2019xrc}
\bibitem{Asadi:2019xrc} 
  P.~Asadi and D.~Shih,
  Maximizing the Impact of New Physics in $b\rightarrow c \tau \nu$ Anomalies,
  Phys.\ Rev.\ D {\bf 100}, no. 11, 115013 (2019)
  %doi:10.1103/PhysRevD.100.115013
  [arXiv:1905.03311 [hep-ph]].
  %%CITATION = doi:10.1103/PhysRevD.100.115013;%%
  %8 citations counted in INSPIRE as of 08 Apr 2020

%\cite{Murgui:2019czp}
\bibitem{Murgui:2019czp} 
  C.~Murgui, A.~Peñuelas, M.~Jung and A.~Pich,
  Global fit to $b \to c \tau \nu$ transitions,
  JHEP {\bf 1909}, 103 (2019)
  %doi:10.1007/JHEP09(2019)103
  [arXiv:1904.09311 [hep-ph]].
  %%CITATION = doi:10.1007/JHEP09(2019)103;%%
  %41 citations counted in INSPIRE as of 08 Apr 2020

%\cite{Mandal:2020htr}
\bibitem{Mandal:2020htr}
R.~Mandal, C.~Murgui, A.~Peñuelas and A.~Pich, The role of right-handed neutrinos in $b \to c \tau \bar{\nu}$ anomalies, JHEP \textbf{08}, 022 (2020)
%doi:10.1007/JHEP08(2020)022
[arXiv:2004.06726 [hep-ph]].
%7 citations counted in INSPIRE as of 22 Aug 2020

  %\cite{Cheung:2020sbq}
\bibitem{Cheung:2020sbq} 
  K.~Cheung, Z.~R.~Huang, H.~D.~Li, C.~D.~Lü, Y.~N.~Mao and R.~Y.~Tang,
  Revisit to the $b\to c\tau\nu$ transition: in and beyond the SM,
  arXiv:2002.07272 [hep-ph].
  
%\cite{Sahoo:2019hbu}
\bibitem{Sahoo:2019hbu} 
  S.~Sahoo and R.~Mohanta,
  Investigating the role of new physics in $b \to c \tau \bar \nu_\tau$ transitions,
  arXiv:1910.09269 [hep-ph].
  %%CITATION = ARXIV:1910.09269;%%
  %2 citations counted in INSPIRE as of 08 Apr 2020


%\cite{Shi:2019gxi}
\bibitem{Shi:2019gxi} 
  R.~X.~Shi, L.~S.~Geng, B.~Grinstein, S.~Jäger and J.~Martin Camalich,
  Revisiting the new-physics interpretation of the $b\to c\tau\nu$ data,
  JHEP {\bf 1912}, 065 (2019)
  %doi:10.1007/JHEP12(2019)065
  [arXiv:1905.08498 [hep-ph]].
  %%CITATION = doi:10.1007/JHEP12(2019)065;%%
  %23 citations counted in INSPIRE as of 08 Apr 2020
  
  \bibitem{Bardhan:2019ljo}
  D.~Bardhan and D.~Ghosh,
  $B$ -meson charged current anomalies: The post-Moriond 2019 status,
  Phys.\ Rev.\ D {\bf 100} (2019) no.1,  011701
  %doi:10.1103/PhysRevD.100.011701
  [arXiv:1904.10432 [hep-ph]].
 
 
%\cite{Blanke:2018yud,Blanke:2019qrx}
\bibitem{Blanke:2018yud} 
M.~Blanke, A.~Crivellin, S.~de Boer, T.~Kitahara, M.~Moscati, U.~Nierste and I.~Nisandzic, Impact of polarization observables and $ B_c\to \tau \nu$ on new physics explanations of the $b\to c \tau \nu$ anomaly,'
Phys. Rev. D \textbf{99}, no.7, 075006 (2019)
%doi:10.1103/PhysRevD.99.075006
[arXiv:1811.09603 [hep-ph]].

\bibitem{Blanke:2019qrx}
M.~Blanke, A.~Crivellin, T.~Kitahara, M.~Moscati, U.~Nierste and I.~Nisandzic, Addendum to “Impact of polarization observables and $B_c\to \tau \nu$ on new physics explanations of the $b\to c \tau \nu$ anomaly",
Phys. Rev. D \textbf{100}, 035035 (2019)
%doi:10.1103/PhysRevD.100.035035
[arXiv:1905.08253 [hep-ph]].


%\cite{Alok:2019uqc}
\bibitem{Alok:2019uqc}
A.~K.~Alok, D.~Kumar, S.~Kumbhakar and S.~Uma Sankar, Solutions to $R_D$-$R_{D^*}$ in light of Belle 2019 data,
Nucl. Phys. B \textbf{953}, 114957 (2020)
%doi:10.1016/j.nuclphysb.2020.114957
[arXiv:1903.10486 [hep-ph]].
%18 citations counted in INSPIRE as of 25 Jun 2020

%\cite{Huang:2018nnq}
\bibitem{Huang:2018nnq}
Z.~R.~Huang, Y.~Li, C.~D.~Lu, M.~A.~Paracha and C.~Wang, Footprints of New Physics in $b\to c\tau\nu$ Transitions, Phys. Rev. D \textbf{98}, no.9, 095018 (2018)
%doi:10.1103/PhysRevD.98.095018
[arXiv:1808.03565 [hep-ph]].
%50 citations counted in INSPIRE as of 25 Jun 2020

%---------------------------------------------------
%\cite{Crivellin:2012ye}
\bibitem{Crivellin:2012ye} 
  A.~Crivellin, C.~Greub and A.~Kokulu,
  Explaining $B\to D\tau\nu$, $B\to D^*\tau\nu$ and $B\to \tau\nu$ in a 2HDM of type III,
  Phys.\ Rev.\ D {\bf 86}, 054014 (2012)
  %doi:10.1103/PhysRevD.86.054014
  [arXiv:1206.2634 [hep-ph]].
  %%CITATION = doi:10.1103/PhysRevD.86.054014;%%
  %263 citations counted in INSPIRE as of 08 Apr 2020

%\cite{Crivellin:2013wna}
\bibitem{Crivellin:2013wna} 
  A.~Crivellin, A.~Kokulu and C.~Greub,
  Flavor-phenomenology of two-Higgs-doublet models with generic Yukawa structure,
  Phys.\ Rev.\ D {\bf 87}, no. 9, 094031 (2013)
  %doi:10.1103/PhysRevD.87.094031
  [arXiv:1303.5877 [hep-ph]].
  %%CITATION = doi:10.1103/PhysRevD.87.094031;%%
  %217 citations counted in INSPIRE as of 08 Apr 2020

%\cite{Celis:2012dk}
\bibitem{Celis:2012dk} 
  A.~Celis, M.~Jung, X.~Q.~Li and A.~Pich,
  Sensitivity to charged scalars in $\boldsymbol{B\to D^{(*)}\tau\nu_\tau}$ and $\boldsymbol{B\to\tau\nu_\tau}$ decays,
  JHEP {\bf 1301}, 054 (2013)
  %doi:10.1007/JHEP01(2013)054
  [arXiv:1210.8443 [hep-ph]].
  %%CITATION = doi:10.1007/JHEP01(2013)054;%%
  %228 citations counted in INSPIRE as of 08 Apr 2020

%\cite{Celis:2016azn}
\bibitem{Celis:2016azn} 
  A.~Celis, M.~Jung, X.~Q.~Li and A.~Pich,
  Scalar contributions to $b\to c (u) \tau \nu$ transitions,
  Phys.\ Lett.\ B {\bf 771}, 168 (2017)
  %doi:10.1016/j.physletb.2017.05.037
  [arXiv:1612.07757 [hep-ph]].
  %%CITATION = doi:10.1016/j.physletb.2017.05.037;%%
  %107 citations counted in INSPIRE as of 08 Apr 2020


%\cite{Ko:2012sv}
\bibitem{Ko:2012sv} 
  P.~Ko, Y.~Omura and C.~Yu,
  $B \to D^{*} \tau \nu$ and $B \to \tau \nu$ in chiral U(1)' models with flavored multi Higgs doublets,
  JHEP {\bf 1303}, 151 (2013)
  %doi:10.1007/JHEP03(2013)151
  [arXiv:1212.4607 [hep-ph]].
  %%CITATION = doi:10.1007/JHEP03(2013)151;%%
  %66 citations counted in INSPIRE as of 08 Apr 2020

%\cite{HernandezSanchez:2012eg}
\bibitem{HernandezSanchez:2012eg} 
  J.~Hernandez-Sanchez, S.~Moretti, R.~Noriega-Papaqui and A.~Rosado,
  Off-diagonal terms in Yukawa textures of the Type-III 2-Higgs doublet model and light charged Higgs boson phenomenology,
  JHEP {\bf 1307}, 044 (2013)
  %doi:10.1007/JHEP07(2013)044
  [arXiv:1212.6818 [hep-ph]].
  %%CITATION = doi:10.1007/JHEP07(2013)044;%%
  %32 citations counted in INSPIRE as of 08 Apr 2020

%\cite{Crivellin:2015hha}
\bibitem{Crivellin:2015hha} 
  A.~Crivellin, J.~Heeck and P.~Stoffer,
  A perturbed lepton-specific two-Higgs-doublet model facing experimental hints for physics beyond the Standard Model,
  Phys.\ Rev.\ Lett.\  {\bf 116}, no. 8, 081801 (2016)
  %doi:10.1103/PhysRevLett.116.081801
  [arXiv:1507.07567 [hep-ph]].
  %%CITATION = doi:10.1103/PhysRevLett.116.081801;%%
  %152 citations counted in INSPIRE as of 08 Apr 2020


%\cite{Cline:2015lqp}
\bibitem{Cline:2015lqp} 
  J.~M.~Cline,
  Scalar doublet models confront $\tau$ and $b$ anomalies,
  Phys.\ Rev.\ D {\bf 93}, no. 7, 075017 (2016)
  %doi:10.1103/PhysRevD.93.075017
  [arXiv:1512.02210 [hep-ph]].
  %%CITATION = doi:10.1103/PhysRevD.93.075017;%%
  %45 citations counted in INSPIRE as of 08 Apr 2020

%\cite{Enomoto:2015wbn}
\bibitem{Enomoto:2015wbn} 
  T.~Enomoto and R.~Watanabe,
  Flavor constraints on the Two Higgs Doublet Models of Z$_{2}$ symmetric and aligned types,
  JHEP {\bf 1605}, 002 (2016)
  %doi:10.1007/JHEP05(2016)002
  [arXiv:1511.05066 [hep-ph]].
  %%CITATION = doi:10.1007/JHEP05(2016)002;%%
  %61 citations counted in INSPIRE as of 08 Apr 2020


%\cite{Dhargyal:2016eri}
\bibitem{Dhargyal:2016eri} 
  L.~Dhargyal,
  $R(D^{(*)})$ and $\mathcal{B}r(B \rightarrow \tau\nu_{\tau})$ in a Flipped/Lepton-Specific 2HDM with anomalously enhanced charged Higgs coupling to $\tau$/b,
  Phys.\ Rev.\ D {\bf 93}, no. 11, 115009 (2016)
  %doi:10.1103/PhysRevD.93.115009
  [arXiv:1605.02794 [hep-ph]].
  %%CITATION = doi:10.1103/PhysRevD.93.115009;%%
  %11 citations counted in INSPIRE as of 08 Apr 2020

%\cite{Martinez:2018ynq}
\bibitem{Martinez:2018ynq} 
  R.~Martinez, C.~F.~Sierra and G.~Valencia,
  Beyond $\mathcal{R}(D^{(*)})$ with the general type-III 2HDM for $b\to c\tau\nu$,
  Phys.\ Rev.\ D {\bf 98}, no. 11, 115012 (2018)
  %doi:10.1103/PhysRevD.98.115012
  [arXiv:1805.04098 [hep-ph]].
  %%CITATION = doi:10.1103/PhysRevD.98.115012;%%
  %18 citations counted in INSPIRE as of 08 Apr 2020
  
%\cite{Wang:2016ggf}
\bibitem{Wang:2016ggf} 
  L.~Wang, J.~M.~Yang and Y.~Zhang,
  Probing a pseudoscalar at the LHC in light of $R(D^{(*)})$ and muon g-2 excesses,
  Nucl.\ Phys.\ B {\bf 924}, 47 (2017)
  %doi:10.1016/j.nuclphysb.2017.09.002
  [arXiv:1610.05681 [hep-ph]].
  %%CITATION = doi:10.1016/j.nuclphysb.2017.09.002;%%
  %21 citations counted in INSPIRE as of 08 Apr 2020

%\cite{Chen:2017eby}
\bibitem{Chen:2017eby} 
  C.~H.~Chen and T.~Nomura,
  Charged-Higgs on $R_{D^{(*)}}$, $\tau$ polarization, and FBA,
  Eur.\ Phys.\ J.\ C {\bf 77}, no. 9, 631 (2017)
  %doi:10.1140/epjc/s10052-017-5198-6
  [arXiv:1703.03646 [hep-ph]].
  %%CITATION = doi:10.1140/epjc/s10052-017-5198-6;%%
  %30 citations counted in INSPIRE as of 08 Apr 2020

%\cite{Iguro:2018fni}
\bibitem{Iguro:2018fni} 
  S.~Iguro, Y.~Omura and M.~Takeuchi,
  Test of the $R(D^{(*)})$ anomaly at the LHC,
  Phys.\ Rev.\ D {\bf 99}, no. 7, 075013 (2019)
  %doi:10.1103/PhysRevD.99.075013
  [arXiv:1810.05843 [hep-ph]].
  %%CITATION = doi:10.1103/PhysRevD.99.075013;%%
  %20 citations counted in INSPIRE as of 08 Apr 2020
  

%\cite{Iguro:2017ysu}
\bibitem{Iguro:2017ysu} 
  S.~Iguro and K.~Tobe,
  $R(D^{(*)})$ in a general two Higgs doublet model,
  Nucl.\ Phys.\ B {\bf 925}, 560 (2017)
  %doi:10.1016/j.nuclphysb.2017.10.014
  [arXiv:1708.06176 [hep-ph]].
  %%CITATION = doi:10.1016/j.nuclphysb.2017.10.014;%%
  %55 citations counted in INSPIRE as of 08 Apr 2020


%\cite{Arbey:2017gmh}
\bibitem{Arbey:2017gmh} 
  A.~Arbey, F.~Mahmoudi, O.~Stal and T.~Stefaniak,
  Status of the Charged Higgs Boson in Two Higgs Doublet Models,
  Eur.\ Phys.\ J.\ C {\bf 78}, no. 3, 182 (2018)
  %doi:10.1140/epjc/s10052-018-5651-1
  [arXiv:1706.07414 [hep-ph]].
  %%CITATION = doi:10.1140/epjc/s10052-018-5651-1;%%
  %76 citations counted in INSPIRE as of 08 Apr 2020


%\cite{Chen:2018hqy}
\bibitem{Chen:2018hqy} 
  C.~H.~Chen and T.~Nomura,
  Charged Higgs boson contribution to $B^-_{q} \to \ell \bar \nu$ and $\bar B\to (P, V) \ell \bar\nu$ in a generic two-Higgs doublet model,
  Phys.\ Rev.\ D {\bf 98}, no. 9, 095007 (2018)
  %doi:10.1103/PhysRevD.98.095007
  [arXiv:1803.00171 [hep-ph]].
  %%CITATION = doi:10.1103/PhysRevD.98.095007;%%
  %16 citations counted in INSPIRE as of 08 Apr 2020

%\cite{Hagiwara:2014tsa}
\bibitem{Hagiwara:2014tsa} 
  K.~Hagiwara, M.~M.~Nojiri and Y.~Sakaki,  $CP$ violation in $B \to D\tau \nu_{\tau}$ using multipion tau decays,
  Phys.\ Rev.\ D {\bf 89}, no. 9, 094009 (2014)
  %doi:10.1103/PhysRevD.89.094009
  [arXiv:1403.5892 [hep-ph]].
  %%CITATION = doi:10.1103/PhysRevD.89.094009;%%
  %28 citations counted in INSPIRE as of 08 Apr 2020

%\cite{Lee:2017kbi}
\bibitem{Lee:2017kbi}
J.~P.~Lee, $B\to D^{(*)}\tau\nu_\tau$ in the 2HDM with an anomalous $\tau$ coupling, 
Phys. Rev. D \textbf{96}, no.5, 055005 (2017)
%doi:10.1103/PhysRevD.96.055005
[arXiv:1705.02465 [hep-ph]].
%7 citations counted in INSPIRE as of 24 Jun 2020

 %\cite{Iguro:2018qzf}
\bibitem{Iguro:2018qzf} 
  S.~Iguro and Y.~Omura,
  Status of the semileptonic $B$ decays and muon g-2 in general 2HDMs with right-handed neutrinos,
  JHEP {\bf 1805}, 173 (2018)
  %doi:10.1007/JHEP05(2018)173
  [arXiv:1802.01732 [hep-ph]].

%\cite{Li:2018rax}
\bibitem{Li:2018rax} 
  S.~P.~Li, X.~Q.~Li, Y.~D.~Yang and X.~Zhang,
  $ {R}_{D^{\left(*\right)}},{R}_{K^{\left(*\right)}} $ and neutrino mass in the 2HDM-III with right-handed neutrinos,
  JHEP {\bf 1809}, 149 (2018)
  %doi:10.1007/JHEP09(2018)149
  [arXiv:1807.08530 [hep-ph]].
  %%CITATION = doi:10.1007/JHEP09(2018)149;%%
  %26 citations counted in INSPIRE as of 08 Apr 2020

%----------------------------------------------------------------------

%\cite{Alonso:2016oyd}
\bibitem{Alonso:2016oyd} 
  R.~Alonso, B.~Grinstein and J.~Martin Camalich,
  Lifetime of $B_c^-$ Constrains Explanations for Anomalies in  $B\to D^{(*)}\tau\nu$,
  Phys.\ Rev.\ Lett.\  {\bf 118}, no. 8, 081802 (2017)
  %doi:10.1103/PhysRevLett.118.081802
  [arXiv:1611.06676 [hep-ph]].
  %%CITATION = doi:10.1103/PhysRevLett.118.081802;%%
  %146 citations counted in INSPIRE as of 08 Apr 2020

\bibitem{Akeroyd:2017mhr} 
  A.~G.~Akeroyd and C.~H.~Chen,
  Constraint on the branching ratio of $B_c \to \tau \bar{\nu}$ from LEP1 and consequences for $R(D^{(*)})$ anomaly,
  Phys.\ Rev.\ D {\bf 96}, no. 7, 075011 (2017)
  %doi:10.1103/PhysRevD.96.075011
  [arXiv:1708.04072 [hep-ph]].

%\cite{Iguro:2018vqb}
\bibitem{Iguro:2018vqb} 
  S.~Iguro, T.~Kitahara, Y.~Omura, R.~Watanabe and K.~Yamamoto,
  D$^{*}$ polarization vs. $ {R}_{D^{\left(\ast \right)}} $ anomalies in the leptoquark models,
  JHEP {\bf 1902}, 194 (2019)
  %doi:10.1007/JHEP02(2019)194
  [arXiv:1811.08899 [hep-ph]].
  %%CITATION = doi:10.1007/JHEP02(2019)194;%%
  %29 citations counted in INSPIRE as of 08 Apr 2020

%\cite{Asadi:2018wea}
\bibitem{Asadi:2018wea} 
  P.~Asadi, M.~R.~Buckley and D.~Shih,
  It’s all right(-handed neutrinos): a new W$^{'}$ model for the $ {R}_{D^{{\left(\ast \right)}}} $ anomaly,
  JHEP {\bf 1809}, 010 (2018)
  %doi:10.1007/JHEP09(2018)010
  [arXiv:1804.04135 [hep-ph]].
  %%CITATION = doi:10.1007/JHEP09(2018)010;%%
  %47 citations counted in INSPIRE as of 08 Apr 2020



%\cite{Asadi:2018sym}
\bibitem{Asadi:2018sym} 
  P.~Asadi, M.~R.~Buckley and D.~Shih,
  Asymmetry Observables and the Origin of $R_{D^{(*)}}$ Anomalies,
  Phys.\ Rev.\ D {\bf 99}, no. 3, 035015 (2019)
  %doi:10.1103/PhysRevD.99.035015
  [arXiv:1810.06597 [hep-ph]].
  %%CITATION = doi:10.1103/PhysRevD.99.035015;%%
  %17 citations counted in INSPIRE as of 08 Apr 2020
  
\bibitem{Ligeti:2016npd}
  Z.~Ligeti, M.~Papucci and D.~J.~Robinson,
  New Physics in the Visible Final States of $B\to D^{(*)}\tau\nu$,
  JHEP {\bf 1701} (2017) 083
  %doi:10.1007/JHEP01(2017)083
  [arXiv:1610.02045 [hep-ph]].


\bibitem{Robinson:2018gza}
  D.~J.~Robinson, B.~Shakya and J.~Zupan,
  Right-handed neutrinos and $R(D^{(*)})$,
  JHEP {\bf 1902} (2019) 119
  %doi:10.1007/JHEP02(2019)119
  [arXiv:1807.04753 [hep-ph]].

\bibitem{Greljo:2018ogz}
  A.~Greljo, D.~J.~Robinson, B.~Shakya and J.~Zupan,
  $R(D^{(*)})$ from W$^{\prime}$ and right-handed neutrinos,
  JHEP {\bf 1809} (2018) 169
  %doi:10.1007/JHEP09(2018)169
  [arXiv:1804.04642 [hep-ph]].
  
  \bibitem{Azatov:2018kzb}
  A.~Azatov, D.~Barducci, D.~Ghosh, D.~Marzocca and L.~Ubaldi,
  Combined explanations of B-physics anomalies: the sterile neutrino solution,
  JHEP {\bf 1810} (2018) 092
  %doi:10.1007/JHEP10(2018)092
  [arXiv:1807.10745 [hep-ph]].
  
  \bibitem{Heeck:2018ntp}
  J.~Heeck and D.~Teresi,
  Pati-Salam explanations of the B-meson anomalies,
  JHEP {\bf 1812} (2018) 103
  %doi:10.1007/JHEP12(2018)103
  [arXiv:1808.07492 [hep-ph]].
  
  \bibitem{Babu:2018vrl}
  K.~S.~Babu, B.~Dutta and R.~N.~Mohapatra,
  A theory of R(D$^{*}$, D) anomaly with right-handed currents,
  JHEP {\bf 1901} (2019) 168
  %doi:10.1007/JHEP01(2019)168
  [arXiv:1811.04496 [hep-ph]].
   
  \bibitem{He:2017bft}
  X.~G.~He and G.~Valencia,
  Lepton universality violation and right-handed currents in $b \to c \tau \nu$,
  Phys.\ Lett.\ B {\bf 779} (2018) 52
  %doi:10.1016/j.physletb.2018.01.073
  [arXiv:1711.09525 [hep-ph]].
  
  \bibitem{Gomez:2019xfw}
  J.~D.~Gómez, N.~Quintero and E.~Rojas,
  Charged current $b \to c \tau \bar{\nu}_\tau$ anomalies in a general $W^\prime$ boson scenario,
  Phys.\ Rev.\ D {\bf 100} (2019) no.9,  093003
  %doi:10.1103/PhysRevD.100.093003
  [arXiv:1907.08357 [hep-ph]].
  
  \bibitem{Alguero:2020ukk} 
  M.~Algueró, S.~Descotes-Genon, J.~Matias and M.~Novoa Brunet,
  Symmetries in $B \to D^* \ell \nu$ angular observables,
  arXiv:2003.02533 [hep-ph].

%\cite{Dutta:2013qaa}
\bibitem{Dutta:2013qaa}
R.~Dutta, A.~Bhol and A.~K.~Giri, Effective theory approach to new physics in $b \to u$ and $b \to c$ leptonic and semileptonic decays,
Phys. Rev. D \textbf{88}, no.11, 114023 (2013)
%doi:10.1103/PhysRevD.88.114023
[arXiv:1307.6653 [hep-ph]].
%47 citations counted in INSPIRE as of 25 Jun 2020

%\cite{Dutta:2017xmj}
\bibitem{Dutta:2017xmj}
R.~Dutta and A.~Bhol, $B_c \to (J/\psi,\,\eta_c)\tau\nu$ semileptonic decays within the standard model and beyond,''
Phys. Rev. D \textbf{96}, no.7, 076001 (2017)
%doi:10.1103/PhysRevD.96.076001
[arXiv:1701.08598 [hep-ph]].
%56 citations counted in INSPIRE as of 25 Jun 2020

%\cite{Dutta:2017wpq}
\bibitem{Dutta:2017wpq}
R.~Dutta, Exploring $R_D$, $R_{D^{\ast}}$ and $R_{J/\Psi}$ anomalies,
[arXiv:1710.00351 [hep-ph]].
%29 citations counted in INSPIRE as of 25 Jun 2020

%\cite{Dutta:2018jxz}
\bibitem{Dutta:2018jxz}
R.~Dutta and N.~Rajeev, Signature of lepton flavor universality violation in $B_s \to D_s\tau\nu$  semileptonic decays,
Phys. Rev. D \textbf{97}, no.9, 095045 (2018)
%doi:10.1103/PhysRevD.97.095045
[arXiv:1803.03038 [hep-ph]].
%12 citations counted in INSPIRE as of 25 Jun 2020

\bibitem{PDG2020}
 P. A. Zyla \textit{et al.} [Particle Data Group], Review of Particle Physics, to be published in Prog. Theor. Exp. Phys. 2020, 083C01 (2020).


%\cite{Tanaka:1994ay}
\bibitem{Tanaka:1994ay} 
  M.~Tanaka,
  Charged Higgs effects on exclusive semitauonic $B$ decays,
  Z.\ Phys.\ C {\bf 67}, 321 (1995)
  %doi:10.1007/BF01571294
  [hep-ph/9411405].
  %%CITATION = doi:10.1007/BF01571294;%%
  %169 citations counted in INSPIRE as of 08 Apr 2020


%\cite{Tanaka:2010se}
\bibitem{Tanaka:2010se} 
  M.~Tanaka and R.~Watanabe,
  Tau longitudinal polarization in $B \to D \tau \nu$ and its role in the search for charged Higgs boson,
  Phys.\ Rev.\ D {\bf 82}, 034027 (2010)
  %doi:10.1103/PhysRevD.82.034027
  [arXiv:1005.4306 [hep-ph]].
  %%CITATION = doi:10.1103/PhysRevD.82.034027;%%
  %115 citations counted in INSPIRE as of 08 Apr 2020


%\cite{Hou:1992sy}
\bibitem{Hou:1992sy} 
  W.~S.~Hou,
  Enhanced charged Higgs boson effects in $B^- \to \tau \bar{\nu}, \mu \bar{\nu}$ and $b \to \tau \bar{\nu} + X$,
  Phys.\ Rev.\ D {\bf 48}, 2342 (1993).
  %doi:10.1103/PhysRevD.48.2342
  %%CITATION = doi:10.1103/PhysRevD.48.2342;%%
  %463 citations counted in INSPIRE as of 08 Apr 2020




%\cite{Kamenik:2008tj}
\bibitem{Kamenik:2008tj} 
  J.~F.~Kamenik and F.~Mescia,
  $B \to  D \tau \nu$ Branching Ratios: Opportunity for Lattice QCD and Hadron Colliders,
  Phys.\ Rev.\ D {\bf 78}, 014003 (2008)
  %doi:10.1103/PhysRevD.78.014003
  [arXiv:0802.3790 [hep-ph]].
  %%CITATION = doi:10.1103/PhysRevD.78.014003;%%
  %171 citations counted in INSPIRE as of 08 Apr 2020


%\cite{Nierste:2008qe}
\bibitem{Nierste:2008qe} 
  U.~Nierste, S.~Trine and S.~Westhoff,
  Charged-Higgs effects in a new B ---> D tau nu differential decay distribution,
  Phys.\ Rev.\ D {\bf 78}, 015006 (2008)
  %doi:10.1103/PhysRevD.78.015006
  [arXiv:0801.4938 [hep-ph]].
  %%CITATION = doi:10.1103/PhysRevD.78.015006;%%
  %160 citations counted in INSPIRE as of 08 Apr 2020


%\cite{Fajfer:2012vx}
\bibitem{Fajfer:2012vx} 
  S.~Fajfer, J.~F.~Kamenik and I.~Nisandzic,
  On the $B \to D^* \tau \bar \nu_{\tau}$ Sensitivity to New Physics,
  Phys.\ Rev.\ D {\bf 85}, 094025 (2012)
  %doi:10.1103/PhysRevD.85.094025
  [arXiv:1203.2654 [hep-ph]].
  %%CITATION = doi:10.1103/PhysRevD.85.094025;%%
  %486 citations counted in INSPIRE as of 08 Apr 2020

%\cite{Misiak:2017bgg}
\bibitem{Misiak:2017bgg} 
  M.~Misiak and M.~Steinhauser,
  Weak radiative decays of the B meson and bounds on $M_{H^\pm }$ in the Two-Higgs-Doublet Model,
  Eur.\ Phys.\ J.\ C {\bf 77}, no. 3, 201 (2017)
  %doi:10.1140/epjc/s10052-017-4776-y
  [arXiv:1702.04571 [hep-ph]].
  %%CITATION = doi:10.1140/epjc/s10052-017-4776-y;%%
  %149 citations counted in INSPIRE as of 08 Apr 2020


%---------------------------------------------------------
\end{thebibliography}
%\newpage

\end{document}